\documentclass[10pt,aps,prd,reprint,nofootinbib,superscriptaddress,preprintnumbers]{revtex4-2}

\usepackage{amsmath}
\usepackage{graphicx}
\usepackage[usenames,dvipsnames]{xcolor}
\usepackage[colorlinks,pdfusetitle,breaklinks,linktocpage=true]{hyperref}
\hypersetup{colorlinks=true,allcolors=[rgb]{1,0.56,0}}
\usepackage[capitalise]{cleveref}
\usepackage{float}
\usepackage{orcidlink}
\usepackage{xspace}
\usepackage{tabularx}
\usepackage{booktabs}
\usepackage[dvipsnames]{xcolor}

\Crefname{appsec}{Appendix}{Appendices}

\DeclareMathOperator{\sign}{sgn}

\newcommand{\snudd}{\texttt{SNuDD}\xspace}
\newcommand{\cevns}{CE$\nu$NS\xspace}
\newcommand{\eves}{E$\nu$ES\xspace}
\newcommand{\tnyr}{{\rm tonne\text{-}years}\xspace}
\newcommand{\usim}{\mathord{\sim}}
\newcommand{\nsi}{\mathcal{E}}
\newcommand{\dd}{\mathop{}\!\mathrm{d}}

\date{\today}

\begin{document}

\title{Direct Detection of Solar Neutrinos to Probe Tensions and Break Degeneracies in Long-Baseline Neutrino Experiments}

\author{Dorian W.~P.~Amaral\,\orcidlink{0000-0002-1414-932X}}
\email{damaral@ifae.es}
\affiliation{Institut de F\`{ı}sica d’Altes Energies (IFAE), The Barcelona Institute of Science and Technology,
Campus UAB, 08193 Bellaterra (Barcelona), Spain}

\author{David Cerde\~no\,\orcidlink{0000-0002-7649-1956}}
\email{davidg.cerdeno@ift.csic.es}
\affiliation{Instituto de F\'isica Te\'orica IFT-UAM/CSIC, 
Cantoblanco, E-28049, Madrid, Spain}

\author{Andrew Cheek\,\orcidlink{0000-0002-8773-831X}}
\email{acheek@sjtu.edu.cn}
\affiliation{Tsung-Dao Lee Institute \& School of Physics and Astronomy, Shanghai Jiao Tong
University, Shanghai 200240, China}
\affiliation{Key Laboratory for Particle Astrophysics and Cosmology (MOE) \& Shanghai Key Laboratory for Particle Physics and Cosmology, Shanghai Jiao Tong University, Shanghai 200240, China}

\author{Valeria Costa\,\orcidlink{0009-0000-8129-5741}}
\email{valeria.costa@ift.csic.es}
\affiliation{Instituto de F\'isica Te\'orica IFT-UAM/CSIC, 
Cantoblanco, E-28049, Madrid, Spain}
\affiliation{Departamento de Física Te\'orica, Universidad Aut\'onoma de Madrid, Cantoblanco, E-28049, Madrid, Spain}

\author{Peter B.~Denton\,\orcidlink{0000-0002-5209-872X}}
\email{pdenton@bnl.gov}
\affiliation{High Energy Theory Group, Physics Division, Brookhaven National Laboratory, Upton, NY 11973, USA}

\author{Patrick Foldenauer\,\orcidlink{0000-0003-4334-4228}}
\email{patrick.foldenauer@csic.es}
\affiliation{Instituto de F\'isica Te\'orica IFT-UAM/CSIC, 
Cantoblanco, E-28049, Madrid, Spain}

\begin{abstract}
Long-baseline (LBL) accelerator experiments are sensitive not only to the standard neutrino oscillation parameters but also to new physics. 
Currently, there exists a tension between NOvA and T2K data that can be explained by vector non-standard neutrino interactions (NSI) with complex couplings, which in turn introduce degeneracies in LBL measurements. We demonstrate that dark matter direct detection experiments can use their measurements of solar neutrinos to test the NSI resolution of this tension and to break these degeneracies. 
Using recent solar neutrino data from LZ, XENONnT, and PandaX-4T, we show that dark matter direct detection experiments already constrain part of the NSI solution of the NOvA--T2K tension. These experiments can also probe parts of the NSI parameter space that even DUNE will be blind to. Our analysis combines nuclear recoils from $^8$B neutrinos and electron recoils from $pp$ neutrinos. 
A future xenon experiment, such as XLZD or PandaX-xT, will be able to cover a significant fraction of the NSI solution of the NOvA--T2K tension.
\end{abstract}

\preprint{IFT-UAM/CSIC-26-154}
\maketitle

\section{Introduction}

The discovery of neutrino oscillations in atmospheric neutrinos, later confirmed in solar neutrinos, indicates a clear need for new physics beyond the Standard Model (SM)~\cite{Super-Kamiokande:1998kpq,SNO:2002tuh}. These flavor oscillations imply that neutrinos must have a mass;
a wide variety of mechanisms exist to explain this, with some also positing new, non-standard interactions with SM particles \cite{Babu:2019mfe}.
These non-standard interactions \cite{Wolfenstein:1977ue} (see also \cite{Farzan:2017xzy,Proceedings:2019qno}) modify the propagation of neutrino states through matter beyond the standard Wolfenstein matter effect \cite{Wolfenstein:1977ue} and tend to scale with density and neutrino energy relative to regular oscillations.
As most of the six standard oscillation parameters are now determined to modest significance \cite{Esteban:2024eli,Capozzi:2025ovi,Denton:2025jkt}, there is increasing power to probe new physics that further modifies oscillations beyond the standard three-flavor picture.

Long-baseline (LBL) accelerator neutrino oscillation experiments will provide the most precise measurements of the remaining unknown oscillation parameters: the CP violating phase $\delta$, the octant of $\theta_{23}$, and the atmospheric mass ordering $\sign(\Delta m^2_{31})$.
The current state of the art, NOvA \cite{NOvA:2004blv} and T2K \cite{T2K:2011qtm}, have modest sensitivity to each of these at the $\usim3\sigma$ level.
The next generation LBL accelerator experiments DUNE \cite{DUNE:2020ypp} and HyperK \cite{Hyper-Kamiokande:2018ofw} aim to measure these parameters at the $\usim5\sigma$ level, with atmospheric and reactor neutrinos providing complementary information.
NOvA and DUNE benefit from significant matter effects, which are minimal in T2K and HyperK.
All of these LBL accelerator experiments have excellent capabilities to probe new physics in neutrino oscillations, especially in combination, due to the different strengths of their matter effects.

Presently, NOvA and T2K data mildly disagree in their determination of $\delta$ \cite{NOvA:2021nfi,T2K:2023smv,T2K:2025yoy,NOvA:2025tmb}. This leads either to a curious swap in the preferred mass ordering from normal to inverted at low significance \cite{Kelly:2020fkv} or, at slightly higher significance, to an improved fit in the presence of new physics via neutral-current, flavor-changing, CP-violating, vector non-standard interactions (NSI)\footnote{In this work, NSI  always refer to \textit{vector} NSI.} \cite{Denton:2020uda} (see also \cite{Chatterjee:2020kkm,Chatterjee:2024kbn,Yu:2024nkc}).
A number of studies of the data from both experiments have been performed assuming different new physics scenarios; however, none explain the tension better than NSI \cite{Chatterjee:2020yak,Forero:2021azc,Rahaman:2021leu,deGouvea:2022kma,Majhi:2022wyp,Denton:2022pxt,Cherchiglia:2023ojf,Li:2023mil}.\footnote{In addition, NOvA has completed a study of flavor-changing CP-violating vector NSI with only their data and found no evidence of new physics \cite{NOvA:2024lti}, which highlights the importance of having two complementary LBL accelerator neutrino experiments at different baselines and energies to reliably probe new physics in oscillations.}

The majority of studies discussing NSI in LBL focus on the SM charged-current interactions in the far detector, where the new physics only affects propagation.
Recently there has been interest in using neutral-current interactions at LBL experiments that are sensitive to vector NSI in propagation and detection simultaneously \cite{Gehrlein:2026ier}.
In addition, most studies focus on real vector NSI due to the complexity of including complex CP violating phases in numerical studies as well as the small amount of appearance data until recent years, see e.g.~\cite{Biggio:2009nt,Escrihuela:2011cf,Gonzalez-Garcia:2011vlg,Gonzalez-Garcia:2013usa,Coloma:2017egw,Esteban:2018ppq,Coloma:2019mbs,Dutta:2020che,Coloma:2023ixt}.

In addition to this tension, LBL experiments also have several important degeneracies in the presence of new physics that are challenging or impossible to break.
One key degeneracy arises when multiple NSI parameters are considered at the same time. Specifically, when considering complex $\varepsilon_{e\tau}$ alongside the flavor-conserving (diagonal) NSI parameters, large degenerate solutions exist \cite{Friedland:2004ah,Friedland:2006pi,Kikuchi:2008vq,Coloma:2011rq,Coloma:2015kiu,Liao:2016hsa,Coloma:2016gei,Chatterjee:2021wac,Denton:2025kvy} that require an experiment beyond DUNE; it has been shown that a future Neutrino Factory could address this \cite{Denton:2025kvy}.
Additional degeneracies such as the diagonal degeneracy and the LMA-Dark degeneracy, which together can be described by $(\varepsilon_{ee},\varepsilon_{\mu\mu},\varepsilon_{\tau\tau})=(x-2,x,x)$ for any $x$, are impossible to probe in oscillation experiments \cite{Bakhti:2014pva,Coloma:2016gei,Denton:2018xmq,Denton:2022nol}.
Moreover, different measurements will be crucial to break degeneracies and ensure that precision experiments like DUNE and HyperK can probe NSI parameters down to small and perturbative levels.

Several approaches exist to realizing such independent probes.
Atmospheric neutrinos at $\mathcal O(10)\,\mathrm{GeV}$ scales, where regular oscillations occur, and at $\mathcal O(1)\,\mathrm{TeV}$ scales, where only new physics plays a role, have excellent sensitivity to NSI parameters that specifically contain only $\nu_\mu$ and $\nu_\tau$ \cite{Gonzalez-Garcia:2011vlg,Super-Kamiokande:2011dam,Salvado:2016uqu,IceCubeCollaboration:2021euf,IceCube:2022ubv}.
Another possibility is to probe NSI with solar neutrinos via standard charged-current detection \cite{Gonzalez-Garcia:2013usa,Super-Kamiokande:2022lyl,Coloma:2023ixt}, which leverages the high density in the Sun to overcome the low energy of solar neutrinos.
Finally, NSI can be probed via scattering, which is generically sensitive to the same NSI models as oscillations \cite{Scholberg:2005qs,Coloma:2017egw} but with the important caveat that the mediator mass becomes important \cite{Coloma:2017egw,Shoemaker:2017lzs,Liao:2017uzy,Papoulias:2017qdn,Denton:2018xmq,Farzan:2018gtr,Abdullah:2018ykz,Dutta:2019eml,Denton:2022nol}.

Constraints from scattering experiments such as NuTeV and CHARM disfavor many NSI scenarios for mediators heavier than $\usim 1\,\mathrm{GeV}$ \cite{CHARM:1986vuz,NuTeV:2001whx,Coloma:2017egw}, while
new physics scenarios with mediators lighter than $\usim5\,\mathrm{MeV}$ are disfavored by cosmological constraints \cite{Kamada:2015era,Huang:2017egl}.
To probe the intermediate regime, we must measure coherent elastic neutrino--nucleus scattering (\cevns) \cite{Freedman:1973yd}, a phenomenon first detected by COHERENT via pion decay at rest ($\pi$-DAR) \cite{COHERENT:2017ipa}.
Scattering experiments are not sensitive to the diagonal and LMA-Dark degeneracies, but do have their own degeneracy due to interference with the SM weak interaction \cite{Coloma:2017egw}; while this can be ameliorated by using multiple detector materials with different neutron fractions, it is generally challenging.
Thus, it is difficult for scattering experiments to comprehensively probe, in a model-independent fashion, the relevant anomalies and degeneracies present in oscillation data.\footnote{Although, see the discussion about LMA-Dark and the diagonal degeneracy in \cite{Denton:2022nol}.}

Dark matter direct detection (DM-DD) experiments provide a complementary way to examine the complex NSI motivated by the LBL tension and the degeneracies present in LBL neutrino experiments using solar neutrinos. Originally developed to search for the paradigmatic weakly interacting massive particle, direct detection experiments also provide an exceptional platform for studying neutrino physics. This shifts the role that neutrinos play in these detectors from an unavoidable background (the \textit{neutrino fog} \cite{Monroe:2007xp,Vergados:2008jp,Strigari:2009bq,Gutlein:2010tq,OHare:2021utq,Carew:2023qrj}) to an invaluable signal of solar physics, Standard Model electroweak physics, and new physics in the neutrino sector~\cite{Harnik:2012ni,Baudis:2013qla,Billard:2014yka,Cerdeno:2016sfi,Bertuzzo:2017tuf,Cui:2017ytb,Dutta:2017nht,Newstead:2018muu,Essig:2018tss,AristizabalSierra:2019ykk,Dutta:2019oaj,Amaral:2020tga,Abdullah:2020iiv,Amaral:2021rzw,deGouvea:2021ymm}.
Direct detection experiments based on dual-phase xenon time projection chambers such as LZ~\cite{LZ:2019sgr}, XENONnT~\cite{XENON:2024wpa}, and PandaX-4T~\cite{PandaX:2014mem}, have demonstrated  to be particularly well-suited to this, finding the first indications/observations of solar neutrino \cevns~\cite{LZ:2025igz,XENON:2026ydt,PandaX:2024muv}, as well as recently measuring solar neutrino \eves~\cite{PandaX:2026qdp,XENON:2026rbz}. Next-generation detectors, such as XLZD~\cite{XLZD:2024nsu} and PandaX-xT~\cite{PANDA-X:2024dlo}, will further establish these detectors as powerful solar neutrino observatories.

Furthermore, DM-DD probes of NSI provide a different means of probing neutrino physics compared to neutrino scattering experiments measured via \cevns from a $\pi$-DAR source. Concretely, DM-DD experiments are sensitive to two channels, nuclear recoils (NRs) and electron recoils (ERs), which provide complementary flavor information. Projected bounds and limits on new neutrino interactions from DM-DD have already been derived with these signal channels \cite{Amaral:2023tbs,DeRomeri:2024iaw,AtzoriCorona:2025gyz,DeRomeri:2026prc,Amaral:2026gml,Leone:2026knr}.

In this work, we connect searches for NSI in solar neutrino measurements at DM-DD experiments to regions of the NSI parameter space of interest to LBL experiments. These regions are motivated by the current tension between NOvA and T2K data, as well as by degeneracies that oscillation experiments such as DUNE will find difficult or impossible to break. We explore how recent DM-DD data from solar \cevns and \eves can be used to constrain these regions, and we project the sensitivity of a future xenon-based detector.

\section{Vector Non-Standard Neutrino Interaction Framework}

Neutrino NSI modify the expected signals measured at DM-DD experiments due to both propagation and scattering effects. 
Below, we set up the NSI parametrization we use in this work and provide an abridged account of how NSI impact the physics underlying these phenomena. 
Ultimately, we will be interested in the differential solar neutrino scattering rate, which is governed by the quantum statistical expectation value of the (generalized) neutrino scattering cross section $\mathrm{d}\zeta/\mathrm{d} E_R$,
\begin{equation}
    \frac{\mathrm{d}R}{\mathrm{d}E_R}\propto {\rm Tr}\left[\rho\, \frac{\mathrm{d}\zeta}{\mathrm{d}E_R}\right]\,,
\end{equation}
where $\rho$ is the neutrino density matrix.
For a more complete treatment, please refer to Refs.~\cite{Amaral:2023tbs, Amaral:2026gml}.

\subsection{NSI Parametrization}

Neutral-current vector NSI can be described by the effective Lagrangian~\cite{Wolfenstein:1977ue, Davidson:2003ha, Miranda:2015dra, Farzan:2017xzy, Proceedings:2019qno}
\begin{equation}\label{eq:lagr}
	\!\!\mathcal{L}_{\rm NSI} = - 2 \sqrt{2}\, G_F \sum_{f, \alpha, \beta,P} \varepsilon^{fP}_{\alpha\beta}\, \left(\bar{\nu}_\alpha \gamma_\mu P_L \nu_\beta\right) \left(\bar{f} \gamma^\mu P f\right)\,,
\end{equation}
where $G_F$ is the Fermi constant and the sum runs over all SM fermions, neutrino flavours $\alpha,\beta \in \{e, \mu, \tau\}$, and chiral projectors $P\in\{P_L, P_R\}$. Since ordinary matter is composed of first-generation fermions, in practice we take $f \in \{e, u, d\}$. The dimensionless coupling coefficients $\varepsilon^{fP}_{\alpha\beta}$ measure the strength of the NSI relative to $G_F$. Assuming that the neutrino flavor structure of the NSI factorizes from the charged-fermion couplings, these coefficients may be decomposed as
\begin{equation}\label{eq:NSIcoeff}
 \varepsilon^{fP}_{\alpha\beta} \equiv \varepsilon^{\eta,\varphi}_{\alpha\beta}\,\xi^{fP}\,,
\end{equation}
where $\varepsilon^{\eta,\varphi}_{\alpha\beta}$ specifies the flavor-dependent NSI matrix elements, while $\xi^{fP}$ encodes the relative coupling to the fermion $f$ with chiral projection $P$. For vector interactions, we may define the NSI coefficients
\begin{equation}
    \varepsilon^f_{\alpha\beta} \equiv \varepsilon^{fL}_{\alpha\beta} + \varepsilon^{fR}_{\alpha\beta} = \xi^f \varepsilon_{\alpha \beta}^{\eta,\varphi}\,,
\end{equation}
with $\xi^f \equiv \xi^{fL} + \xi^{fR}$. 

To describe the impact of NSI on neutrino propagation and scattering, we consider NSI with electrons, protons, and neutrons. We define the total proton and neutron NSI as
\begin{equation}
    \varepsilon_{\alpha\beta}^{p} \equiv 2 \varepsilon_{\alpha\beta}^{u} + \varepsilon_{\alpha\beta}^{d} \quad \text{and} \quad \varepsilon_{\alpha\beta}^{n} \equiv 2 \varepsilon_{\alpha\beta}^{d} + \varepsilon_{\alpha\beta}^{u}\,,
\end{equation}
as well as the relative NSI strengths
\begin{equation}
\begin{split}
    \xi^e &\equiv \sqrt{5} \cos\eta \sin\varphi\,, \\
    \xi^p &\equiv \sqrt{5} \cos\eta \cos\varphi\,, \\
    \xi^n &\equiv \sqrt{5}\sin\eta\,,
\end{split}
\label{eq:xi_defs}
\end{equation}
following the parametrization introduced in Ref.~\cite{Amaral:2023tbs}. The angles $\eta$ and $\varphi$ characterize the relative strength of the proton--neutron and proton--electron NSI, respectively, with $\varphi=0$ corresponding to purely nucleon NSI and $\eta = 0$ to purely charged NSI.
In this work, we study the three representative NSI configurations listed in~\cref{tab:bps}.

\begin{table}[t]
\renewcommand{\arraystretch}{1.5}
    \centering
    \begin{tabular*}{1\columnwidth}{@{\extracolsep{\fill}}lccc}
    \toprule\midrule
    Configuration & $\eta$ & $\varphi$ & Interaction Weight $(p, n, e)$ \\
    \midrule
    {\bf C1} & $\pi/4$ & $0$ & $(1, 1, 0)$ \\
    {\bf C2} &  $0$ & $\pi/2$ & $(0, 0, 1)$ \\
    {\bf C3} & $\pi/4$ & $\pi/4$ & $(1, \sqrt{2}, 1)$\\
    \midrule\bottomrule   
    \end{tabular*}
    \caption{The three NSI configurations {\bf C1--C3} that we consider in this work. The relative strengths of the NSI couplings to protons, neutrons, and electrons, $(p, n, e)$, are shown given the angles $\eta$ and $\varphi$. \textbf{C1} gives nucleon-only interactions of equal strength, \textbf{C2} gives electron-only interactions, and \textbf{C3} gives comparable couplings to nucleons and electrons.}
    \label{tab:bps}
\end{table}

Of greater relevance to neutrino oscillations are the matrix elements of the NSI matter Hamiltonian 
\begin{equation}\label{eq:epsdef}
    \nsi_{\alpha\beta}(x) \equiv \sum_{f} \frac{N_f(x)}{N_e(x)} \varepsilon_{\alpha \beta}^f =\xi(x)\, \varepsilon_{\alpha \beta}^{\eta, \varphi}\,,
\end{equation}
where $\xi(x)=\left(\xi^p + \xi^e\right)+Y_{n}(x)\, \xi^{n}$ and $Y_n(x)$ denotes the neutron--to--electron number ratio. Since Hermiticity of the Hamiltonian requires that $\mathcal{E}_{\alpha\beta} = (\mathcal{E}_{\beta\alpha})^*$, the NSI elements can in general be complex; we parametrize them as
\begin{equation}\label{eq:plot}
    \nsi_{\alpha\beta} = |\nsi_{\alpha\beta}| e^{i \phi_{\alpha \beta}}\,,
\end{equation}
where $\phi_{\alpha \beta}$ is a flavor-dependent phase.

\subsection{Effect on Neutrino Propagation}

Neutrino propagation through matter is modified in the presence of NSI due to an additional contribution to the matter Hamiltonian. This can be derived from the Lagrangian in \cref{eq:lagr} and is given by
\begin{equation}
\mathcal{H}_{\mathrm{NSI}}(x) = V_\mathrm{cc}
\left(\begin{matrix}
\mathcal{E}_{ee} & \mathcal{E}_{e\mu} & \mathcal{E}_{e\tau} \\
\mathcal{E}_{e\mu}^* & \mathcal{E}_{\mu\mu} & \mathcal{E}_{\mu\tau} \\
\mathcal{E}_{e\tau}^* & \mathcal{E}_{\mu\tau}^* & \mathcal{E}_{\tau\tau} \\
\end{matrix}
\right)\,,
\label{eq:hamiltonian_nsi}
\end{equation}
where $V_\mathrm{cc}(x) \equiv \sqrt{2} G_F N_e(x)$ is the CC potential in matter, with $N_{e}(x)$ the spatially-dependent electron number density. 
To apply these NSI to solar neutrino propagation, we reduce the problem to an effective $2$-neutrino picture. This approximation is reasonable because the third mass eigenstate effectively decouples in the Sun since both $\Delta m_{31}^2 \gg \Delta m_{21}^2$~\cite{Esteban:2024eli} and $\Delta m_{31}^2 \gg 2 E_\nu V_{\mathrm{cc}}$, with $E_\nu$ the neutrino energy. This renders the Hamiltonian effectively block-diagonal, and the remaining $2\times2$ system is then described by a total effective Hamiltonian of $\mathcal{H}^\mathrm{eff} = \mathcal{H}_{\mathrm{vac}}^\mathrm{eff}+\mathcal{H}_{\mathrm{mat}}^\mathrm{eff}+\mathcal{H}_{\mathrm{NSI}}^\mathrm{eff}$, where $\mathcal{H}_{\mathrm{vac}}^\mathrm{eff}$ and $\mathcal{H}_{\mathrm{mat}}^\mathrm{eff}$ are the SM effective vacuum and matter Hamiltonians~\cite{Amaral:2026gml}. The effective NSI Hamiltonian is given by
\begin{equation}\label{eq:ham_eff}
\mathcal{H}_{\mathrm{NSI}}^\mathrm{eff}(x) = \xi(x)\left(\begin{matrix}
-\varepsilon_{D}^{\eta, \varphi} & \varepsilon_{N}^{\eta, \varphi} \\[1ex]
(\varepsilon_{N}^{\eta, \varphi})^* & \varepsilon_{D}^{\eta, \varphi}
\end{matrix}\right)\,,
\end{equation}
where the diagonal and non-diagonal elements $\varepsilon_{D}^{\eta, \varphi}$ and $\varepsilon_{N}^{\eta, \varphi}$ are linear combinations of $\varepsilon_{\alpha \beta}^{\eta, \varphi}$, defined as
\begin{align}
\label{eq:nsi_eff}
    \varepsilon_{D}^{\eta, \varphi} &\equiv c_{13}\, s_{13} \mathrm{Re}\left[\left(s_{23}\, \varepsilon_{e \mu}^{\eta, \varphi}+c_{23} \,\varepsilon_{e \tau}^{\eta, \varphi}\right)\right] \\
    &-\left(1+s_{13}^{2}\right) c_{23}\, s_{23}\, \mathrm{Re}\left[\varepsilon_{\mu \tau}^{\eta, \varphi}\right] \nonumber\\
    &-\frac{c_{13}^{2}}{2}\left(\varepsilon_{e e}^{\eta, \varphi}-\varepsilon_{\mu \mu}^{\eta, \varphi}\right)+\frac{s_{23}^{2}-s_{13}^{2}\, c_{23}^{2}}{2}\left(\varepsilon_{\tau \tau}^{\eta, \varphi}-\varepsilon_{\mu \mu}^{\eta, \varphi}\right)\,, \nonumber\\[1.5ex]
    \varepsilon_{N}^{\eta, \varphi} &\equiv c_{13}\left(c_{23} \,\varepsilon_{e \mu}^{\eta, \varphi}-s_{23}\, \varepsilon_{e \tau}^{\eta, \varphi}\right) \nonumber\\
    &+ s_{13}\left[s_{23}^{2}\, \varepsilon_{\mu \tau}^{\eta, \varphi}-c_{23}^{2}\, \varepsilon_{\mu \tau}^{\eta, \varphi}+c_{23}\, s_{23}\left(\varepsilon_{\tau \tau}^{\eta, \varphi}-\varepsilon_{\mu \mu}^{\eta, \varphi}\right)\right]\,, \nonumber
\end{align}
with $s_{ab} \equiv \sin \theta_{ab}$ and $c_{ab} \equiv \cos \theta_{ab}$. From these expressions, it becomes clear why experiments sensitive only to propagation effects can have degeneracies to certain combinations of NSI parameters: If changing these parameters holds the differences $\varepsilon_{e e}^{\eta, \varphi}-\varepsilon_{\mu \mu}^{\eta, \varphi}$ or $\varepsilon_{\tau\tau}^{\eta, \varphi}-\varepsilon_{\mu \mu}^{\eta, \varphi}$ fixed, NSI do not modify neutrino propagation even when these coefficients are individually non-zero.

With this effective Hamiltonian, we can evolve the neutrino states using the Schrödinger equation. Solving this differential equation simplifies significantly if the effective Hamiltonian can be treated as a diagonal matrix, which is true if neutrinos propagate adiabatically out of the Sun.  Quantitatively, the validity of this assumption is controlled by the dimensionless adiabaticity parameter $\gamma$,\footnote{See Eq.~(A.9) of Ref.~\cite{Amaral:2026gml} for a definition of the adiabaticity parameter $\gamma$.} which represents the ratio of the diagonal to the off-diagonal entries of the total Hamiltonian $\mathcal{H}^\mathrm{eff}$. 
If $\gamma \gg 1$, then the adiabatic approximation holds.

\begin{figure}[t]
    \centering
    \includegraphics[width=\linewidth]{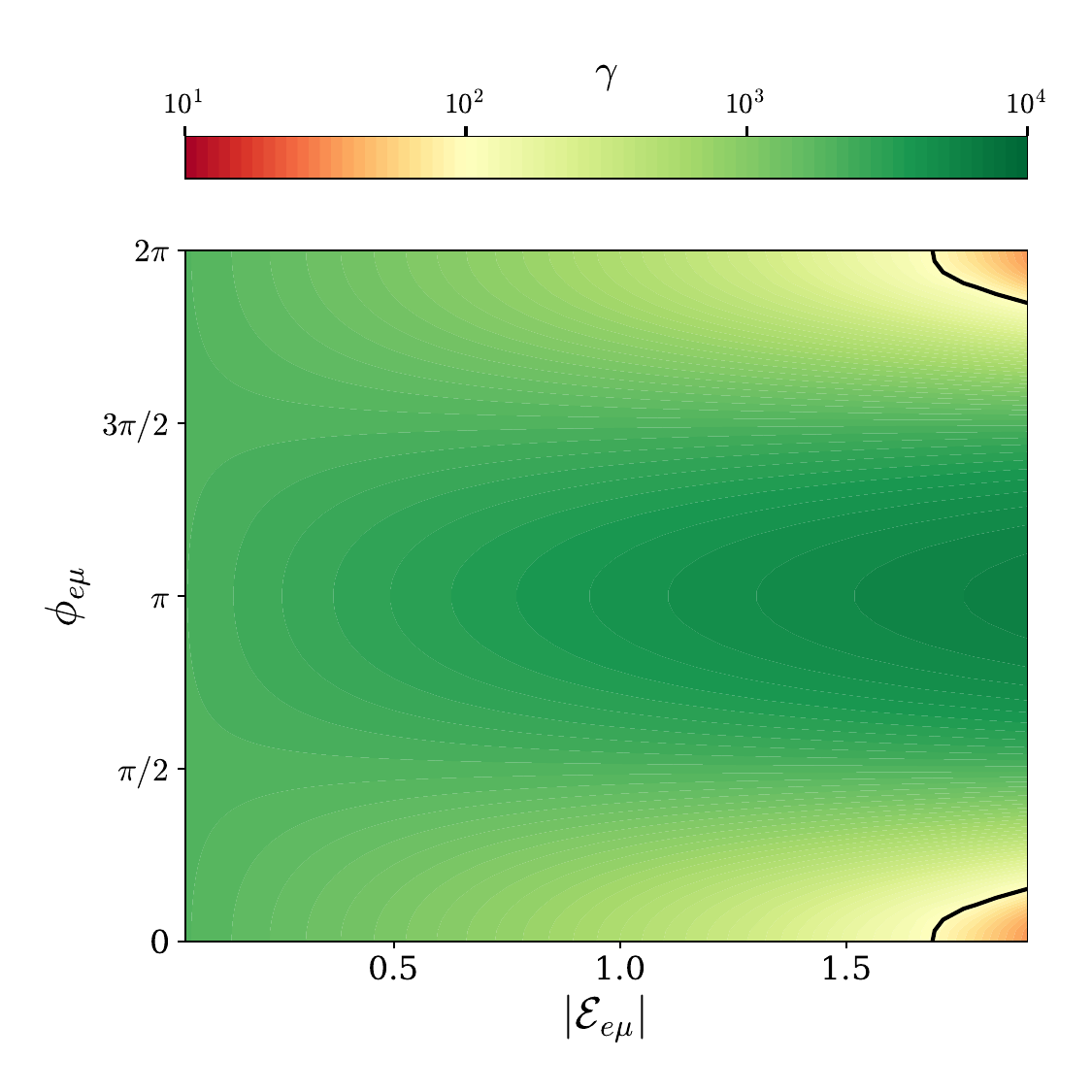}
    \caption{The behavior of the adiabaticity parameter $\gamma$ with varying NSI modulus $|\nsi_{e\mu}|$ and complex phase $\phi_{e\mu}$. Regions where $\gamma \geq 100$, demarcated by the black line, are deemed to satisfy the adiabatic condition. This condition thus holds for the parameter values we scan over when performing our data analyses. The approximation is also valid for other choices of NSI parameters. }
    \label{fig:emu_ad}
\end{figure}

In~\cref{fig:emu_ad}, we demonstrate the behavior of the adiabaticity parameter for complex NSI by varying the phase and modulus of the effective NSI coefficient $\nsi_{\alpha \beta}$, given in~\cref{eq:epsdef,eq:plot}. We set $\gamma = 100$ as the threshold below which the adiabatic assumption breaks down. We show the result for the choice $\alpha\beta = e\mu$; other flavor combinations yield similar results. When 
deriving our limits
in \cref{sec:results}, we restrict ourselves to the domain
in which the adiabatic approximation remains valid. We are therefore justified in adopting the adiabatic approximation to significantly simplify our numerical treatment of solar neutrino propagation without loss of accuracy.

After exiting the Sun, neutrinos decohere on their journey in vacuum, reaching the Earth as an incoherent sum of mass eigenstates. Neutrinos can be detected on the Earth both at daytime or nighttime, having to pass through the Earth in the latter case. We model the  journey through the Earth following Ref.~\cite{Amaral:2026gml} using PREM~\cite{DZIEWONSKI1981297}, which describes the Earth as a series of spherical shells of constant density. Across these shells, the adiabatic assumption no longer holds and the Hamiltonian must instead be numerically diagonalized in each shell. 
Combining all of these steps, we recover the full solar neutrino density matrix at the detector,
which is needed as an input to compute the solar neutrino scattering rate in the detector.

\subsection{Effect on Neutrino Scattering}
\label{subsec:xsec}

In addition to modifying neutrino propagation, NSI also affect the neutrino scattering cross section with nuclei and electrons at the detection point. To capture the effects of flavor correlations in the scattering matrix elements, we must work with the generalized cross section. This was pointed out in Ref.~\cite{Coloma:2022umy} and was further expanded on in Ref.~\cite{Amaral:2023tbs}. 

To compute the generalized CE$\nu$NS cross section, we require the square of the amplitude summed over all possible final neutrino flavors $\gamma$~\cite{Amaral:2023tbs,Amaral:2026gml}:
\begin{align}
    &\left(\frac{\mathrm{d} \zeta_{\nu N}}{\mathrm{d} E_R}\right)_{\alpha\beta}
    = \frac{G_F^2\, m_N}{\pi}\left(1-\frac{m_N\, E_R}{2 E_\nu^2}\right)\ F^2(E_R) \\ 
    &  \times
    \left[ \frac{1}{4} \, Q_{\nu N}^2\, \delta_{\alpha\beta} - Q_{\nu N} \, \operatorname{Re}(G^\mathrm{NSI}_{\alpha\beta}) + \sum_\gamma (G^\mathrm{NSI}_{\alpha\gamma})^*G^{\mathrm{NSI}}_{\gamma\beta}\right]\,, \nonumber
    \label{eq:sig_CEVNS_gen}
\end{align}
where $F(E_R)$ is the Helm form factor~\cite{Helm:1956zz} and  ${Q_{\nu N} = N - (1 -4\,\sin^2\theta_W)\,Z}$ is the SM coherence factor. We have also used the Hermiticity of the NSI nucleus coupling, defined by
\begin{align}
\begin{split}
    G^\mathrm{NSI}_{\alpha\beta} & \equiv \left(2\,\varepsilon^{u}_{\alpha\beta} + \varepsilon^{d}_{\alpha\beta} \right) Z +  \left(\varepsilon^{u}_{\alpha\beta} + 2\,\varepsilon^{d}_{\alpha\beta} \right) N \\ 
    & = (\xi^p\, Z + \xi^n \, N)\ \varepsilon_{\alpha\beta}^{\eta,\varphi} \,.
\end{split}
\end{align}

For \eves, we instead have the generalized cross section~\cite{Amaral:2023tbs,Amaral:2026gml}
\begin{align}\label{eq:xsec_el}
        \left(\frac{\mathrm{d}\zeta_{\nu e}}{\mathrm{d} E_R}\right)_{\alpha\beta} &=  \frac{2 \, G_F^2\, m_e}{\pi} \sum_\gamma \Bigg\{(G^L_{\alpha\gamma})^*G^{L}_{\gamma\beta}  \nonumber\\ & + (G^R_{\alpha\gamma})^* G^{R}_{\gamma\beta}\left(1-\frac{E_R}{E_\nu}\right)^2 \\ &
        - \big[(G^L_{\alpha\gamma})^* G^{R}_{\gamma\beta} + (G^R_{\alpha\gamma})^* G^{L}_{\gamma\beta} \big] \frac{m_e\, E_R}{2 E_\nu^2} \Bigg\} \,, \nonumber 
\end{align}
where we have defined the generalised neutrino-electron couplings as
\begin{equation}
    G^P_{\alpha\beta} \equiv
    g^e_{P\alpha}\delta_{\alpha\beta} + \varepsilon^{eP}_{\alpha\beta}\,,~~~
    (G^P_{\alpha\beta})^* \equiv
    g^e_{P\alpha}\delta_{\alpha\beta} + (\varepsilon^{eP}_{\alpha\beta})^*\,,
\end{equation}
with $P\in\{L,R\}$ denoting the chirality of the scattered electron. 
For vector NSI, we have $\varepsilon_L=\varepsilon_R$.
The SM electroweak neutrino-electron couplings are given in terms of the $Z$-coupling $g^f_P = T^3_f - \sin^2\theta_w\, Q^\mathrm{EM}_f$ as
\begin{equation}
    g^e_{P\alpha} = \begin{cases}  1 + g^e_L  & \text{if}\ \alpha=e \ \text{and} \ P=L \,, \\
    g^e_P  & \text{otherwise} \,.
    \end{cases} 
\end{equation}

\section{NSI at Long-Baseline and Direct Detection Experiments}
\label{sec:exps}

We highlight the phenomenology of NSI in the LBL experiments NOvA, T2K, and DUNE, which motivates the regions of parameter space we study. We then turn to solar neutrinos at DM-DD experiments, focusing on the xenon-based detectors LZ, XENONnT, and PandaX-4T.

\subsection{Benchmarks from Long-Baseline Experiments}

There are several key regions of NSI parameter space with complex CP violating phases motivated by neutrino oscillation measurements. 
One is related to existing measurements, and the other is relevant for future sensitivities. We target these \textit{benchmarks} with direct detection measurements of solar neutrinos.
Since LBL experiments are only sensitive to NSI through the effect of in-medium neutrino oscillations driven by the NSI matter Hamiltonian of~\cref{eq:hamiltonian_nsi}, 
they are governed the effective combined NSI elements of~\cref{eq:epsdef} entering the Hamiltonian,
\begin{equation}\label{eq:nsi_lbl}
    \nsi_{\alpha \beta} = \sqrt{5}\,  \left[  \cos \eta\, (\cos \varphi + \sin \varphi) +  Y_n \sin  \eta \right] \,  \varepsilon_{\alpha \beta}^{\eta, \varphi}\,,
\end{equation}
with a constant neutron fraction, $Y_n=1.05$, for Earth's crust, and where we have used \cref{eq:xi_defs}.

\textbf{Benchmark I:} The first benchmark is related to a tension in the NOvA and T2K data \cite{NOvA:2021nfi,T2K:2023smv,T2K:2025yoy,NOvA:2025tmb}.
While the data sets are each internally consistent and are also consistent with each other and other results on many measurements (including the mass ordering, for which both experiments prefer the normal ordering (NO)), they point to different values of the standard CP violating phase $\delta$ along with the octant of $\theta_{23}$.
T2K prefers $\delta$ near $3\pi/2$ and the upper octant of $\theta_{23}$, while NOvA data disfavors this same region.
This tension can be slightly ameliorated by flipping the mass ordering to the inverted ordering (IO): both  experiments then prefer $\delta$ near $3\pi/2$ and the upper octant, and the statistical penalty for flipping the mass ordering is mild \cite{Kelly:2020fkv}.
However, a better fit to the data is found in the presence of new physics, specifically neutral-current, flavor-changing, CP-violating NSI, in particular with $\nsi_{e\mu}$ or $\nsi_{e\tau}$ \cite{Denton:2020uda,Chatterjee:2020kkm}.
In both cases the data prefers $|\nsi_{e\beta}|\sim0.2$ in the Earth's crust with phase $\phi_{e\beta}\sim3\pi/2$, along with the NO and $\delta\sim3\pi/2$, for $\beta\neq e$.
This parameter space is at the edge of the constraints by IceCube \cite{IceCubeCollaboration:2021euf}.

\textbf{Benchmark II:} The second benchmark is relevant for next generation experiments, such as DUNE and HyperK.
This benchmark involves a strong degeneracy in NSI analyses that appears when two or more NSI parameters are considered at once, along with complex phases; this important degeneracy is missed if focusing on only a single NSI parameter at a time.
Parts of this degeneracy space have been discussed in numerous contexts over the years, and it is a dramatic effect~\cite{Friedland:2004ah,Friedland:2006pi,Kikuchi:2008vq,Coloma:2011rq,Coloma:2015kiu,Liao:2016hsa,Coloma:2016gei,Chatterjee:2021wac,Denton:2025kvy}. For example, when considered one at a time, DUNE will have sensitivity to $\nsi_{e\tau}$ at the $\usim0.05$ level and to $\nsi_{ee}-\nsi_{\mu\mu}$ and $\nsi_{\tau\tau}-\nsi_{\mu\mu}$ at the $\usim0.1$ level.
However, when two NSI parameters are considered at once, the sensitivities fall off precipitously (much more than when multiple other new physics parameters are turned on at once): $\nsi_{e\tau}$ can only be probed down to the $\usim0.5$ and the diagonal NSI parameters $\nsi_{ee}-\nsi_{\mu\mu}$, and $\nsi_{\tau\tau}-\nsi_{\mu\mu}$ can be as large as $2$.
These degeneracies represent a weakening of constraints by an order of magnitude or more.
At these levels, even with DUNE, we cannot know for sure whether NSI are perturbatively small, as is often assumed in many analyses.

\subsection{Dark Matter Direct Detection Experiments}
\label{subsec:dd}

In a direct detection experiment, solar neutrinos are observed through the scattering events they induce. The expected signal is governed by the differential rate spectrum, which captures the rate of scattering events per unit recoil energy $E_R$ per unit detector mass:
\begin{align}\label{eq:trace}
    \frac{\mathrm{d} R}{\mathrm{d} E_R}(E_R)=n_T \int_{E_\nu^{\min }} \frac{\mathrm{d} \phi_\nu}{\mathrm{d} E_\nu} \operatorname{Tr}\left[\rho \frac{\mathrm{d} \zeta}{\mathrm{~d} E_R}\right]\,\mathrm{d} E_\nu\,.
\end{align}
Here, $n_T$ is the  number of scattering targets per unit mass, $\mathrm{d}\phi_\nu/\mathrm{d} E_\nu$ is the neutrino flux spectrum at the source (which we take from the B16-GS98 standard solar model~\cite{Bahcall:2005va}), $\rho(E_\nu)$ is the neutrino flavor density matrix evolved up to the interaction point, and $\dd \zeta / \dd E_R$ is the generalized scattering cross section matrix we introduced in \cref{subsec:xsec}. Both $n_T$ and the minimum neutrino energy required to cause a recoil of specified energy depend on the target; the latter is given by $E_\nu^\mathrm{min}(E_R) = \big(E_R + \sqrt{E_R^2 + 2 m_T E_R}\big)  /2$, where $m_T$ is the mass of the nucleus or electron, depending on whether we are interested in the \cevns or \eves spectrum, respectively.

\begin{figure}[t]
    \centering
    \includegraphics[width=\linewidth]{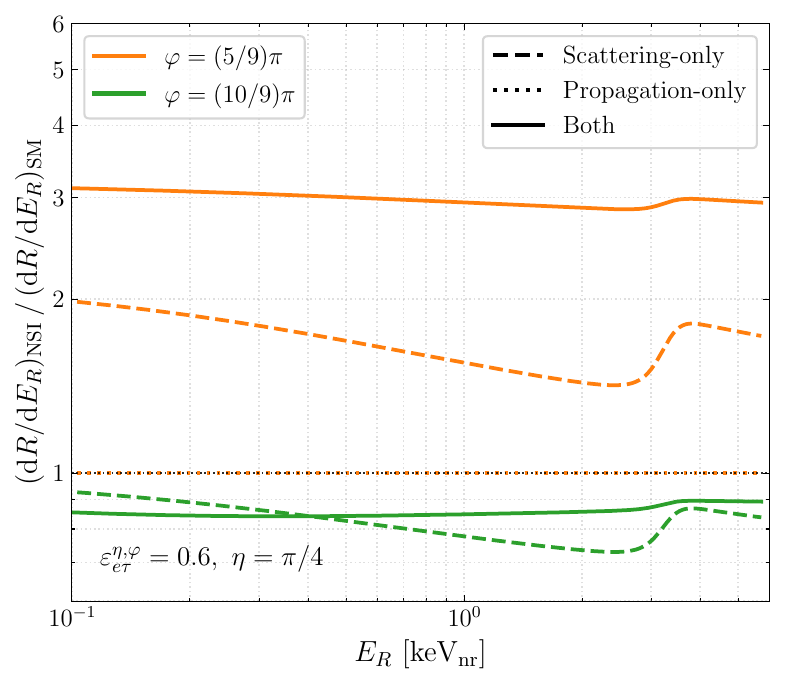}
    \caption{The relative impact of NSI on the \cevns differential rate spectrum compared to the SM result when including only propagation effects (dotted), only scattering effects (dashed), or both simultaneously (solid). The benchmark values $\varepsilon^{\eta,\varphi}_{e\tau}=0.6$ and $\eta=\pi/4$ are used for all lines, with $\varphi=(10/9)\pi$ and $(5/9)\pi$ for the green and orange lines, respectively.}
    \label{fig:rel_NSI}
\end{figure}

In \cref{fig:rel_NSI}, we show the impact of non-zero NSI on the NR signal, split into its two origins: modified propagation effects in the solar (and terrestrial) medium through $\rho$ and changes to the scattering cross section at the detection point via  $\dd\zeta/\dd E_R$. For comparison, we also show their net impact, which is what an experiment ultimately observes. The rate is shown normalized to the SM NR rate. We use the benchmark values $\varepsilon^{\eta,\varphi}_{e\tau}=0.6$ and $\eta=\pi/4$ for all lines, with $\varphi=(10/9)\pi$ and $(5/9)\pi$.
If NSI are included only in the propagation, the rate is identical to the SM prediction at all recoil energies. This is because SM \cevns is flavor-universal, so it is insensitive to how the flux is distributed among flavors at the detector. In the scattering-only case, the $\varphi=(5/9)\pi$ line shows an enhancement, while the $\varphi=(10/9)\pi$ line shows a suppression at the $10$--$30\%$ level. The kink at $3\text{--}4\,{\rm keV_{nr}}$ is a result of the transition from the $^8$B to the \textit{hep} neutrino flux dominating the signal.

We compute the \cevns and \eves spectra in our analysis using the open source Python package \texttt{SNuDD}~\cite{Amaral:2023tbs,Amaral:2026gml}, which is tailored to direct detection experiments and accounts for both modified propagation and scattering.  To compare with data, \snudd folds in detector-specific effects, such as energy-dependent resolution and efficiency functions. Further information on how \snudd computes these spectra---including details such as the propagation treatment, how the rate is mapped to a count, and the microscopic handling of ERs---may be found in Ref.~\cite{Amaral:2026gml}. 
We calculate the recoil spectra for the LZ, XENONnT, and PandaX-4T experiments, as well as for a future xenon experiment based on the planned XLZD and PandaX-xT detectors. We detail the relevant specifications of these experiments and the data sets we use in our analysis below; additional details and validations of the code against experimental data may be found in Sec.~4.1 of Ref.~\cite{Amaral:2026gml}.

\textbf{LZ}~\cite{LZ:2019sgr}\textbf{:} We perform both an NR and an ER analysis. For the NR analysis, we use the data reported in Ref.~\cite{LZ:2025igz}, featuring an exposure of $5.7$ \tnyr. We take the efficiency function from the same study, reaching $50\%$ of its maximum at $1.88\,{\rm keV}_{\rm nr}$, and the energy resolution from Ref.~\cite{LUX:2016rfb}. For the ER analysis, we use the 2022 and 2024 datasets given in Refs.~\cite{LZ:2022lsv, LZ:2024zvo}, which correspond to exposures of $0.9$ \tnyr and $3.3$ \tnyr, respectively. Our analysis is based on the 1D Region of Interest (ROI) analysis reported in Ref.~\cite{LZ:2025zpw}. For this latter analysis, we use a common efficiency function taken from Ref.~\cite{xu_2025_17660135}. We take the backgrounds from Ref.~\cite{LZ:2025zpw}, excluding solar neutrinos, and assume the background uncertainty scales as the quadrature sum of the component backgrounds.
We use the full energy range of Ref.~\cite{XENON:2026rbz} of $[1,20]\,{\rm keV}_{ee}$ in bins of $1\,{\rm keV}_{ee}$ in our analysis.

\textbf{XENONnT}~\cite{XENON:2024wpa}\textbf{:} We perform only an NR analysis, using the data  of Refs.~\cite{XENON:2026ydt,dacheng_xu_2026_20576156}, which correspond an exposure of $6.77$ \tnyr.  We take the efficiency curve from the same studies, reaching $50\%$ of its maximum at $2.21\,{\rm keV}_{\rm nr}$, and the energy resolution from Ref.~\cite{XENON:2020rca}. Due to the $\usim2\sigma$ tension between the B16-GS98 prediction and the recently measured value of the $pp$ flux reported by the XENONnT collaboration~\cite{XENON:2026rbz}, we do not show the results of an ER analysis here, but instead provide a detailed discussion in~\cref{app:xe_pp}. This analysis yields a preferred region of NSI parameter space that is in tension with the bounds we derive for LZ.

\textbf{PandaX-4T}~\cite{PandaX:2014mem}\textbf{:} We perform both an NR and an ER analysis. For the NR analysis, we use the data from Ref.~\cite{PandaX:2024muv}, with a reported exposure of $1.20$ \tnyr.  We take the efficiency function from the same study, reaching $50\%$ of its maximum at $1.61\,{\rm keV}_{\rm nr}$, and the energy resolution from Ref.~\cite{PandaX:2022ood}. For the ER analysis, we replicate Ref.~\cite{PandaX:2026qdp}, which uses two data-taking periods with a $1.6\,{\rm tonne}$ fiducial mass: a first period of $233.8$ live days with a relatively high $^{39}$Ar background, and a second of $204.2$ live days with a lower one. Since the energy range for the analysis is quite high, $20\text{--}1000\,\mathrm{keV_\mathrm{ee}}$, we take a flat efficiency function of $1$.
We restrict our analysis to the energy range of $[20, 260]\,\mathrm{keV_{ee}}$ most relevant for solar neutrinos.
The bin width is $1\, {\rm keV_{ee}}$ and the energy resolution is taken to be the same as for the NR analysis~\cite{PandaX:2022ood}. To obtain the individual background contributions, we have digitally extracted the data from the stacked histograms of Fig.~3 in Ref.~\cite{PandaX:2026qdp}, and we treat each background independently with its associated uncertainty, $\sigma_k$, from the fitted values reported in their Table~1. 
As detailed in~\cref{app:xe_pp}, our reproduction of the $pp$ neutrino flux measurement matches the predictions of the collaborations.

\textbf{Future}~\cite{XLZD:2024nsu,PANDA-X:2024dlo}\textbf{:} To project the ultimate reach of a multi-ton xenon direct detection experiment, we consider a future realization based on the proposed XLZD and PandaX-xT detectors. Both experiments plan to achieve a baseline exposure of 200 \tnyr, and we consider both the NR and ER channels. 
For our NR projection, we use the same resolution as that for our XENONnT study, but we anticipate that energy thresholds will undergo a modest improvement. We therefore take as our nominal efficiency the function provided by LZ in Ref.~~\cite{LZ:2025igz}, and we model an optimistic efficiency as a Heaviside step function given by $0.15\,\Theta(E_R - 1\,\mathrm{keV_\mathrm{nr}})$, which reaches a maximum of $15\%$ at high energies. For the ER analysis, we use the efficiency function and energy binning from the more recent LZ study in Ref.~\cite{LZ:2025zpw} and take the resolution function from Ref.~\cite{LUX:2016rfb} for consistency. We assume the projected reduced backgrounds presented in Ref.~\cite{DARWIN:2020bnc}. 
For both analyses, we use an Asimov data set as a proxy for real experimental data. In this case, the data is assumed to follow the SM expectation exactly and should capture the expected median sensitivity~\cite{Cowan:2010js}.

For all of our NR analyses, we assume a background-free measurement. The systematics are then dominated by the theoretical uncertainty in the $\mathrm{^8B}$ flux from the B16-GS98 standard solar model, which currently stands at $12\%$~\cite{Vinyoles:2016djt}. For our future realization, we assume that this will be improved to $6\%$ for our optimistic benchmark, still above the current experimental uncertainty of $4\%$~\cite{SNO:2011hxd}. 
For our ER analyses, experimental background systematics dominate instead.

\begin{figure*}
    \centering
    \includegraphics[width=0.99\textwidth]{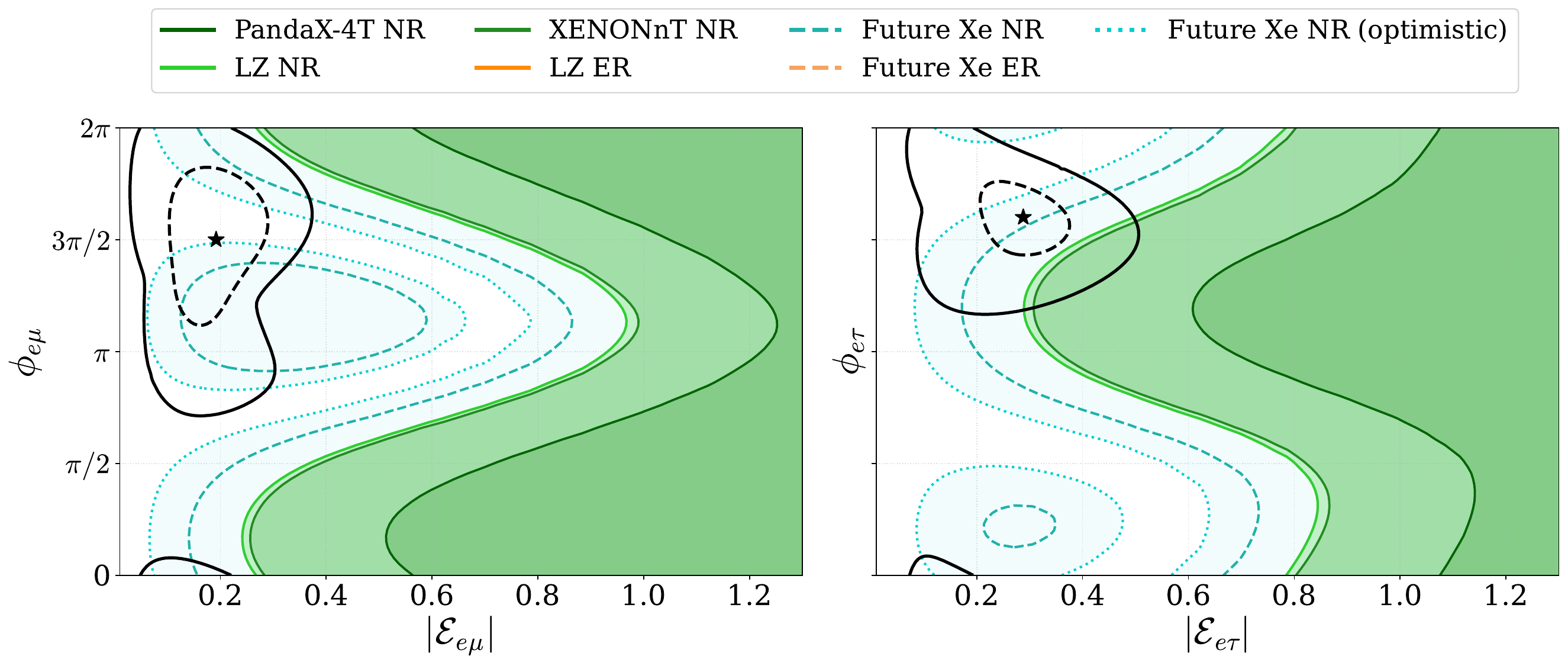}
    \includegraphics[width=0.99\textwidth]{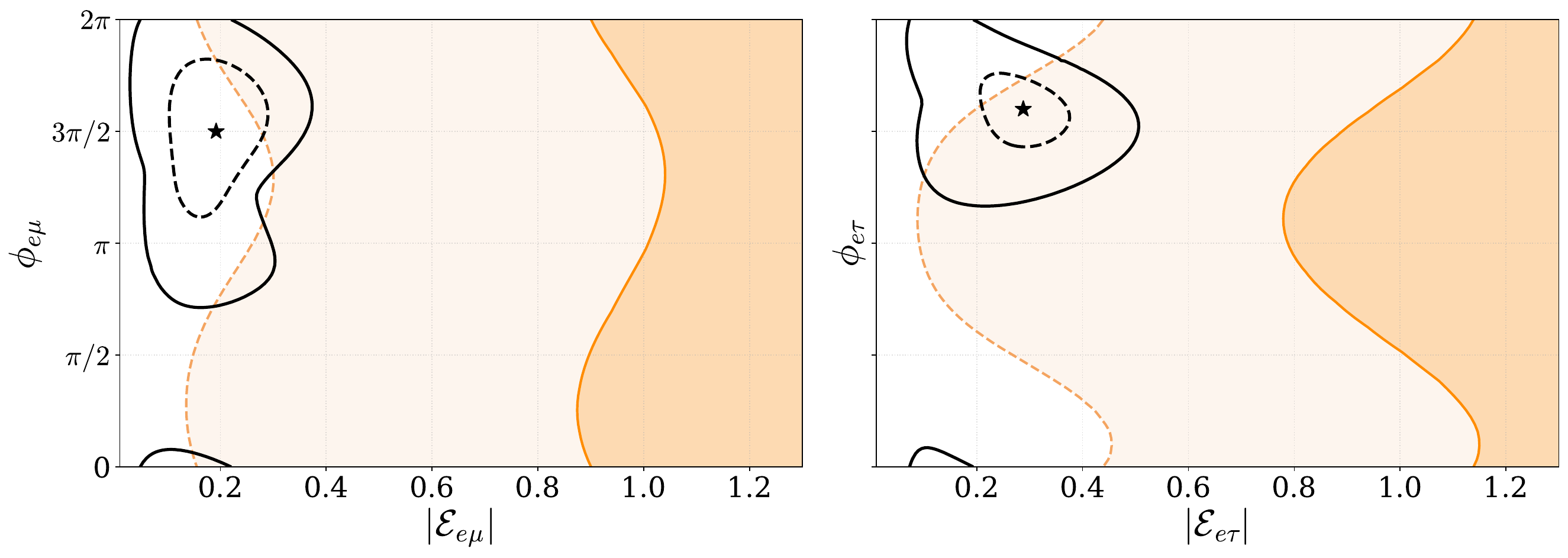}
    \includegraphics[width=0.99\textwidth]{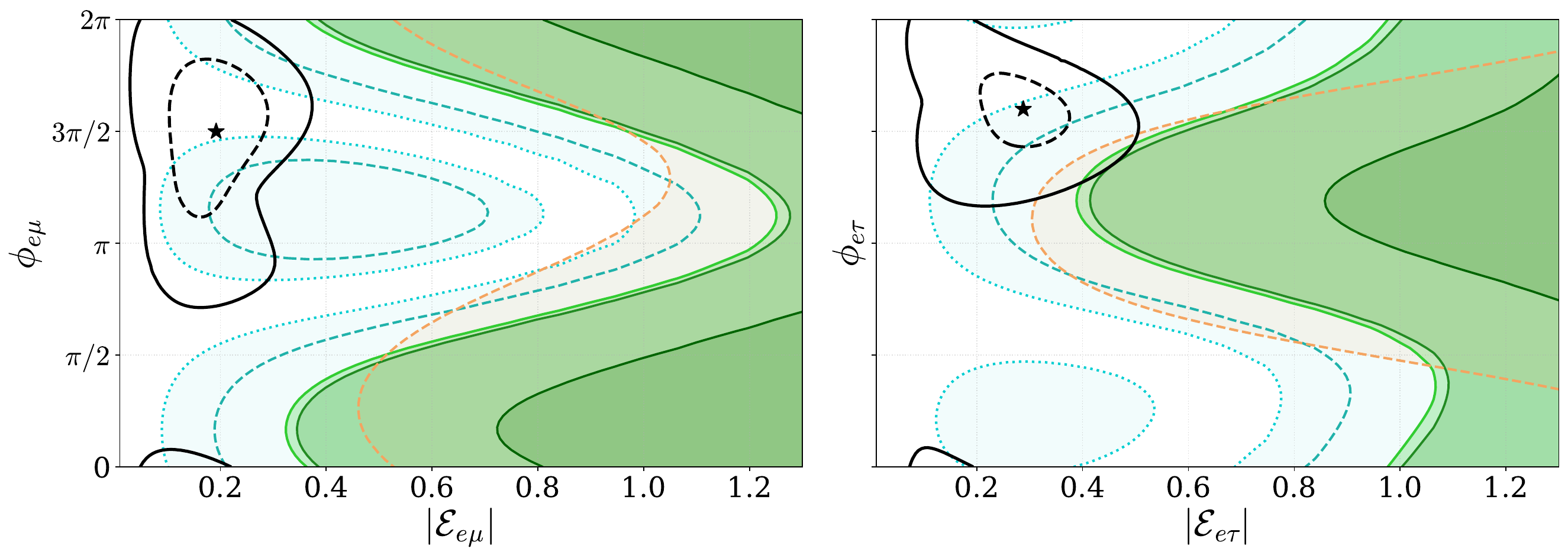}

    \caption{The $90\%$ CL limits on the complex NSI modulus $|\mathcal{E}_{\alpha\beta}|$ and phase $\phi_{\alpha\beta}$ for the flavor indices $\alpha\beta = e\mu$ (\textbf{left}) and $\alpha\beta = e\tau$ (\textbf{right}). Shown are the results for the three parameter configurations given in \cref{tab:bps}, corresponding to the nucleon-only configuration C1 (\textbf{top}), electron-only configuration C2 (\textbf{middle}), and configuration C3 of comparable strength to both (\textbf{bottom}). The limits from the NR analyses are shown in shades of green using current data from the LZ, XENONnT, and PandaX-4T experiments. A projection is also shown for a future realization of a xenon experiment, for which we consider a nominal setup and an optimistic setup that improves on both threshold and the modeling uncertainty in the $\mathrm{^8 B}$ flux. In orange, the results from the ER analyses are shown for the same experiments. The black contours indicate the $68\%$ (dashed) and $90\%$ (solid) preferred regions for the NSI parameters that help to resolve the NOvA--T2K tension, with the star representing the best-fit point~\cite{Denton:2020uda}.}
    \label{Fig:complex}
\end{figure*}

\section{Direct Detection Constraints and Sensitivities}
\label{sec:results}

We present the results for the three parameter configurations defined in~\cref{tab:bps}, corresponding to NSI that are nucleon-only, electron-only, and of comparable strength to both. For each of these configurations, we compute the $90\%$ confidence level (CL) constraints for the NR and ER channels given current direct detection data and project the reach of a future xenon experiment. We compare our results to two LBL benchmarks, showing how DM-DD experiments can help resolve the NOvA--T2K tension and probe key degeneracies in the NSI parameter space. For ease of comparison with LBL experiments, we express our results in terms of the NSI Hamiltonian matrix elements $\mathcal{E}_{\alpha\beta}$ via the prescription given in~\cref{eq:nsi_lbl}.

In our statistical analyses, we use standard profile log-likelihood ratio tests, as described in  Refs.~\cite{Amaral:2023tbs,Amaral:2026gml}. 
We construct the likelihood function 
as the product of a Poisson term, which accounts for the binned event counts, and Gaussian terms, which are intended to capture experimental and theoretical uncertainties via nuisance pull parameters:
\begin{align}
    \!\!\!\!&\mathcal{L}(\boldsymbol{\theta}; a, b) \equiv \mathrm{Gauss}\left(a \big| 0, \sigma_{a}\right)
     \prod_k^{N_\mathrm{bkgs}}\mathrm{Gauss}\left(b_k \big| 0, \sigma_{b_k}\right) \\ 
    &\times \prod_i^{N_\mathrm{bins}} \mathrm{Po}\left[N^i_\mathrm{obs}\Bigg|(1 + a) N^i_\nu(\boldsymbol{\theta}) + \sum_k^{N_\mathrm{bkgs}} (1 + b_k) N^i_{\mathrm{bkg},k} \right]\,. \nonumber
 \end{align}
Here, $N^i_\mathrm{obs}$ is the number of observed neutrino events, $N^i_\nu$ is the number of expected neutrino events for the model parameters $\boldsymbol{\theta}$, and $N^i_{\mathrm{bkg},k}$ is the number background events of the $k^\mathrm{th}$ background component, all of which fall in the $i^\mathrm{th}$ energy bin. The pull parameters $a$ and $b_k$ govern the effect of uncertainties in the neutrino flux and  the $k=1,\dots, N_\mathrm{bkgs}$ detector backgrounds, with standard deviations $\sigma_a$ and $\sigma_{b_k}$, respectively. 
For our NR analysis, we assume a background-free experiment assuming $a$ dominates, setting $\sigma_a = 0.12$ to reflect the uncertainty in the ${}^8$B neutrino flux. We also employ an assumed improved uncertainty of  $\sigma_a = 0.06$ in the case of an optimistic future xenon experiment. For our ER analysis, the effect of the $b_k$ dominate, and $\sigma_b$ are the fractional background uncertainties.

To set our limits, we perform a profile log-likelihood ratio test using the two-sided test statistic
\begin{equation}
    \label{eq:test_stat}
    t(\boldsymbol{\theta}) \equiv -2 \ln\left[\frac{\mathcal{L}(\boldsymbol{\theta};\hat{a},\, \hat{b})}{\mathcal{L}(\hat{\hat{\boldsymbol{\theta}}}; \hat{\hat{a}},\,\hat{\hat{b}})}\right]\,.
\end{equation}
In the above, single hats denote the maximum likelihood
estimators of the likelihood conditioned on the model parameters $\boldsymbol{\theta}$ being tested, while double hats represent the maximum likelihood estimators of the unconstrained likelihood.

\subsection{Benchmark I: The NOvA--T2K Tension}
\label{sec:res_tension}

The latest data from NOvA and T2K point to incompatible preferred values for the CP phase $\delta$ and mixing angle $\theta_{23}$, leading to a tension between these experiments. This discrepancy can be resolved with complex valued NSI, and we show how the NSI parameters $\nsi_{e\mu}= |\nsi_{e\mu}|e^{i \phi_{e\mu}}$ (left column) and $\nsi_{e\tau}= |\nsi_{e\tau}|e^{i \phi_{e\tau}}$ (right column) can achieve this in \cref{Fig:complex} by the closed black contours~\cite{Denton:2020uda}. We also show the constraints and projections we compute for DM-DD experiments for the three NSI configurations C1-C3 defined in~\cref{tab:bps}.

Among the current experiments, LZ (light green) provides the strongest limits on the configuration C1 of nucleon-only NSI (top row), followed by XENONnT (green) and PandaX-4T (dark green).
We also show the projected sensitivity of a next-generation global observatory based on XLZD or PandaX-xT. As we described in \cref{subsec:dd}, we consider this future experiment to have either a nominal threshold 2.21 $\mathrm{keV_\mathrm{nr}}$ (modeled after XENONnT) and the current 12\% uncertainty on the $^8$B flux (blue dashed), or an assumed improved threshold of $1\,\mathrm{keV_{nr}}$ and a 6\% $^8$B flux uncertainty. In the case of $\mathcal{E}_{e\tau}$, current constraints from LZ and XENONnT already rule out part of the parameter space where the NOvA--T2K tension can be explained. A future xenon observatory will be able to significantly improve over current limits and almost fully probe the $1\sigma$ preferred region in the optimistic setup. For $\mathcal{E}_{e\mu}$, on the other hand, current direct detection limits are not sensitive to the preferred region. Nonetheless, a future xenon experiment will have sensitivity to a large fraction of it.

For configuration C2 of electron-only NSI (middle row), we show the constraints from the recent LZ ER analysis~\citep{LZ:2025zpw} (solid orange) and sensitivity projection for the future XLZD or PandaX-xT experiments (orange dashed).
The corresponding limits from the PandaX-4T ER analysis~\cite{PandaX:2026qdp} are weaker and outside of the plotted region.
For this configuration, current constraints are not sensitive to the region of parameter space where the tension can be explained. However, future XLZD or PandaX-xT will be able to probe part of the relevant parameter space of both $\nsi_{e\mu}$ and $\nsi_{e\tau}$.

Finally, for configuration C3 of comparable electron and nucleon NSI (bottom row), we can see the complementarity of NR and ER sensitivities of DM-DD experiments. Since for comparable electron and nucleon NSI \cevns is still the dominant signal, we see that current ER limits are not competitive with the NR limits from LZ, PandaX and XENONnT (solid green lines). However, as can be seen by comparing the light blue and orange projections, ER will provide complementary sensitivity at a future large-exposure xenon observatory like XLZD and PandaX-xT.

\subsection{Benchmark II: LBL Degeneracies}

\begin{figure*}
    \centering
    \includegraphics[width=0.99\linewidth]{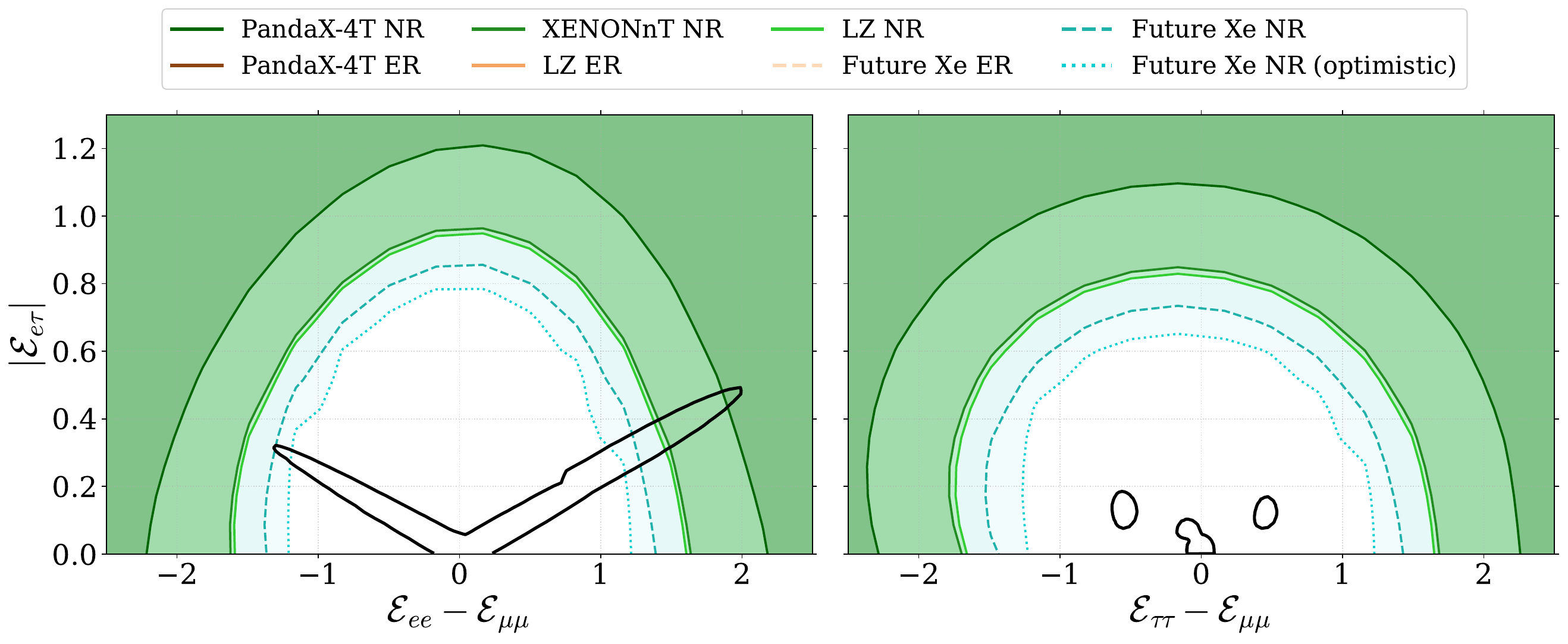} \\
    \includegraphics[width=0.99\linewidth]{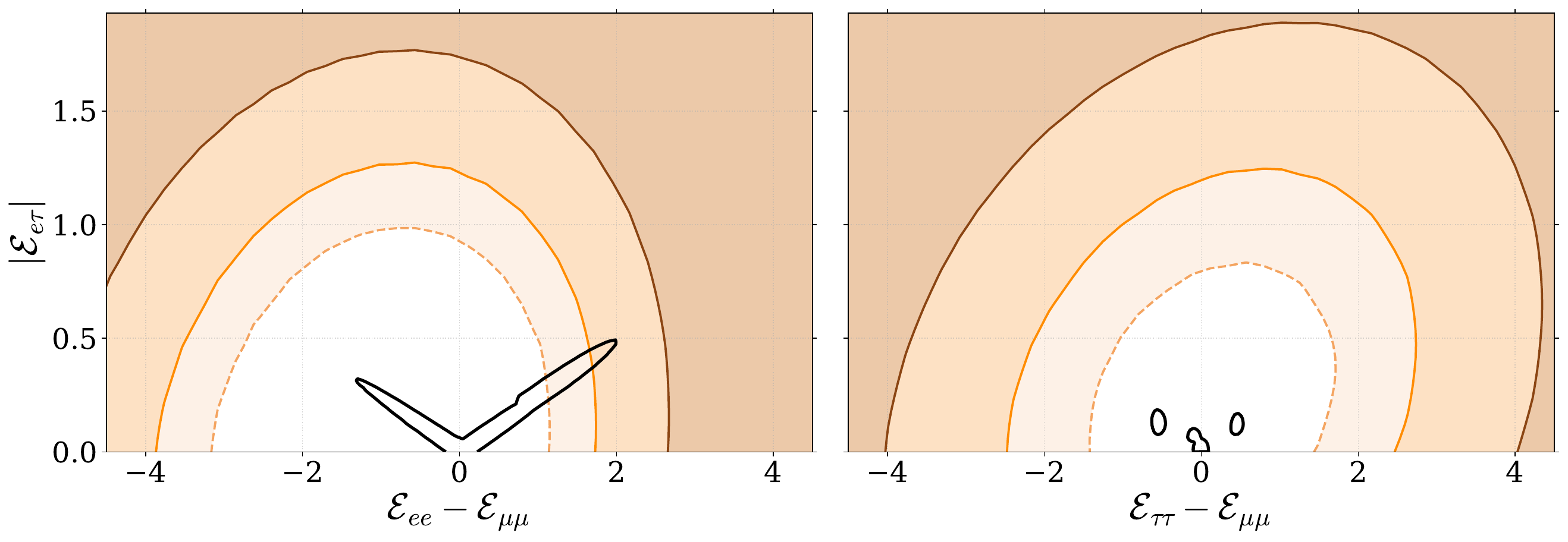} \\
    \includegraphics[width=0.99\linewidth]{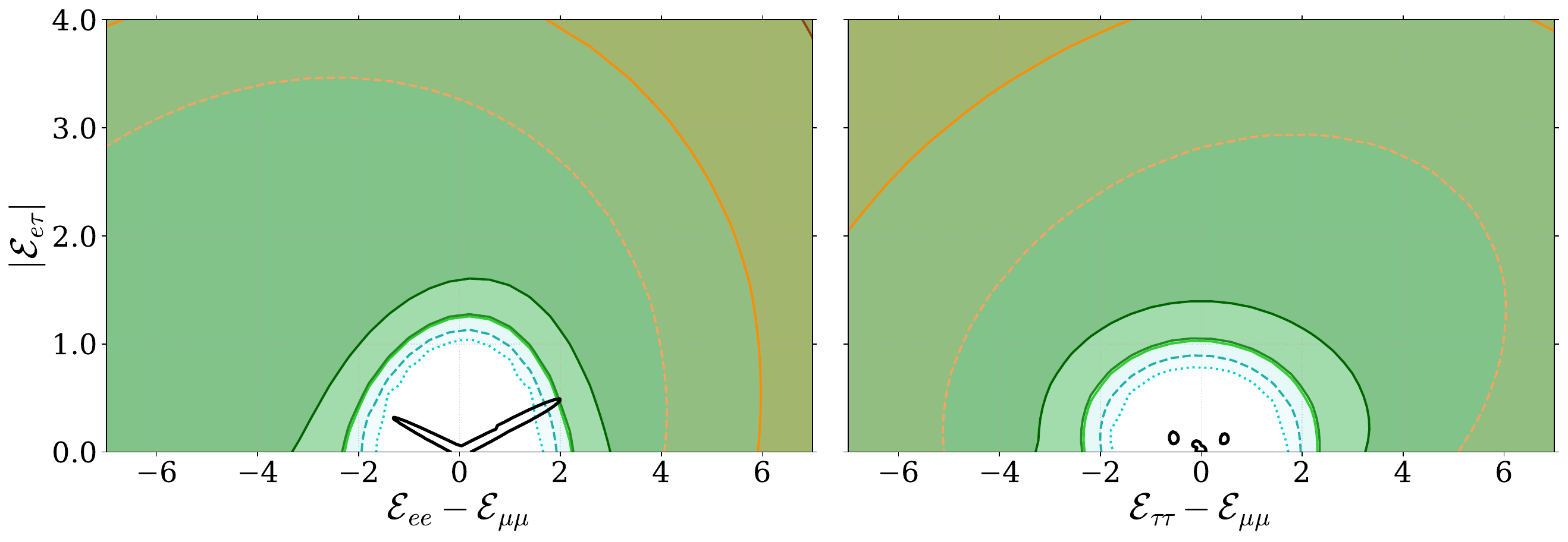}%
    \caption{The $90\%$ CL limits for the modulus $|\mathcal{E}_{e \tau}|$ with  $\nsi_{ee}-\nsi_{\mu\mu}$ (\textbf{left}) and $\nsi_{\tau\tau}-\nsi_{\mu \mu}$ (\textbf{right}) following the profiling procedure outlined in \cref{eq:q_prof}. The black contours show the $90\%$ CL projected constraints from a $20$-year run of DUNE~\cite{Denton:2025kvy}.  The remaining details are identical to those of \cref{Fig:complex}. 
    }
    \label{Fig:marg}
\end{figure*}

LBL experiments probe NSI exclusively through forward coherent scattering during flavor evolution. Since a global shift of the Hamiltonian proportional to the identity matrix leaves oscillation probabilities invariant, LBL searches are inherently sensitive only to differences between diagonal couplings, for example $\nsi_{ee} - \nsi_{\mu\mu}$. This is also indicated by the expressions for the elements of the effective NSI matrix given in \cref{eq:nsi_eff} governing matter oscillations.

In contrast, DM-DD experiments sensitive to both \cevns and \eves of solar neutrinos break this degeneracy: While flavor propagation through the Sun and the Earth still depends on coupling differences, the target scattering cross section depends directly on the absolute magnitudes of the individual couplings $\nsi_{\alpha\beta}$. 
In~\cref{Fig:marg}, we show a comparison of the sensitivities of LBL and DM-DD experiments. We explore the LBL degeneracies arising in the 2D parameter space of the off-diagonal $\nsi_{e\tau}$ and the diagonal combinations $\nsi_{ee} - \nsi_{\mu\mu}$ (left column) and $\nsi_{\tau\tau} - \nsi_{\mu\mu}$ (right column)~\cite{Denton:2025kvy}.

To enable an unbiased, model-independent comparison of DM-DD sensitivities to the
oscillation-accessible parameters, we employ a profile likelihood approach rather than fixing any of the redundant diagonal parameters.
Concretely, for each value of the difference $\Delta\nsi_{\alpha\beta} \equiv \nsi_{\alpha\alpha} - \nsi_{\beta\beta}$
in the two-dimensional parameter plane spanned by $( |\nsi_{e\tau}|, \Delta\nsi_{\alpha\beta})$, we profile the likelihood over $\nsi_{\beta\beta}$,
\begin{equation}
    t_{\text{marg}}( |\nsi_{e\tau}|, \Delta\nsi_{\alpha\beta}) = \min_{\nsi_{\beta\beta}} \ t \left( |\nsi_{e\tau}|, \Delta\nsi_{\alpha\beta}, \nsi_{\beta\beta} \right)\,.
    \label{eq:q_prof}
\end{equation}
The $90\%$ CL upper limits shown in \cref{Fig:marg} are then drawn when the resulting profile log likelihood ratio becomes $t_{\text{marg}} \le 4.61$ ($2$ degrees of freedom).

In the nucleon-only configuration C1 (top row), we show the limits computed with PandaX-4T (dark green), XENONnT (green), and LZ (light green), from least to most constraining. Similarly, we show projections for a future large-exposure xenon experiment based on XLZD or PandaX-xT, both for the nominal parameters (blue dashed) and for the optimistic scenario (blue dotted). 
These limits from DM-DD experiments already constrain part of the degeneracy region of the DUNE sensitivity projection (black solid) for the diagonal combination $\nsi_{ee}-\nsi_{\mu\mu}$ (left plot), even following marginalization over one  of the diagonal NSI. 
For the second diagonal combination $\nsi_{\tau\tau}-\nsi_{\mu\mu}$ (right plot), 
the post-marginalization limits from DM-DD experiments are not competitive with the DUNE future sensitivities and cannot break the degeneracies. This is even the case with the improved sensitivity of our future xenon realization. 

For configuration C2 of pure electron NSI (middle row), we show the limits from ER analyses of both PandaX-4T (brown solid) and LZ (orange solid), as well as the sensitivity projections for a future xenon experiment. For this configuration, the limits from LZ similarly already exclude part of the degenerate region in the DUNE sensitivity for the diagonal combination $\nsi_{ee}-\nsi_{\mu\mu}$ (left plot), and we see that a future xenon experiment will further improve on this. For the second combination  $\nsi_{\tau\tau}-\nsi_{\mu\mu}$ (right plot), DM-DD constraints and future sensitivities will not be able to probe the degeneracy region, analogously to the previous NR limits.

Finally, for configuration C3 of comparable electron and nucleon NSI (bottom row), we show the limits and sensitivity projections from both NR and ER analyses at DD experiments. 
The current NR limits of LZ and XENONnT (green lines) already reach the degenerate region of the  DUNE sensitivity projection for the diagonal combination NSI $\nsi_{ee}-\nsi_{\mu\mu}$ (left plot). 
Future sensitivities (blue dashed/dotted) will be able to constrain part of this region. For this configuration, the post-marginalization ER limits are not competitive with the NR constraints. 
Note, however, that DM-DD experiments are sensitive to the individual diagonal NSI elements $\nsi_{\alpha\alpha}$.  If instead of marginalizing we were to fix one of the diagonal couplings (e.g.~$\nsi_{\mu\mu}=0$), the limits from NR and ER become significantly more stringent and complementary to each other, constraining different parts of the DUNE sensitivity regions (see~\cref{app:2D_fix} for details). 
Finally, as in the previous NSI configurations, DM-DD experiments are unable to set competitive bounds to the projected DUNE sensitivity for the diagonal NSI combination $\nsi_{\tau\tau}-\nsi_{\mu\mu}$ (right plot).

\section{Discussion and Conclusions}

Connections between the standard neutrino oscillation program and dark matter direct detection (DM-DD) are now a reality. Already with current data, we see that existing DM-DD measurements from LZ, XENONnT, and PandaX-4T probe NSI parameter space that is relevant for the tension between NOvA and T2K data.
In the future, DUNE and other LBL accelerator experiments will have excellent sensitivity to NSI, but some degeneracies will remain when realistic scenarios including multiple NSI parameters are studies at the same time; we have shown that these future degenerate regions of parameter space are beginning to be probed with \emph{current} DM-DD measurements and will be further constrained as measurements improve in coming years.
This clearly shows that DM-DD measurements and sensitivities must be considered alongside present and future studies of NSI in studies of more traditional oscillation data sets.

We emphasize that DM-DD, while primarily focused on the NR channel which is NR and thus flavor blind, also has complementary flavor information from ER which is more sensitive to electron neutrinos than muon and tau neutrinos.
While the flavor information in DM-DD will not match that from the environments of LBL accelerator experiments, it still provides some handles.

We note that the recent XENONnT measurement of solar $pp$ neutrinos have a slight excess of ER events~\cite{XENON:2026rbz}. In~\cref{app:xe_pp} we show the relevant parameter space in the context of NSI parameters and for a specific UV-complete model alongside other terrestrial, astrophysical, and cosmological constraints.
While this hint is in slight tension with some of the constraints, this result demonstrates that it is in principle possible to discover NSI by a careful analysis of data at a wide variety of scales.

Conversely, in order to have any chance of going below the neutrino fog in DM-DD, precise predictions of the solar neutrino event rate, including flux, propagation, and scattering, are necessary.
The possible presence of new physics on the neutrino side makes the prediction considerably less certain and more challenging to robustly probe \cite{Bertuzzo:2017tuf,Boehm:2018sux,DeRomeri:2025nkx}.
This requires precise bounds on new physics parameters in neutrinos in order to make clean predictions for DM-DD experiments.
This further highlights the important interplay between neutrino oscillation and DM experiments going in both directions.

\section*{Acknowledgments}

DA has been supported by ERC grant ERC-2024-SYG 101167211 by the European Union.
Views and opinions expressed are however those of the author(s) only and do not necessarily reflect those of the European Union, European Research Council Executive Agency, or other awarding body. Neither the European Union nor the granting authority can be held responsible for them. AC acknowledges the support of S.~Ge, funded by the NSFC (Grant Nos.~12425506, 12375101, 12090060, and 12090064). 
The work of PF and VC was supported by a fellowship from La Caixa Foundation (ID 100010434 with code LCF/BQ/PR25/12110019). 
DC, PF, and VC acknowledge support from the Spanish Agencia Estatal de Investigaci\'on through the grants PID2024-155874NB-C22 (TheDeAs) and CEX2020-001007-S, funded by 
MCIN/AEI/10.13039/501100011033.
The work of PD is supported by the US Department of Energy under Grant Contract DE-SC0012704.
This work is partially funded by the European Commission – NextGenerationEU, through Momentum CSIC Programme: Develop Your Digital Talent. We acknowledge High Performance Computing support by Emilio Ambite, staff hired under the Generation D initiative, promoted by Red.es, an organisation attached to the Spanish Ministry for Digital Transformation and the Civil Service, for the attraction and retention of talent through grants and training contracts, financed by the Recovery, Transformation and Resilience Plan through the EU’s Next Generation funds.

\appendix

\crefalias{section}{appsec}
\crefalias{subsection}{appsec}

\section{First results of solar $pp$ induced electron recoil events}
\label{app:xe_pp}

Recently, the PandaX-4T~\cite{PandaX:2026qdp} collaboration reported the first evidence ($2.2\,\sigma$) of measuring the solar $pp$ neutrinos flux with a DM-DD experiment. Shortly after, the  XENONnT~\cite{XENON:2026rbz} collaboration reported the first observation with a significance of $5.0\,\sigma$. Both collaborations report a central value for the $pp$ flux, $\Phi_{pp}$, that is larger compared to previous measurements using Gallium~\cite{SAGE:2009eeu}, or at Borexino~\cite{BOREXINO:2018ohr}. Furthermore, when both measurements are compared to Standard Solar Models (SSM)~\cite{Grevesse:1998bj,Asplund:2009fu,Haxton:2012wfz,Vinyoles:2016djt}, the DM-DD measurements are higher. 
The value inferred by the PandaX-4T collaboration of $\Phi_{pp}=(8.5\pm3.5)\times 10^{10}\,{\rm cm}^{-2}{\rm s}^{-1}$ is in agreement with the Borexino result at the $1\sigma$ level. Whereas the XENONnT result of $\Phi_{pp}=\left(10.2\pm2.0\right)\times10^{10}\,{\rm cm}^{-2}{\rm s}^{-1}$ is in mild tension at $2.6\sigma$.

Due to this tension of the XENONnT ER data, we have not included these results in our fits in~\cref{Fig:complex,Fig:marg}. However, in this section we take a closer look at the XENONnT ER results and interpret them in context of the PandaX-4T and LZ ER data.

\begin{figure}[t]
    \centering
    \includegraphics[width=\linewidth]{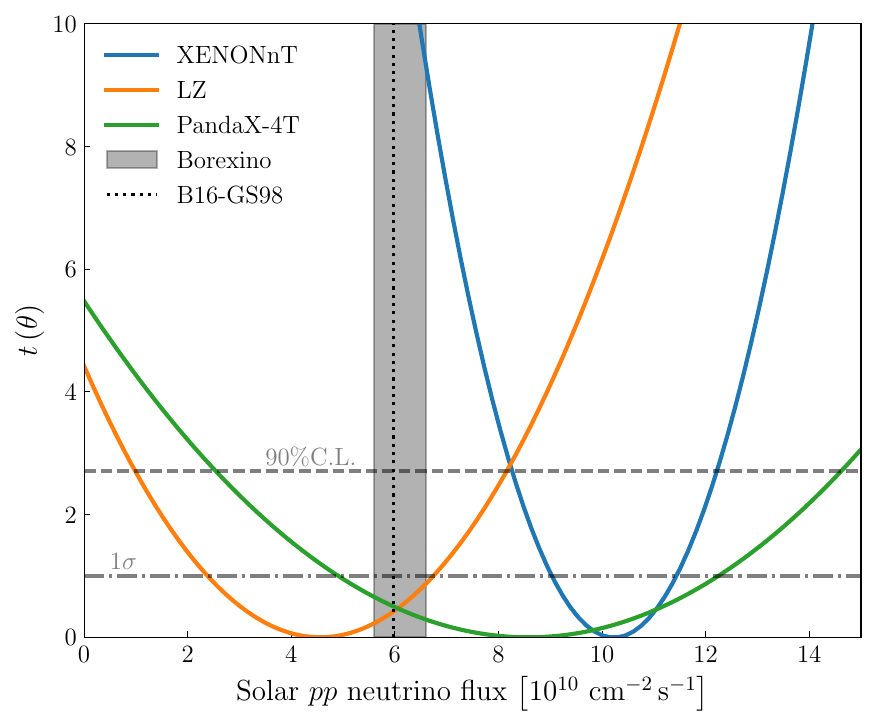}
    \caption{The log likelihood ratio test statistic, $t(\theta)$ (see~\cref{eq:test_stat}), as a function of the solar $pp$ flux in our analyses of the the recent ER data of LZ~\cite{LZ:2025zpw}, PandaX-4T~\cite{PandaX:2026qdp}, and XENONnT~\cite{XENON:2026rbz}. 
    The gray band represents the previous result of the Borexino collaboration~\cite{BOREXINO:2018ohr}, while the black dotted line denotes the SSM prediction~\cite{Vinyoles:2016djt}.
    }
    \label{fig:fluxes}
\end{figure}

First, we validate our analysis strategy. For LZ and PandaX-4T our method is described in~\cref{subsec:dd}, but now the parameter of interest, $\mathbf{\theta}$, is the flux $\Phi_{pp}$ and not the NSI parameters. We use the same two-side test statistic as in~\cref{eq:test_stat} to fit $\Phi_{pp}$ given the data, the backgrounds and the SM solar neutrino signal as calculated by \snudd. The 1D result is shown in~\cref{fig:fluxes} for LZ (orange), PandaX-4T (green) and XENONnT (blue). 

\begin{figure}[t]
    \centering
    \includegraphics[width=\linewidth]{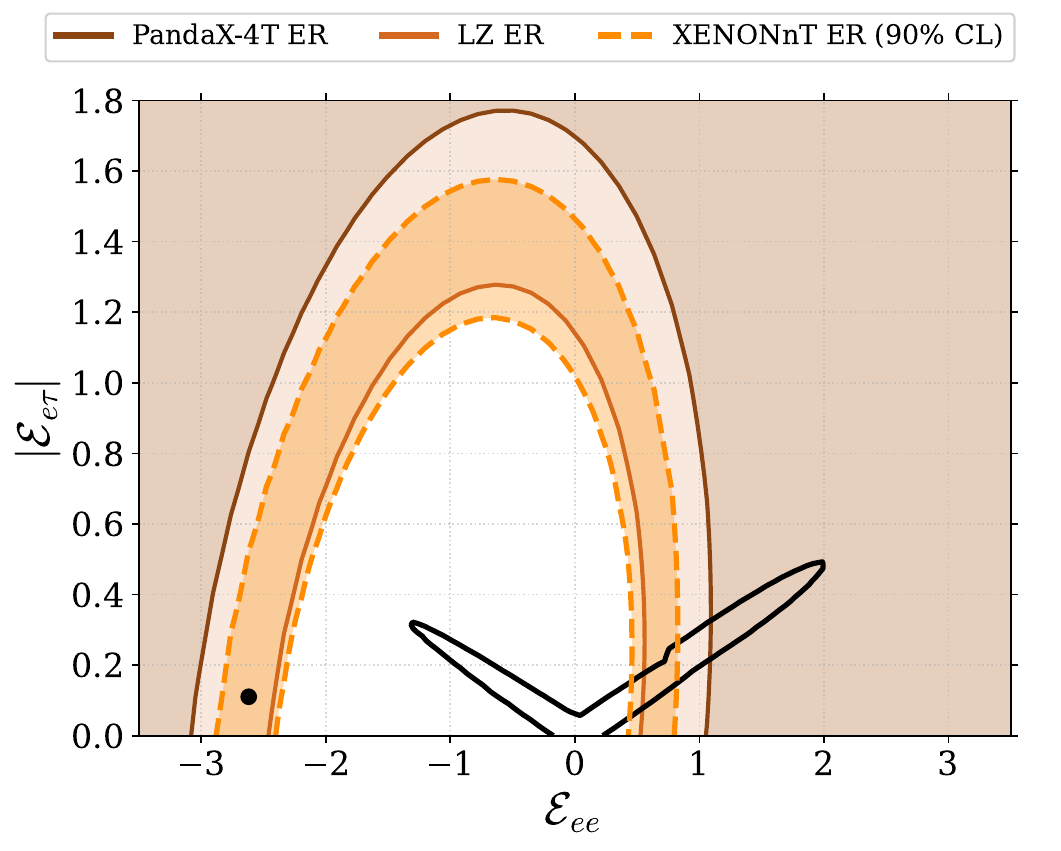}
    \caption{2D parameter constraints and confidence regions in the $(\nsi_{ee}-\nsi_{\mu\mu}, \vert{}\nsi_{e\tau}\vert{})$ plane for the ER channel under the benchmark ($\nsi_{\mu\mu} = 0$) with mixing angles $\eta = 0$ and $\varphi = \pi/2$. Solid lines denote the 90\% CL exclusion limits ($\Delta\chi^2 = 4.61$) for PandaX-4T (dark brown) and LZ (dark orange), with the corresponding excluded parameter space shaded. For XENONnT, the orange contour and shaded interior represent the 90\% CL preferred confidence region around the profiled best-fit point (black dot). The solid black curve represents the projected 90\% CL sensitivity from the DUNE oscillation experiment.}
    \label{fig:xe_nsi_fit}
\end{figure}

The method utilized to replicate the XENONnT result~\cite{XENON:2026rbz} is similar to what we did for LZ and PandaX-4T but with the relevant experimental details changed. For example, XENONnT reported two data sets, the so-called SR0 and SR1b with exposures of $1.09\,{\rm tonne\, yr}$ and $1.37\,{\rm tonne\, yr}$ respectively. 
Similarly, the signal efficiency function has been taken from Fig.~2 of Ref.~\cite{XENON:2026rbz} for both SR0 and SR1b and the energy resolution function used is from Ref.~\cite{XENON:2020rca}. 
In our analysis, we use the full data set of Ref.~\cite{XENON:2026rbz} in the range of $[1,140]$~keV binned in bins of 1 keV for both signal regions.
We have extracted each individual background from Fig.~5 of Ref.~\cite{XENON:2026rbz} and determine their relative uncertainty from the fits reported in Table~IV.
In our likelihood fit, we assign each background component $k$ a separate Gaussian likelihood with pull parameter $a_k$ and Gaussian error $\sigma_k$ taken from the fit from the XENONnT collaboration.

We find that our results from the data are in good agreement with the extracted values reported by the collaborations: $\Phi_{pp}={(8.6\pm3.6)\times 10^{10}}\,{\rm cm}^{-2}{\rm s}^{-1}$ for PandaX-4T and $\Phi_{pp}={(10.3\pm1.2)\times 10^{10}}\,{\rm cm}^{-2}{\rm s}^{-1}$ for XENONnT.
Note that our inferred value of the XENONnT $pp$ flux has a smaller error bar than the value quoted by the collaboration. We believe this is mainly due to the fact that we use the fitted backgrounds of the collaboration and treat them as independent, which might overestimate the accuracy of the $pp$ flux measurement.
We note that LZ has not used their data to extract the $pp$ flux from their signal template fit in~\cite{LZ:2025zpw}. Using their ER data, we find a flux value of $\Phi_{pp}={(4.6\pm 2.2)\times 10^{10}}\,{\rm cm}^{-2}{\rm s}^{-1}$. 
Due to the relatively large exposure of the LZ ER data set and comparably low backgrounds, this value has a relatively small error bar compared to the value inferred from PandaX-4T. 
Nevertheless, we note that the central value of the LZ result lies close to the Borexino and SSM values and is in agreement statistically.

\subsection{Electron NSI}

\begin{figure*}[t]
    \centering
    \includegraphics[width=\textwidth]{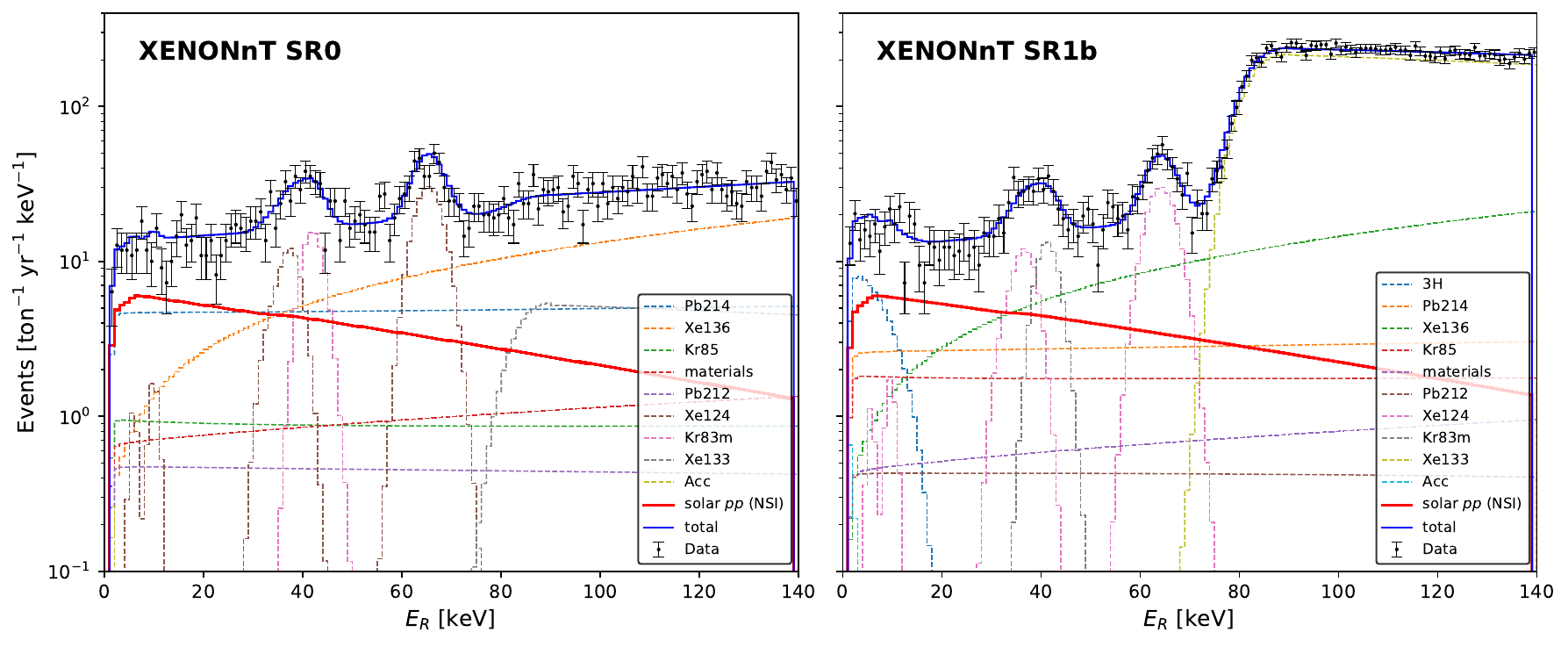}
    \caption{
    Histograms of XENONnT electron recoil data (black dots) and reconstructed binned  backgrounds (dashed lines) of Ref.~\cite{XENON:2026rbz}, together with $pp$ neutrino signal prediction for NSI best fit point  ($\nsi_{ee}=-2.62$, $|\nsi_{e\tau}|=0.11$) for SR0 (\textbf{left}) and SR1b (\textbf{right}). The blue solid line denotes the sum of all background components and the $pp$ signal.}
    \label{fig:xe_hist}
\end{figure*}

The reported tension by the XENONnT collaboration can be interpreted as a signal of new physics. In~\cref{fig:xe_nsi_fit}, we show an interpretation in terms of NSI in the 2D plane of $(|\nsi_{e\tau}|, \nsi_{ee})$ for a pure electron NSI configuration ($\eta=0, \varphi=\pi/2$) relevant to the LBL programme at DUNE. The light orange band, delineated by the dashed lines, represents the 90\%~CL preferred region by the recent XENONnT ER data~\cite{XENON:2026rbz}, while the black dot represents the NSI best fit point ($\nsi_{ee}=-2.62$, $|\nsi_{e\tau}|=0.11$).
The solid brown and red lines represent the 90\%~CL upper limits from the PandaX-4T and LZ ER data sets, respectively.  The solid black line shows the DUNE sensitivity projection. 
We can see that most of the preferred  region by XENONnT is covered by the LZ upper limit and hence is in tension with it. However, a small sliver of the preferred region  is not yet excluded by the LZ limit, part of which could be independently tested at the future DUNE experiment, should this tension persist or even strengthen further.

In~\cref{fig:xe_hist}, we show the extracted binned XENONnT ER data (black dots with error bars), as well as the extracted backgrounds (dashed lines) together with the $pp$ neutrino signal histogram (red solid line) for the NSI best fit point of~\cref{fig:xe_nsi_fit} in our analysis. The blue solid line represents the summed histogram of the backgrounds and $pp$ signal.

\subsection{Vector mediator}

Alternatively, we can interpret the enhanced XENONnT solar $pp$ flux  in the context of a   leptophilic vector mediator, rather than the  effective NSI operators of mass dimension six. Leptophilic vector mediators naturally arise from the gauging of anomaly-free combination of the accidental global symmetries $U(1)_{B}$, $U(1)_{L_e}$, $U(1)_{L_\mu}$, and $U(1)_{L_\tau}$~\cite{Foot:1990mn,He:1990pn,He:1991qd,Foot:1994vd}. Most of these models are severely constrained by astrophysics, beam dump, collider or laboratory neutrino experiments in regions of parameter space where, these mediators could lead to appreciably enhanced solar neutrino signals at direct detection experiments. However, in the case of $U(1)_{L_\mu-L_\tau}$, there is a small region of parameter space for mediator masses of $m_{A'}\sim10$ MeV, where solar neutrino observations lead to the most stringent constraints~\cite{DeRomeri:2026prc}. This motivates us to interpret the XENONnT data in terms of a novel light $U(1)_{L_\mu-L_\tau}$ gauge boson.

For concreteness, we consider a minimal extension of the SM by the Lagrangian,
\begin{equation}
    \mathcal{L}\supset -\frac{1}{4} F'_{\alpha\beta}F'^{\alpha\beta} - \frac{\epsilon_A}{2} F_{\alpha\beta}F'^{\alpha\beta} +\frac{m_{A'}^2}{2 } A'_\alpha A'^\alpha - g_{\mu\tau}\, j_{\mu\tau}^\alpha A'_{\alpha}\,,
\end{equation}
where we assume  the kinetic mixing to arise at one-loop from the $SU(2)_L$ lepton doublets and right-handed singlets  running in the loop~\cite{Bauer:2022nwt}. We assume the $U(1)_{L_\mu-L_\tau}$ symmetry to be broken in the infrared (IR) regime.
The relevant gauge current  in the IR is given by
\begin{align}
j_{\mu\tau}^\alpha &= \bar \mu \gamma^\alpha \mu
          + \bar \nu_\mu\gamma^\alpha P_L \nu_\mu 
          - \bar \tau \gamma^\alpha \tau -\bar\nu_\tau\gamma^\alpha P_L\nu_\tau\,,
\label{eq:lilj_currents}
\end{align}
and the loop-induced kinetic mixing is approximately given by
\begin{align}
\epsilon_A &=\frac{e\, g_{\mu\tau}}{6\pi^2}\ \log\left(\frac{m_\mu}{m_\tau}\right)  \ \sim \ - \frac{g_{\mu\tau}}{70} \,.
\end{align}

\begin{figure}[t]
    \centering
    \includegraphics[width=\linewidth]{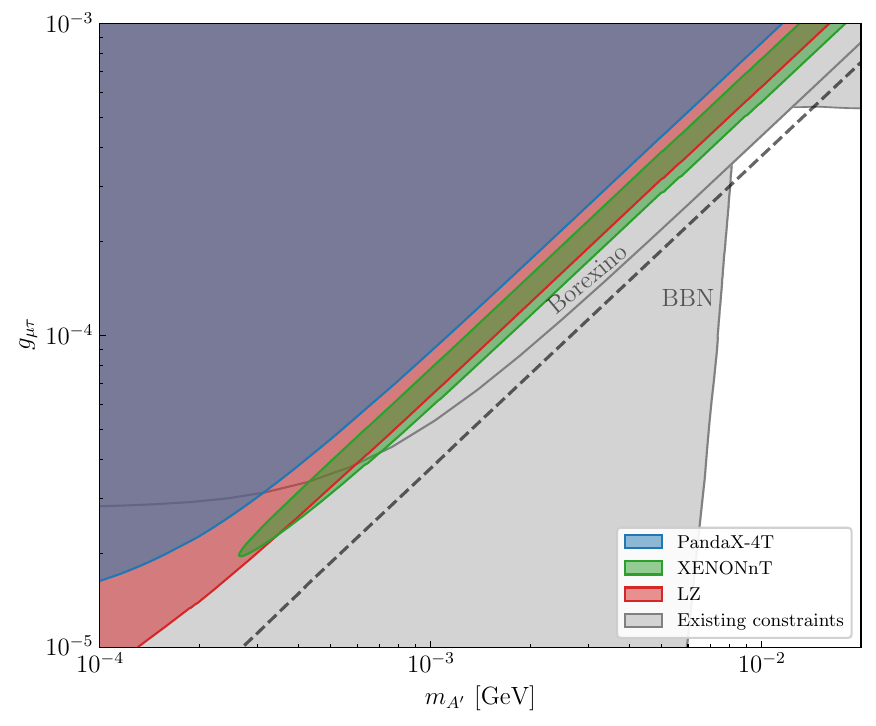}
    \caption{Best fit regions of the observed $pp$ neutrino flux in electron recoils at XENONnT~\cite{XENON:2026rbz} within $U(1)_{L_\mu-L_\tau}$. We also show the upper bounds derived from LZ and PandaX-4T~\cite{PandaX:2026qdp}.
    Also shown are the limits from NA64$\mu$~\cite{NA64:2024klw} and $N_{\rm eff}$ during BBN~\cite{Escudero:2019gzq}. The dashed line shows the sensitivity of white dwarf cooling~\cite{Foldenauer:2024cdp}.}
    \label{fig:lmultau}
\end{figure}

In~\cref{fig:lmultau}, we show the region in the $U(1)_{L_\mu-L_\tau}$ parameter space
preferred at 90\%~CL by the observed $pp$ flux  in XENONnT~\cite{XENON:2026rbz} (green area).
We also show the corresponding 90\%~CL upper limits from LZ (red area) and PandaX-4T (blue area). In gray, we show existing limits 
from $N_{\rm eff}$ during BBN~\cite{Escudero:2019gzq} and NA64$\mu$~\cite{NA64:2024klw}.
Analogous to the NSI case, we see that most of the region preferred by XENONnT is excluded by the LZ result. We also show the upper limit from the Borexino data~\cite{Amaral:2020tga,Gninenko:2020xys}, which covers all of the preferred region in the MeV region. This is consistent with the result of the XENONnT collaboration, who note that their extracted value of the $pp$ flux is in tension with Borexino data.

\section{Complementarity of electron and nuclear recoil data}
\label{app:2D_fix}

\begin{figure}[t]
    \centering
    \includegraphics[width=\linewidth]{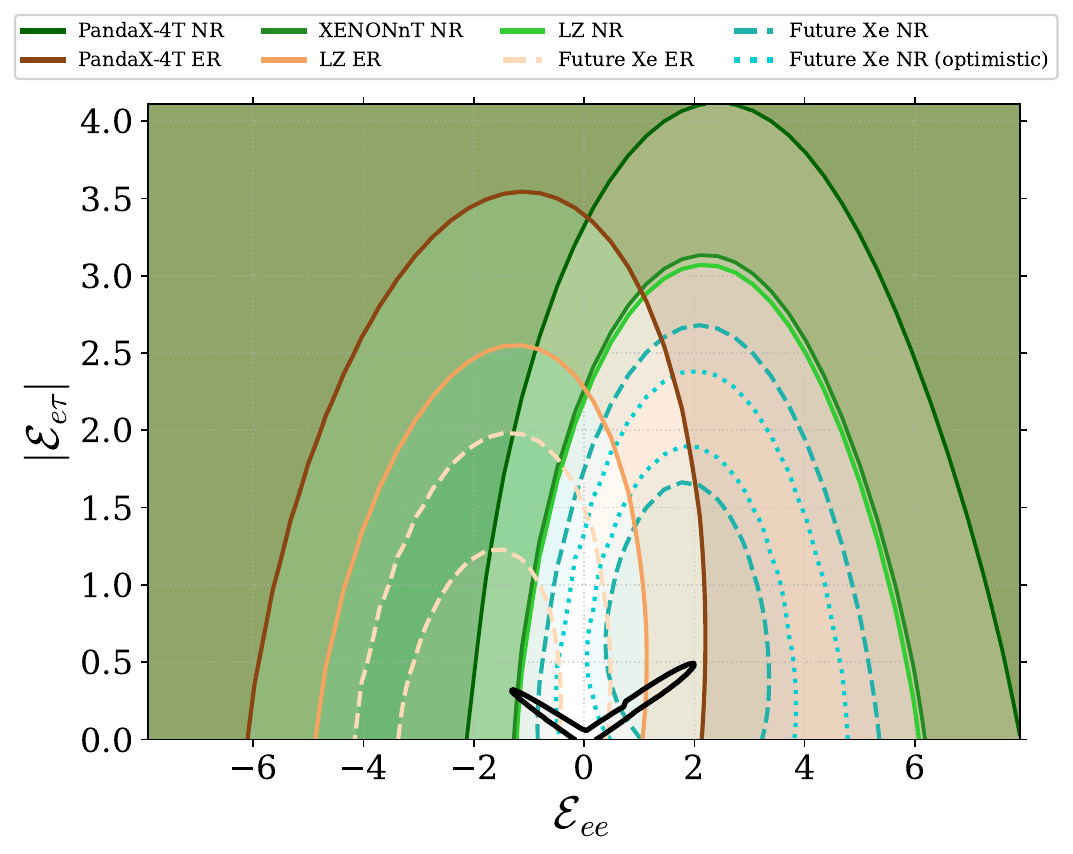}%
    \caption{
    The 90\% CL limits in the 2D NSI parameter space of ($\nsi_{ee}-\nsi_{\mu\mu}$, $|\nsi_{e\tau}|$) plane, for  $\eta = 0, \ \varphi = \pi/4$, and setting $\nsi_{\mu\mu}=0$. The green solid lines represent the limits from the NR analyses of LZ, XENONnT, and PandaX-4T, while NR projections for a future xenon experiment are shown by the blue dashed (nominal) and dotted (optimistic) lines. Conversely, we show the limits from ER analyses of LZ, and PandaX-4T by the orange contours and ER projections for a future xenon experiment by the orange dashed contours.}
    \label{fig:complementary}
\end{figure}

In~\cref{fig:complementary}, we show limits from both NR and ER data of DM-DD experiments in the 2D parameter space of $|\nsi_{e\tau}|$ versus $\Delta\nsi_{e\mu}= \nsi_{ee}-\nsi_{\mu\mu}$ for a configuration of equal NSI couplings with protons and electrons ($\eta=0, \varphi=\pi/4$). In contrast to the results presented in~\cref{Fig:marg}, we are not marginalizing over $\nsi_{\mu\mu}$, but rather set it to zero, $\nsi_{\mu\mu}=0$.
In this case, we can see the complementarity of NR and ER data from DM-DD experiments to constrain NSI. The ER limits from LZ (solid orange line) set rather tight constraints in the positive domain of $\nsi_{ee}$, already ruling out parameter space that can be tested at DUNE. Conversely, the NR limits  from LZ and XENONnT (solid green lines) constrain the negative domain of $\nsi_{ee}$, equally constraining part of the sensitivity region of DUNE.

Hence, we can exploit the complementarity between NR and ER channels in direct detection experiments as a unique tool to resolve degeneracies in the NSI parameter space that oscillation experiments cannot break on their own.

\clearpage

\bibliography{bibliography}

\begin{thebibliography}{142}%
\makeatletter
\providecommand \@ifxundefined [1]{%
 \@ifx{#1\undefined}
}%
\providecommand \@ifnum [1]{%
 \ifnum #1\expandafter \@firstoftwo
 \else \expandafter \@secondoftwo
 \fi
}%
\providecommand \@ifx [1]{%
 \ifx #1\expandafter \@firstoftwo
 \else \expandafter \@secondoftwo
 \fi
}%
\providecommand \natexlab [1]{#1}%
\providecommand \enquote  [1]{``#1''}%
\providecommand \bibnamefont  [1]{#1}%
\providecommand \bibfnamefont [1]{#1}%
\providecommand \citenamefont [1]{#1}%
\providecommand \href@noop [0]{\@secondoftwo}%
\providecommand \href [0]{\begingroup \@sanitize@url \@href}%
\providecommand \@href[1]{\@@startlink{#1}\@@href}%
\providecommand \@@href[1]{\endgroup#1\@@endlink}%
\providecommand \@sanitize@url [0]{\catcode `\\12\catcode `\$12\catcode
  `\&12\catcode `\#12\catcode `\^12\catcode `\_12\catcode `\%12\relax}%
\providecommand \@@startlink[1]{}%
\providecommand \@@endlink[0]{}%
\providecommand \url  [0]{\begingroup\@sanitize@url \@url }%
\providecommand \@url [1]{\endgroup\@href {#1}{\urlprefix }}%
\providecommand \urlprefix  [0]{URL }%
\providecommand \Eprint [0]{\href }%
\providecommand \doibase [0]{https://doi.org/}%
\providecommand \selectlanguage [0]{\@gobble}%
\providecommand \bibinfo  [0]{\@secondoftwo}%
\providecommand \bibfield  [0]{\@secondoftwo}%
\providecommand \translation [1]{[#1]}%
\providecommand \BibitemOpen [0]{}%
\providecommand \bibitemStop [0]{}%
\providecommand \bibitemNoStop [0]{.\EOS\space}%
\providecommand \EOS [0]{\spacefactor3000\relax}%
\providecommand \BibitemShut  [1]{\csname bibitem#1\endcsname}%
\let\auto@bib@innerbib\@empty
\bibitem [{\citenamefont {Fukuda}\ \emph {et~al.}(1998)\citenamefont {Fukuda}
  \emph {et~al.}}]{Super-Kamiokande:1998kpq}%
  \BibitemOpen
  \bibfield  {author} {\bibinfo {author} {\bibfnamefont {Y.}~\bibnamefont
  {Fukuda}} \emph {et~al.} (\bibinfo {collaboration} {Super-Kamiokande}),\
  }\bibfield  {title} {\bibinfo {title} {{Evidence for oscillation of
  atmospheric neutrinos}},\ }\href
  {https://doi.org/10.1103/PhysRevLett.81.1562} {\bibfield  {journal} {\bibinfo
   {journal} {Phys. Rev. Lett.}\ }\textbf {\bibinfo {volume} {81}},\ \bibinfo
  {pages} {1562} (\bibinfo {year} {1998})},\ \Eprint
  {https://arxiv.org/abs/hep-ex/9807003} {arXiv:hep-ex/9807003} \BibitemShut
  {NoStop}%
\bibitem [{\citenamefont {Ahmad}\ \emph {et~al.}(2002)\citenamefont {Ahmad}
  \emph {et~al.}}]{SNO:2002tuh}%
  \BibitemOpen
  \bibfield  {author} {\bibinfo {author} {\bibfnamefont {Q.~R.}\ \bibnamefont
  {Ahmad}} \emph {et~al.} (\bibinfo {collaboration} {SNO}),\ }\bibfield
  {title} {\bibinfo {title} {{Direct evidence for neutrino flavor
  transformation from neutral current interactions in the Sudbury Neutrino
  Observatory}},\ }\href {https://doi.org/10.1103/PhysRevLett.89.011301}
  {\bibfield  {journal} {\bibinfo  {journal} {Phys. Rev. Lett.}\ }\textbf
  {\bibinfo {volume} {89}},\ \bibinfo {pages} {011301} (\bibinfo {year}
  {2002})},\ \Eprint {https://arxiv.org/abs/nucl-ex/0204008}
  {arXiv:nucl-ex/0204008} \BibitemShut {NoStop}%
\bibitem [{\citenamefont {Babu}\ \emph {et~al.}(2020)\citenamefont {Babu},
  \citenamefont {Dev}, \citenamefont {Jana},\ and\ \citenamefont
  {Thapa}}]{Babu:2019mfe}%
  \BibitemOpen
  \bibfield  {author} {\bibinfo {author} {\bibfnamefont {K.~S.}\ \bibnamefont
  {Babu}}, \bibinfo {author} {\bibfnamefont {P.~S.~B.}\ \bibnamefont {Dev}},
  \bibinfo {author} {\bibfnamefont {S.}~\bibnamefont {Jana}},\ and\ \bibinfo
  {author} {\bibfnamefont {A.}~\bibnamefont {Thapa}},\ }\bibfield  {title}
  {\bibinfo {title} {{Non-Standard Interactions in Radiative Neutrino Mass
  Models}},\ }\href {https://doi.org/10.1007/JHEP03(2020)006} {\bibfield
  {journal} {\bibinfo  {journal} {JHEP}\ }\textbf {\bibinfo {volume} {03}},\
  \bibinfo {pages} {006}},\ \Eprint {https://arxiv.org/abs/1907.09498}
  {arXiv:1907.09498 [hep-ph]} \BibitemShut {NoStop}%
\bibitem [{\citenamefont {Wolfenstein}(1978)}]{Wolfenstein:1977ue}%
  \BibitemOpen
  \bibfield  {author} {\bibinfo {author} {\bibfnamefont {L.}~\bibnamefont
  {Wolfenstein}},\ }\bibfield  {title} {\bibinfo {title} {{Neutrino
  Oscillations in Matter}},\ }\href {https://doi.org/10.1103/PhysRevD.17.2369}
  {\bibfield  {journal} {\bibinfo  {journal} {Phys. Rev. D}\ }\textbf {\bibinfo
  {volume} {17}},\ \bibinfo {pages} {2369} (\bibinfo {year}
  {1978})}\BibitemShut {NoStop}%
\bibitem [{\citenamefont {Farzan}\ and\ \citenamefont
  {Tortola}(2018)}]{Farzan:2017xzy}%
  \BibitemOpen
  \bibfield  {author} {\bibinfo {author} {\bibfnamefont {Y.}~\bibnamefont
  {Farzan}}\ and\ \bibinfo {author} {\bibfnamefont {M.}~\bibnamefont
  {Tortola}},\ }\bibfield  {title} {\bibinfo {title} {{Neutrino oscillations
  and Non-Standard Interactions}},\ }\href
  {https://doi.org/10.3389/fphy.2018.00010} {\bibfield  {journal} {\bibinfo
  {journal} {Front. in Phys.}\ }\textbf {\bibinfo {volume} {6}},\ \bibinfo
  {pages} {10} (\bibinfo {year} {2018})},\ \Eprint
  {https://arxiv.org/abs/1710.09360} {arXiv:1710.09360 [hep-ph]} \BibitemShut
  {NoStop}%
\bibitem [{Pro(2019)}]{Proceedings:2019qno}%
  \BibitemOpen
  \href {https://doi.org/10.21468/SciPostPhysProc.2.001} {\emph {\bibinfo
  {title} {{Neutrino Non-Standard Interactions: A Status Report}}}},\
  Vol.~\bibinfo {volume} {2}\ (\bibinfo {year} {2019})\ \Eprint
  {https://arxiv.org/abs/1907.00991} {arXiv:1907.00991 [hep-ph]} \BibitemShut
  {NoStop}%
\bibitem [{\citenamefont {Esteban}\ \emph {et~al.}(2024)\citenamefont
  {Esteban}, \citenamefont {Gonzalez-Garcia}, \citenamefont {Maltoni},
  \citenamefont {Martinez-Soler}, \citenamefont {Pinheiro},\ and\ \citenamefont
  {Schwetz}}]{Esteban:2024eli}%
  \BibitemOpen
  \bibfield  {author} {\bibinfo {author} {\bibfnamefont {I.}~\bibnamefont
  {Esteban}}, \bibinfo {author} {\bibfnamefont {M.~C.}\ \bibnamefont
  {Gonzalez-Garcia}}, \bibinfo {author} {\bibfnamefont {M.}~\bibnamefont
  {Maltoni}}, \bibinfo {author} {\bibfnamefont {I.}~\bibnamefont
  {Martinez-Soler}}, \bibinfo {author} {\bibfnamefont {J.~P.}\ \bibnamefont
  {Pinheiro}},\ and\ \bibinfo {author} {\bibfnamefont {T.}~\bibnamefont
  {Schwetz}},\ }\bibfield  {title} {\bibinfo {title} {{NuFit-6.0: updated
  global analysis of three-flavor neutrino oscillations}},\ }\href
  {https://doi.org/10.1007/JHEP12(2024)216} {\bibfield  {journal} {\bibinfo
  {journal} {JHEP}\ }\textbf {\bibinfo {volume} {12}},\ \bibinfo {pages}
  {216}},\ \Eprint {https://arxiv.org/abs/2410.05380} {arXiv:2410.05380
  [hep-ph]} \BibitemShut {NoStop}%
\bibitem [{\citenamefont {Capozzi}\ \emph {et~al.}(2025)\citenamefont
  {Capozzi}, \citenamefont {Lisi}, \citenamefont {Marcone}, \citenamefont
  {Marrone},\ and\ \citenamefont {Palazzo}}]{Capozzi:2025ovi}%
  \BibitemOpen
  \bibfield  {author} {\bibinfo {author} {\bibfnamefont {F.}~\bibnamefont
  {Capozzi}}, \bibinfo {author} {\bibfnamefont {E.}~\bibnamefont {Lisi}},
  \bibinfo {author} {\bibfnamefont {F.}~\bibnamefont {Marcone}}, \bibinfo
  {author} {\bibfnamefont {A.}~\bibnamefont {Marrone}},\ and\ \bibinfo {author}
  {\bibfnamefont {A.}~\bibnamefont {Palazzo}},\ }\bibfield  {title} {\bibinfo
  {title} {{Updated bounds on the (1,2) neutrino oscillation parameters after
  first JUNO results}},\ }\href@noop {} {\  (\bibinfo {year} {2025})},\ \Eprint
  {https://arxiv.org/abs/2511.21650} {arXiv:2511.21650 [hep-ph]} \BibitemShut
  {NoStop}%
\bibitem [{\citenamefont {Denton}(2025)}]{Denton:2025jkt}%
  \BibitemOpen
  \bibfield  {author} {\bibinfo {author} {\bibfnamefont {P.~B.}\ \bibnamefont
  {Denton}},\ }\bibfield  {title} {\bibinfo {title} {{Neutrino Oscillations in
  the Three Flavor Paradigm}},\ }\href@noop {} {\  (\bibinfo {year} {2025})},\
  \Eprint {https://arxiv.org/abs/2501.08374} {arXiv:2501.08374 [hep-ph]}
  \BibitemShut {NoStop}%
\bibitem [{\citenamefont {Ayres}\ \emph {et~al.}(2004)\citenamefont {Ayres}
  \emph {et~al.}}]{NOvA:2004blv}%
  \BibitemOpen
  \bibfield  {author} {\bibinfo {author} {\bibfnamefont {D.~S.}\ \bibnamefont
  {Ayres}} \emph {et~al.} (\bibinfo {collaboration} {NOvA}),\ }\bibfield
  {title} {\bibinfo {title} {{NOvA: Proposal to Build a 30 Kiloton Off-Axis
  Detector to Study $\nu_{\mu} \to \nu_e$ Oscillations in the NuMI Beamline}},\
  }\href@noop {} {\  (\bibinfo {year} {2004})},\ \Eprint
  {https://arxiv.org/abs/hep-ex/0503053} {arXiv:hep-ex/0503053} \BibitemShut
  {NoStop}%
\bibitem [{\citenamefont {Abe}\ \emph {et~al.}(2011)\citenamefont {Abe} \emph
  {et~al.}}]{T2K:2011qtm}%
  \BibitemOpen
  \bibfield  {author} {\bibinfo {author} {\bibfnamefont {K.}~\bibnamefont
  {Abe}} \emph {et~al.} (\bibinfo {collaboration} {T2K}),\ }\bibfield  {title}
  {\bibinfo {title} {{The T2K Experiment}},\ }\href
  {https://doi.org/10.1016/j.nima.2011.06.067} {\bibfield  {journal} {\bibinfo
  {journal} {Nucl. Instrum. Meth. A}\ }\textbf {\bibinfo {volume} {659}},\
  \bibinfo {pages} {106} (\bibinfo {year} {2011})},\ \Eprint
  {https://arxiv.org/abs/1106.1238} {arXiv:1106.1238 [physics.ins-det]}
  \BibitemShut {NoStop}%
\bibitem [{\citenamefont {Abi}\ \emph {et~al.}(2020)\citenamefont {Abi} \emph
  {et~al.}}]{DUNE:2020ypp}%
  \BibitemOpen
  \bibfield  {author} {\bibinfo {author} {\bibfnamefont {B.}~\bibnamefont
  {Abi}} \emph {et~al.} (\bibinfo {collaboration} {DUNE}),\ }\bibfield  {title}
  {\bibinfo {title} {{Deep Underground Neutrino Experiment (DUNE), Far Detector
  Technical Design Report, Volume II: DUNE Physics}}\ }\href
  {https://doi.org/10.2172/1599307} {10.2172/1599307} (\bibinfo {year}
  {2020}),\ \Eprint {https://arxiv.org/abs/2002.03005} {arXiv:2002.03005
  [hep-ex]} \BibitemShut {NoStop}%
\bibitem [{\citenamefont {Abe}\ \emph {et~al.}(2018)\citenamefont {Abe} \emph
  {et~al.}}]{Hyper-Kamiokande:2018ofw}%
  \BibitemOpen
  \bibfield  {author} {\bibinfo {author} {\bibfnamefont {K.}~\bibnamefont
  {Abe}} \emph {et~al.} (\bibinfo {collaboration} {Hyper-Kamiokande}),\
  }\bibfield  {title} {\bibinfo {title} {{Hyper-Kamiokande Design Report}},\
  }\href@noop {} {\  (\bibinfo {year} {2018})},\ \Eprint
  {https://arxiv.org/abs/1805.04163} {arXiv:1805.04163 [physics.ins-det]}
  \BibitemShut {NoStop}%
\bibitem [{\citenamefont {Acero}\ \emph {et~al.}(2022)\citenamefont {Acero}
  \emph {et~al.}}]{NOvA:2021nfi}%
  \BibitemOpen
  \bibfield  {author} {\bibinfo {author} {\bibfnamefont {M.~A.}\ \bibnamefont
  {Acero}} \emph {et~al.} (\bibinfo {collaboration} {NOvA}),\ }\bibfield
  {title} {\bibinfo {title} {{Improved measurement of neutrino oscillation
  parameters by the NOvA experiment}},\ }\href
  {https://doi.org/10.1103/PhysRevD.106.032004} {\bibfield  {journal} {\bibinfo
   {journal} {Phys. Rev. D}\ }\textbf {\bibinfo {volume} {106}},\ \bibinfo
  {pages} {032004} (\bibinfo {year} {2022})},\ \Eprint
  {https://arxiv.org/abs/2108.08219} {arXiv:2108.08219 [hep-ex]} \BibitemShut
  {NoStop}%
\bibitem [{\citenamefont {Abe}\ \emph {et~al.}(2023)\citenamefont {Abe} \emph
  {et~al.}}]{T2K:2023smv}%
  \BibitemOpen
  \bibfield  {author} {\bibinfo {author} {\bibfnamefont {K.}~\bibnamefont
  {Abe}} \emph {et~al.} (\bibinfo {collaboration} {T2K}),\ }\bibfield  {title}
  {\bibinfo {title} {{Measurements of neutrino oscillation parameters from the
  T2K experiment using $3.6\times 10^{21}$ protons on target}},\ }\href
  {https://doi.org/10.1140/epjc/s10052-023-11819-x} {\bibfield  {journal}
  {\bibinfo  {journal} {Eur. Phys. J. C}\ }\textbf {\bibinfo {volume} {83}},\
  \bibinfo {pages} {782} (\bibinfo {year} {2023})},\ \Eprint
  {https://arxiv.org/abs/2303.03222} {arXiv:2303.03222 [hep-ex]} \BibitemShut
  {NoStop}%
\bibitem [{\citenamefont {Abe}\ \emph {et~al.}(2025)\citenamefont {Abe} \emph
  {et~al.}}]{T2K:2025yoy}%
  \BibitemOpen
  \bibfield  {author} {\bibinfo {author} {\bibfnamefont {K.}~\bibnamefont
  {Abe}} \emph {et~al.} (\bibinfo {collaboration} {T2K}),\ }\bibfield  {title}
  {\bibinfo {title} {{Results from the T2K Experiment on Neutrino Mixing
  Including a New Far Detector {\ensuremath{\mu}}-like Sample}},\ }\href
  {https://doi.org/10.1103/gh5j-5cwv} {\bibfield  {journal} {\bibinfo
  {journal} {Phys. Rev. Lett.}\ }\textbf {\bibinfo {volume} {135}},\ \bibinfo
  {pages} {261801} (\bibinfo {year} {2025})},\ \Eprint
  {https://arxiv.org/abs/2506.05889} {arXiv:2506.05889 [hep-ex]} \BibitemShut
  {NoStop}%
\bibitem [{\citenamefont {Abubakar}\ \emph {et~al.}(2026)\citenamefont
  {Abubakar} \emph {et~al.}}]{NOvA:2025tmb}%
  \BibitemOpen
  \bibfield  {author} {\bibinfo {author} {\bibfnamefont {S.}~\bibnamefont
  {Abubakar}} \emph {et~al.} (\bibinfo {collaboration} {NOvA}),\ }\bibfield
  {title} {\bibinfo {title} {{Precision Measurement of Neutrino Oscillation
  Parameters with 10 Years of Data from the NOvA Experiment}},\ }\href
  {https://doi.org/10.1103/x53y-2b86} {\bibfield  {journal} {\bibinfo
  {journal} {Phys. Rev. Lett.}\ }\textbf {\bibinfo {volume} {136}},\ \bibinfo
  {pages} {011802} (\bibinfo {year} {2026})},\ \Eprint
  {https://arxiv.org/abs/2509.04361} {arXiv:2509.04361 [hep-ex]} \BibitemShut
  {NoStop}%
\bibitem [{\citenamefont {Kelly}\ \emph {et~al.}(2021)\citenamefont {Kelly},
  \citenamefont {Machado}, \citenamefont {Parke}, \citenamefont
  {Perez-Gonzalez},\ and\ \citenamefont {Funchal}}]{Kelly:2020fkv}%
  \BibitemOpen
  \bibfield  {author} {\bibinfo {author} {\bibfnamefont {K.~J.}\ \bibnamefont
  {Kelly}}, \bibinfo {author} {\bibfnamefont {P.~A.~N.}\ \bibnamefont
  {Machado}}, \bibinfo {author} {\bibfnamefont {S.~J.}\ \bibnamefont {Parke}},
  \bibinfo {author} {\bibfnamefont {Y.~F.}\ \bibnamefont {Perez-Gonzalez}},\
  and\ \bibinfo {author} {\bibfnamefont {R.~Z.}\ \bibnamefont {Funchal}},\
  }\bibfield  {title} {\bibinfo {title} {{Neutrino mass ordering in light of
  recent data}},\ }\href {https://doi.org/10.1103/PhysRevD.103.013004}
  {\bibfield  {journal} {\bibinfo  {journal} {Phys. Rev. D}\ }\textbf {\bibinfo
  {volume} {103}},\ \bibinfo {pages} {013004} (\bibinfo {year} {2021})},\
  \Eprint {https://arxiv.org/abs/2007.08526} {arXiv:2007.08526 [hep-ph]}
  \BibitemShut {NoStop}%
\bibitem [{\citenamefont {Denton}\ \emph {et~al.}(2021)\citenamefont {Denton},
  \citenamefont {Gehrlein},\ and\ \citenamefont {Pestes}}]{Denton:2020uda}%
  \BibitemOpen
  \bibfield  {author} {\bibinfo {author} {\bibfnamefont {P.~B.}\ \bibnamefont
  {Denton}}, \bibinfo {author} {\bibfnamefont {J.}~\bibnamefont {Gehrlein}},\
  and\ \bibinfo {author} {\bibfnamefont {R.}~\bibnamefont {Pestes}},\
  }\bibfield  {title} {\bibinfo {title} {{$CP$ -Violating Neutrino Nonstandard
  Interactions in Long-Baseline-Accelerator Data}},\ }\href
  {https://doi.org/10.1103/PhysRevLett.126.051801} {\bibfield  {journal}
  {\bibinfo  {journal} {Phys. Rev. Lett.}\ }\textbf {\bibinfo {volume} {126}},\
  \bibinfo {pages} {051801} (\bibinfo {year} {2021})},\ \Eprint
  {https://arxiv.org/abs/2008.01110} {arXiv:2008.01110 [hep-ph]} \BibitemShut
  {NoStop}%
\bibitem [{\citenamefont {Chatterjee}\ and\ \citenamefont
  {Palazzo}(2021)}]{Chatterjee:2020kkm}%
  \BibitemOpen
  \bibfield  {author} {\bibinfo {author} {\bibfnamefont {S.~S.}\ \bibnamefont
  {Chatterjee}}\ and\ \bibinfo {author} {\bibfnamefont {A.}~\bibnamefont
  {Palazzo}},\ }\bibfield  {title} {\bibinfo {title} {{Nonstandard Neutrino
  Interactions as a Solution to the $NO\nu A$ and T2K Discrepancy}},\ }\href
  {https://doi.org/10.1103/PhysRevLett.126.051802} {\bibfield  {journal}
  {\bibinfo  {journal} {Phys. Rev. Lett.}\ }\textbf {\bibinfo {volume} {126}},\
  \bibinfo {pages} {051802} (\bibinfo {year} {2021})},\ \Eprint
  {https://arxiv.org/abs/2008.04161} {arXiv:2008.04161 [hep-ph]} \BibitemShut
  {NoStop}%
\bibitem [{\citenamefont {Chatterjee}\ and\ \citenamefont
  {Palazzo}(2024)}]{Chatterjee:2024kbn}%
  \BibitemOpen
  \bibfield  {author} {\bibinfo {author} {\bibfnamefont {S.~S.}\ \bibnamefont
  {Chatterjee}}\ and\ \bibinfo {author} {\bibfnamefont {A.}~\bibnamefont
  {Palazzo}},\ }\bibfield  {title} {\bibinfo {title} {{Status of tension
  between NOvA and T2K after Neutrino 2024 and possible role of nonstandard
  neutrino interactions}},\ }\href
  {https://doi.org/10.1103/PhysRevD.110.113002} {\bibfield  {journal} {\bibinfo
   {journal} {Phys. Rev. D}\ }\textbf {\bibinfo {volume} {110}},\ \bibinfo
  {pages} {113002} (\bibinfo {year} {2024})},\ \Eprint
  {https://arxiv.org/abs/2409.10599} {arXiv:2409.10599 [hep-ph]} \BibitemShut
  {NoStop}%
\bibitem [{\citenamefont {Yu}\ \emph {et~al.}(2026)\citenamefont {Yu},
  \citenamefont {Guan}, \citenamefont {Rahaman},\ and\ \citenamefont
  {Ilic}}]{Yu:2024nkc}%
  \BibitemOpen
  \bibfield  {author} {\bibinfo {author} {\bibfnamefont {X.~Y.}\ \bibnamefont
  {Yu}}, \bibinfo {author} {\bibfnamefont {Z.}~\bibnamefont {Guan}}, \bibinfo
  {author} {\bibfnamefont {U.}~\bibnamefont {Rahaman}},\ and\ \bibinfo {author}
  {\bibfnamefont {N.}~\bibnamefont {Ilic}},\ }\bibfield  {title} {\bibinfo
  {title} {{Update on non-unitary mixing in the recent NO{\ensuremath{\nu}}A
  and T2K data}},\ }\href {https://doi.org/10.1016/j.nuclphysb.2026.117303}
  {\bibfield  {journal} {\bibinfo  {journal} {Nucl. Phys. B}\ }\textbf
  {\bibinfo {volume} {1023}},\ \bibinfo {pages} {117303} (\bibinfo {year}
  {2026})},\ \Eprint {https://arxiv.org/abs/2501.00146} {arXiv:2501.00146
  [hep-ph]} \BibitemShut {NoStop}%
\bibitem [{\citenamefont {Chatterjee}\ and\ \citenamefont
  {Palazzo}(2020)}]{Chatterjee:2020yak}%
  \BibitemOpen
  \bibfield  {author} {\bibinfo {author} {\bibfnamefont {S.~S.}\ \bibnamefont
  {Chatterjee}}\ and\ \bibinfo {author} {\bibfnamefont {A.}~\bibnamefont
  {Palazzo}},\ }\bibfield  {title} {\bibinfo {title} {{Interpretation of
  NO$\nu$A and T2K data in the presence of a light sterile neutrino}},\
  }\href@noop {} {\  (\bibinfo {year} {2020})},\ \Eprint
  {https://arxiv.org/abs/2005.10338} {arXiv:2005.10338 [hep-ph]} \BibitemShut
  {NoStop}%
\bibitem [{\citenamefont {Forero}\ \emph {et~al.}(2021)\citenamefont {Forero},
  \citenamefont {Giunti}, \citenamefont {Ternes},\ and\ \citenamefont
  {Tortola}}]{Forero:2021azc}%
  \BibitemOpen
  \bibfield  {author} {\bibinfo {author} {\bibfnamefont {D.~V.}\ \bibnamefont
  {Forero}}, \bibinfo {author} {\bibfnamefont {C.}~\bibnamefont {Giunti}},
  \bibinfo {author} {\bibfnamefont {C.~A.}\ \bibnamefont {Ternes}},\ and\
  \bibinfo {author} {\bibfnamefont {M.}~\bibnamefont {Tortola}},\ }\bibfield
  {title} {\bibinfo {title} {{Nonunitary neutrino mixing in short and
  long-baseline experiments}},\ }\href
  {https://doi.org/10.1103/PhysRevD.104.075030} {\bibfield  {journal} {\bibinfo
   {journal} {Phys. Rev. D}\ }\textbf {\bibinfo {volume} {104}},\ \bibinfo
  {pages} {075030} (\bibinfo {year} {2021})},\ \Eprint
  {https://arxiv.org/abs/2103.01998} {arXiv:2103.01998 [hep-ph]} \BibitemShut
  {NoStop}%
\bibitem [{\citenamefont {Rahaman}(2021)}]{Rahaman:2021leu}%
  \BibitemOpen
  \bibfield  {author} {\bibinfo {author} {\bibfnamefont {U.}~\bibnamefont
  {Rahaman}},\ }\bibfield  {title} {\bibinfo {title} {{Looking for Lorentz
  invariance violation (LIV) in the latest long baseline accelerator neutrino
  oscillation data}},\ }\href {https://doi.org/10.1140/epjc/s10052-021-09598-4}
  {\bibfield  {journal} {\bibinfo  {journal} {Eur. Phys. J. C}\ }\textbf
  {\bibinfo {volume} {81}},\ \bibinfo {pages} {792} (\bibinfo {year} {2021})},\
  \Eprint {https://arxiv.org/abs/2103.04576} {arXiv:2103.04576 [hep-ph]}
  \BibitemShut {NoStop}%
\bibitem [{\citenamefont {de~Gouv{\^e}a}\ \emph
  {et~al.}(2022{\natexlab{a}})\citenamefont {de~Gouv{\^e}a}, \citenamefont
  {Jusino~S{\'a}nchez},\ and\ \citenamefont {Kelly}}]{deGouvea:2022kma}%
  \BibitemOpen
  \bibfield  {author} {\bibinfo {author} {\bibfnamefont {A.}~\bibnamefont
  {de~Gouv{\^e}a}}, \bibinfo {author} {\bibfnamefont {G.}~\bibnamefont
  {Jusino~S{\'a}nchez}},\ and\ \bibinfo {author} {\bibfnamefont {K.~J.}\
  \bibnamefont {Kelly}},\ }\bibfield  {title} {\bibinfo {title} {{Very light
  sterile neutrinos at NOvA and T2K}},\ }\href
  {https://doi.org/10.1103/PhysRevD.106.055025} {\bibfield  {journal} {\bibinfo
   {journal} {Phys. Rev. D}\ }\textbf {\bibinfo {volume} {106}},\ \bibinfo
  {pages} {055025} (\bibinfo {year} {2022}{\natexlab{a}})},\ \Eprint
  {https://arxiv.org/abs/2204.09130} {arXiv:2204.09130 [hep-ph]} \BibitemShut
  {NoStop}%
\bibitem [{\citenamefont {Majhi}\ \emph {et~al.}(2022)\citenamefont {Majhi},
  \citenamefont {Singha}, \citenamefont {Deepthi},\ and\ \citenamefont
  {Mohanta}}]{Majhi:2022wyp}%
  \BibitemOpen
  \bibfield  {author} {\bibinfo {author} {\bibfnamefont {R.}~\bibnamefont
  {Majhi}}, \bibinfo {author} {\bibfnamefont {D.~K.}\ \bibnamefont {Singha}},
  \bibinfo {author} {\bibfnamefont {K.~N.}\ \bibnamefont {Deepthi}},\ and\
  \bibinfo {author} {\bibfnamefont {R.}~\bibnamefont {Mohanta}},\ }\bibfield
  {title} {\bibinfo {title} {{Vector leptoquark $U_3$ and CP violation at T2K,
  NOvA experiments}},\ }\href {https://doi.org/10.1140/epjc/s10052-022-10900-1}
  {\bibfield  {journal} {\bibinfo  {journal} {Eur. Phys. J. C}\ }\textbf
  {\bibinfo {volume} {82}},\ \bibinfo {pages} {919} (\bibinfo {year} {2022})},\
  \Eprint {https://arxiv.org/abs/2205.04269} {arXiv:2205.04269 [hep-ph]}
  \BibitemShut {NoStop}%
\bibitem [{\citenamefont {Denton}\ \emph {et~al.}(2023)\citenamefont {Denton},
  \citenamefont {Giarnetti},\ and\ \citenamefont {Meloni}}]{Denton:2022pxt}%
  \BibitemOpen
  \bibfield  {author} {\bibinfo {author} {\bibfnamefont {P.~B.}\ \bibnamefont
  {Denton}}, \bibinfo {author} {\bibfnamefont {A.}~\bibnamefont {Giarnetti}},\
  and\ \bibinfo {author} {\bibfnamefont {D.}~\bibnamefont {Meloni}},\
  }\bibfield  {title} {\bibinfo {title} {{How to identify different new
  neutrino oscillation physics scenarios at DUNE}},\ }\href
  {https://doi.org/10.1007/JHEP02(2023)210} {\bibfield  {journal} {\bibinfo
  {journal} {JHEP}\ }\textbf {\bibinfo {volume} {02}},\ \bibinfo {pages}
  {210}},\ \Eprint {https://arxiv.org/abs/2210.00109} {arXiv:2210.00109
  [hep-ph]} \BibitemShut {NoStop}%
\bibitem [{\citenamefont {Cherchiglia}\ \emph {et~al.}(2025)\citenamefont
  {Cherchiglia}, \citenamefont {Pasquini}, \citenamefont {Peres}, \citenamefont
  {Rodrigues}, \citenamefont {Rossi},\ and\ \citenamefont
  {Souza}}]{Cherchiglia:2023ojf}%
  \BibitemOpen
  \bibfield  {author} {\bibinfo {author} {\bibfnamefont {A.}~\bibnamefont
  {Cherchiglia}}, \bibinfo {author} {\bibfnamefont {P.}~\bibnamefont
  {Pasquini}}, \bibinfo {author} {\bibfnamefont {O.~L.~G.}\ \bibnamefont
  {Peres}}, \bibinfo {author} {\bibfnamefont {F.~F.}\ \bibnamefont
  {Rodrigues}}, \bibinfo {author} {\bibfnamefont {R.~R.}\ \bibnamefont
  {Rossi}},\ and\ \bibinfo {author} {\bibfnamefont {E.~S.}\ \bibnamefont
  {Souza}},\ }\bibfield  {title} {\bibinfo {title} {{Alleviating the present
  tension between T2K and NO{\ensuremath{\nu}}A with nonstandard neutrino
  interactions}},\ }\href {https://doi.org/10.1103/55qm-zbhv} {\bibfield
  {journal} {\bibinfo  {journal} {Phys. Rev. D}\ }\textbf {\bibinfo {volume}
  {112}},\ \bibinfo {pages} {093004} (\bibinfo {year} {2025})},\ \Eprint
  {https://arxiv.org/abs/2310.18401} {arXiv:2310.18401 [hep-ph]} \BibitemShut
  {NoStop}%
\bibitem [{\citenamefont {Li}\ \emph {et~al.}(2026)\citenamefont {Li},
  \citenamefont {Lin}, \citenamefont {Tang},\ and\ \citenamefont
  {Vihonen}}]{Li:2023mil}%
  \BibitemOpen
  \bibfield  {author} {\bibinfo {author} {\bibfnamefont {X.-Q.}\ \bibnamefont
  {Li}}, \bibinfo {author} {\bibfnamefont {H.-X.}\ \bibnamefont {Lin}},
  \bibinfo {author} {\bibfnamefont {J.}~\bibnamefont {Tang}},\ and\ \bibinfo
  {author} {\bibfnamefont {S.}~\bibnamefont {Vihonen}},\ }\bibfield  {title}
  {\bibinfo {title} {{Ultralight dark matter in long-baseline accelerator
  neutrino oscillations}},\ }\href {https://doi.org/10.1103/twkz-l635}
  {\bibfield  {journal} {\bibinfo  {journal} {Phys. Rev. D}\ }\textbf {\bibinfo
  {volume} {114}},\ \bibinfo {pages} {012009} (\bibinfo {year} {2026})},\
  \Eprint {https://arxiv.org/abs/2312.11704} {arXiv:2312.11704 [hep-ph]}
  \BibitemShut {NoStop}%
\bibitem [{\citenamefont {Acero}\ \emph {et~al.}(2024)\citenamefont {Acero}
  \emph {et~al.}}]{NOvA:2024lti}%
  \BibitemOpen
  \bibfield  {author} {\bibinfo {author} {\bibfnamefont {M.~A.}\ \bibnamefont
  {Acero}} \emph {et~al.} (\bibinfo {collaboration} {NOvA}),\ }\bibfield
  {title} {\bibinfo {title} {{Search for CP-Violating Neutrino Nonstandard
  Interactions with the NOvA Experiment}},\ }\href
  {https://doi.org/10.1103/PhysRevLett.133.201802} {\bibfield  {journal}
  {\bibinfo  {journal} {Phys. Rev. Lett.}\ }\textbf {\bibinfo {volume} {133}},\
  \bibinfo {pages} {201802} (\bibinfo {year} {2024})},\ \Eprint
  {https://arxiv.org/abs/2403.07266} {arXiv:2403.07266 [hep-ex]} \BibitemShut
  {NoStop}%
\bibitem [{\citenamefont {Gehrlein}\ \emph {et~al.}(2026)\citenamefont
  {Gehrlein}, \citenamefont {Hoefken~Zink}, \citenamefont {Machado},\ and\
  \citenamefont {Pinheiro}}]{Gehrlein:2026ier}%
  \BibitemOpen
  \bibfield  {author} {\bibinfo {author} {\bibfnamefont {J.}~\bibnamefont
  {Gehrlein}}, \bibinfo {author} {\bibfnamefont {J.}~\bibnamefont
  {Hoefken~Zink}}, \bibinfo {author} {\bibfnamefont {P.~A.~N.}\ \bibnamefont
  {Machado}},\ and\ \bibinfo {author} {\bibfnamefont {J.~P.}\ \bibnamefont
  {Pinheiro}},\ }\bibfield  {title} {\bibinfo {title} {{Complementarity Between
  Neutrino Neutral and Charged Current Events in the Search for New Physics}},\
  }\href@noop {} {\  (\bibinfo {year} {2026})},\ \Eprint
  {https://arxiv.org/abs/2604.16176} {arXiv:2604.16176 [hep-ph]} \BibitemShut
  {NoStop}%
\bibitem [{\citenamefont {Biggio}\ \emph {et~al.}(2009)\citenamefont {Biggio},
  \citenamefont {Blennow},\ and\ \citenamefont
  {Fernandez-Martinez}}]{Biggio:2009nt}%
  \BibitemOpen
  \bibfield  {author} {\bibinfo {author} {\bibfnamefont {C.}~\bibnamefont
  {Biggio}}, \bibinfo {author} {\bibfnamefont {M.}~\bibnamefont {Blennow}},\
  and\ \bibinfo {author} {\bibfnamefont {E.}~\bibnamefont
  {Fernandez-Martinez}},\ }\bibfield  {title} {\bibinfo {title} {{General
  bounds on non-standard neutrino interactions}},\ }\href
  {https://doi.org/10.1088/1126-6708/2009/08/090} {\bibfield  {journal}
  {\bibinfo  {journal} {JHEP}\ }\textbf {\bibinfo {volume} {08}},\ \bibinfo
  {pages} {090}},\ \Eprint {https://arxiv.org/abs/0907.0097} {arXiv:0907.0097
  [hep-ph]} \BibitemShut {NoStop}%
\bibitem [{\citenamefont {Escrihuela}\ \emph {et~al.}(2011)\citenamefont
  {Escrihuela}, \citenamefont {Tortola}, \citenamefont {Valle},\ and\
  \citenamefont {Miranda}}]{Escrihuela:2011cf}%
  \BibitemOpen
  \bibfield  {author} {\bibinfo {author} {\bibfnamefont {F.~J.}\ \bibnamefont
  {Escrihuela}}, \bibinfo {author} {\bibfnamefont {M.}~\bibnamefont {Tortola}},
  \bibinfo {author} {\bibfnamefont {J.~W.~F.}\ \bibnamefont {Valle}},\ and\
  \bibinfo {author} {\bibfnamefont {O.~G.}\ \bibnamefont {Miranda}},\
  }\bibfield  {title} {\bibinfo {title} {{Global constraints on muon-neutrino
  non-standard interactions}},\ }\href
  {https://doi.org/10.1103/PhysRevD.83.093002} {\bibfield  {journal} {\bibinfo
  {journal} {Phys. Rev. D}\ }\textbf {\bibinfo {volume} {83}},\ \bibinfo
  {pages} {093002} (\bibinfo {year} {2011})},\ \Eprint
  {https://arxiv.org/abs/1103.1366} {arXiv:1103.1366 [hep-ph]} \BibitemShut
  {NoStop}%
\bibitem [{\citenamefont {Gonzalez-Garcia}\ \emph {et~al.}(2011)\citenamefont
  {Gonzalez-Garcia}, \citenamefont {Maltoni},\ and\ \citenamefont
  {Salvado}}]{Gonzalez-Garcia:2011vlg}%
  \BibitemOpen
  \bibfield  {author} {\bibinfo {author} {\bibfnamefont {M.~C.}\ \bibnamefont
  {Gonzalez-Garcia}}, \bibinfo {author} {\bibfnamefont {M.}~\bibnamefont
  {Maltoni}},\ and\ \bibinfo {author} {\bibfnamefont {J.}~\bibnamefont
  {Salvado}},\ }\bibfield  {title} {\bibinfo {title} {{Testing matter effects
  in propagation of atmospheric and long-baseline neutrinos}},\ }\href
  {https://doi.org/10.1007/JHEP05(2011)075} {\bibfield  {journal} {\bibinfo
  {journal} {JHEP}\ }\textbf {\bibinfo {volume} {05}},\ \bibinfo {pages}
  {075}},\ \Eprint {https://arxiv.org/abs/1103.4365} {arXiv:1103.4365 [hep-ph]}
  \BibitemShut {NoStop}%
\bibitem [{\citenamefont {Gonzalez-Garcia}\ and\ \citenamefont
  {Maltoni}(2013)}]{Gonzalez-Garcia:2013usa}%
  \BibitemOpen
  \bibfield  {author} {\bibinfo {author} {\bibfnamefont {M.~C.}\ \bibnamefont
  {Gonzalez-Garcia}}\ and\ \bibinfo {author} {\bibfnamefont {M.}~\bibnamefont
  {Maltoni}},\ }\bibfield  {title} {\bibinfo {title} {{Determination of matter
  potential from global analysis of neutrino oscillation data}},\ }\href
  {https://doi.org/10.1007/JHEP09(2013)152} {\bibfield  {journal} {\bibinfo
  {journal} {JHEP}\ }\textbf {\bibinfo {volume} {09}},\ \bibinfo {pages}
  {152}},\ \Eprint {https://arxiv.org/abs/1307.3092} {arXiv:1307.3092 [hep-ph]}
  \BibitemShut {NoStop}%
\bibitem [{\citenamefont {Coloma}\ \emph {et~al.}(2017)\citenamefont {Coloma},
  \citenamefont {Denton}, \citenamefont {Gonzalez-Garcia}, \citenamefont
  {Maltoni},\ and\ \citenamefont {Schwetz}}]{Coloma:2017egw}%
  \BibitemOpen
  \bibfield  {author} {\bibinfo {author} {\bibfnamefont {P.}~\bibnamefont
  {Coloma}}, \bibinfo {author} {\bibfnamefont {P.~B.}\ \bibnamefont {Denton}},
  \bibinfo {author} {\bibfnamefont {M.~C.}\ \bibnamefont {Gonzalez-Garcia}},
  \bibinfo {author} {\bibfnamefont {M.}~\bibnamefont {Maltoni}},\ and\ \bibinfo
  {author} {\bibfnamefont {T.}~\bibnamefont {Schwetz}},\ }\bibfield  {title}
  {\bibinfo {title} {{Curtailing the Dark Side in Non-Standard Neutrino
  Interactions}},\ }\href {https://doi.org/10.1007/JHEP04(2017)116} {\bibfield
  {journal} {\bibinfo  {journal} {JHEP}\ }\textbf {\bibinfo {volume} {04}},\
  \bibinfo {pages} {116}},\ \Eprint {https://arxiv.org/abs/1701.04828}
  {arXiv:1701.04828 [hep-ph]} \BibitemShut {NoStop}%
\bibitem [{\citenamefont {Esteban}\ \emph {et~al.}(2018)\citenamefont
  {Esteban}, \citenamefont {Gonzalez-Garcia}, \citenamefont {Maltoni},
  \citenamefont {Martinez-Soler},\ and\ \citenamefont
  {Salvado}}]{Esteban:2018ppq}%
  \BibitemOpen
  \bibfield  {author} {\bibinfo {author} {\bibfnamefont {I.}~\bibnamefont
  {Esteban}}, \bibinfo {author} {\bibfnamefont {M.~C.}\ \bibnamefont
  {Gonzalez-Garcia}}, \bibinfo {author} {\bibfnamefont {M.}~\bibnamefont
  {Maltoni}}, \bibinfo {author} {\bibfnamefont {I.}~\bibnamefont
  {Martinez-Soler}},\ and\ \bibinfo {author} {\bibfnamefont {J.}~\bibnamefont
  {Salvado}},\ }\bibfield  {title} {\bibinfo {title} {{Updated constraints on
  non-standard interactions from global analysis of oscillation data}},\ }\href
  {https://doi.org/10.1007/JHEP08(2018)180} {\bibfield  {journal} {\bibinfo
  {journal} {JHEP}\ }\textbf {\bibinfo {volume} {08}},\ \bibinfo {pages}
  {180}},\ \bibinfo {note} {[Addendum: JHEP 12, 152 (2020)]},\ \Eprint
  {https://arxiv.org/abs/1805.04530} {arXiv:1805.04530 [hep-ph]} \BibitemShut
  {NoStop}%
\bibitem [{\citenamefont {Coloma}\ \emph {et~al.}(2020)\citenamefont {Coloma},
  \citenamefont {Esteban}, \citenamefont {Gonzalez-Garcia},\ and\ \citenamefont
  {Maltoni}}]{Coloma:2019mbs}%
  \BibitemOpen
  \bibfield  {author} {\bibinfo {author} {\bibfnamefont {P.}~\bibnamefont
  {Coloma}}, \bibinfo {author} {\bibfnamefont {I.}~\bibnamefont {Esteban}},
  \bibinfo {author} {\bibfnamefont {M.~C.}\ \bibnamefont {Gonzalez-Garcia}},\
  and\ \bibinfo {author} {\bibfnamefont {M.}~\bibnamefont {Maltoni}},\
  }\bibfield  {title} {\bibinfo {title} {{Improved global fit to Non-Standard
  neutrino Interactions using COHERENT energy and timing data}},\ }\href
  {https://doi.org/10.1007/JHEP02(2020)023} {\bibfield  {journal} {\bibinfo
  {journal} {JHEP}\ }\textbf {\bibinfo {volume} {02}},\ \bibinfo {pages}
  {023}},\ \bibinfo {note} {[Addendum: JHEP 12, 071 (2020)]},\ \Eprint
  {https://arxiv.org/abs/1911.09109} {arXiv:1911.09109 [hep-ph]} \BibitemShut
  {NoStop}%
\bibitem [{\citenamefont {Dutta}\ \emph {et~al.}(2020)\citenamefont {Dutta},
  \citenamefont {Lang}, \citenamefont {Liao}, \citenamefont {Sinha},
  \citenamefont {Strigari},\ and\ \citenamefont {Thompson}}]{Dutta:2020che}%
  \BibitemOpen
  \bibfield  {author} {\bibinfo {author} {\bibfnamefont {B.}~\bibnamefont
  {Dutta}}, \bibinfo {author} {\bibfnamefont {R.~F.}\ \bibnamefont {Lang}},
  \bibinfo {author} {\bibfnamefont {S.}~\bibnamefont {Liao}}, \bibinfo {author}
  {\bibfnamefont {S.}~\bibnamefont {Sinha}}, \bibinfo {author} {\bibfnamefont
  {L.}~\bibnamefont {Strigari}},\ and\ \bibinfo {author} {\bibfnamefont
  {A.}~\bibnamefont {Thompson}},\ }\bibfield  {title} {\bibinfo {title} {{A
  global analysis strategy to resolve neutrino NSI degeneracies with scattering
  and oscillation data}},\ }\href {https://doi.org/10.1007/JHEP09(2020)106}
  {\bibfield  {journal} {\bibinfo  {journal} {JHEP}\ }\textbf {\bibinfo
  {volume} {09}},\ \bibinfo {pages} {106}},\ \Eprint
  {https://arxiv.org/abs/2002.03066} {arXiv:2002.03066 [hep-ph]} \BibitemShut
  {NoStop}%
\bibitem [{\citenamefont {Coloma}\ \emph {et~al.}(2023)\citenamefont {Coloma},
  \citenamefont {Gonzalez-Garcia}, \citenamefont {Maltoni}, \citenamefont
  {Pinheiro},\ and\ \citenamefont {Urrea}}]{Coloma:2023ixt}%
  \BibitemOpen
  \bibfield  {author} {\bibinfo {author} {\bibfnamefont {P.}~\bibnamefont
  {Coloma}}, \bibinfo {author} {\bibfnamefont {M.~C.}\ \bibnamefont
  {Gonzalez-Garcia}}, \bibinfo {author} {\bibfnamefont {M.}~\bibnamefont
  {Maltoni}}, \bibinfo {author} {\bibfnamefont {J.~P.}\ \bibnamefont
  {Pinheiro}},\ and\ \bibinfo {author} {\bibfnamefont {S.}~\bibnamefont
  {Urrea}},\ }\bibfield  {title} {\bibinfo {title} {{Global constraints on
  non-standard neutrino interactions with quarks and electrons}},\ }\href
  {https://doi.org/10.1007/JHEP08(2023)032} {\bibfield  {journal} {\bibinfo
  {journal} {JHEP}\ }\textbf {\bibinfo {volume} {08}},\ \bibinfo {pages}
  {032}},\ \Eprint {https://arxiv.org/abs/2305.07698} {arXiv:2305.07698
  [hep-ph]} \BibitemShut {NoStop}%
\bibitem [{\citenamefont {Friedland}\ \emph {et~al.}(2004)\citenamefont
  {Friedland}, \citenamefont {Lunardini},\ and\ \citenamefont
  {Maltoni}}]{Friedland:2004ah}%
  \BibitemOpen
  \bibfield  {author} {\bibinfo {author} {\bibfnamefont {A.}~\bibnamefont
  {Friedland}}, \bibinfo {author} {\bibfnamefont {C.}~\bibnamefont
  {Lunardini}},\ and\ \bibinfo {author} {\bibfnamefont {M.}~\bibnamefont
  {Maltoni}},\ }\bibfield  {title} {\bibinfo {title} {{Atmospheric neutrinos as
  probes of neutrino-matter interactions}},\ }\href
  {https://doi.org/10.1103/PhysRevD.70.111301} {\bibfield  {journal} {\bibinfo
  {journal} {Phys. Rev. D}\ }\textbf {\bibinfo {volume} {70}},\ \bibinfo
  {pages} {111301} (\bibinfo {year} {2004})},\ \Eprint
  {https://arxiv.org/abs/hep-ph/0408264} {arXiv:hep-ph/0408264} \BibitemShut
  {NoStop}%
\bibitem [{\citenamefont {Friedland}\ and\ \citenamefont
  {Lunardini}(2006)}]{Friedland:2006pi}%
  \BibitemOpen
  \bibfield  {author} {\bibinfo {author} {\bibfnamefont {A.}~\bibnamefont
  {Friedland}}\ and\ \bibinfo {author} {\bibfnamefont {C.}~\bibnamefont
  {Lunardini}},\ }\bibfield  {title} {\bibinfo {title} {{Two modes of searching
  for new neutrino interactions at MINOS}},\ }\href
  {https://doi.org/10.1103/PhysRevD.74.033012} {\bibfield  {journal} {\bibinfo
  {journal} {Phys. Rev. D}\ }\textbf {\bibinfo {volume} {74}},\ \bibinfo
  {pages} {033012} (\bibinfo {year} {2006})},\ \Eprint
  {https://arxiv.org/abs/hep-ph/0606101} {arXiv:hep-ph/0606101} \BibitemShut
  {NoStop}%
\bibitem [{\citenamefont {Kikuchi}\ \emph {et~al.}(2009)\citenamefont
  {Kikuchi}, \citenamefont {Minakata},\ and\ \citenamefont
  {Uchinami}}]{Kikuchi:2008vq}%
  \BibitemOpen
  \bibfield  {author} {\bibinfo {author} {\bibfnamefont {T.}~\bibnamefont
  {Kikuchi}}, \bibinfo {author} {\bibfnamefont {H.}~\bibnamefont {Minakata}},\
  and\ \bibinfo {author} {\bibfnamefont {S.}~\bibnamefont {Uchinami}},\
  }\bibfield  {title} {\bibinfo {title} {{Perturbation Theory of Neutrino
  Oscillation with Nonstandard Neutrino Interactions}},\ }\href
  {https://doi.org/10.1088/1126-6708/2009/03/114} {\bibfield  {journal}
  {\bibinfo  {journal} {JHEP}\ }\textbf {\bibinfo {volume} {03}},\ \bibinfo
  {pages} {114}},\ \Eprint {https://arxiv.org/abs/0809.3312} {arXiv:0809.3312
  [hep-ph]} \BibitemShut {NoStop}%
\bibitem [{\citenamefont {Coloma}\ \emph {et~al.}(2011)\citenamefont {Coloma},
  \citenamefont {Donini}, \citenamefont {Lopez-Pavon},\ and\ \citenamefont
  {Minakata}}]{Coloma:2011rq}%
  \BibitemOpen
  \bibfield  {author} {\bibinfo {author} {\bibfnamefont {P.}~\bibnamefont
  {Coloma}}, \bibinfo {author} {\bibfnamefont {A.}~\bibnamefont {Donini}},
  \bibinfo {author} {\bibfnamefont {J.}~\bibnamefont {Lopez-Pavon}},\ and\
  \bibinfo {author} {\bibfnamefont {H.}~\bibnamefont {Minakata}},\ }\bibfield
  {title} {\bibinfo {title} {{Non-Standard Interactions at a Neutrino Factory:
  Correlations and CP violation}},\ }\href
  {https://doi.org/10.1007/JHEP08(2011)036} {\bibfield  {journal} {\bibinfo
  {journal} {JHEP}\ }\textbf {\bibinfo {volume} {08}},\ \bibinfo {pages}
  {036}},\ \Eprint {https://arxiv.org/abs/1105.5936} {arXiv:1105.5936 [hep-ph]}
  \BibitemShut {NoStop}%
\bibitem [{\citenamefont {Coloma}(2016)}]{Coloma:2015kiu}%
  \BibitemOpen
  \bibfield  {author} {\bibinfo {author} {\bibfnamefont {P.}~\bibnamefont
  {Coloma}},\ }\bibfield  {title} {\bibinfo {title} {{Non-Standard Interactions
  in propagation at the Deep Underground Neutrino Experiment}},\ }\href
  {https://doi.org/10.1007/JHEP03(2016)016} {\bibfield  {journal} {\bibinfo
  {journal} {JHEP}\ }\textbf {\bibinfo {volume} {03}},\ \bibinfo {pages}
  {016}},\ \Eprint {https://arxiv.org/abs/1511.06357} {arXiv:1511.06357
  [hep-ph]} \BibitemShut {NoStop}%
\bibitem [{\citenamefont {Liao}\ \emph {et~al.}(2016)\citenamefont {Liao},
  \citenamefont {Marfatia},\ and\ \citenamefont {Whisnant}}]{Liao:2016hsa}%
  \BibitemOpen
  \bibfield  {author} {\bibinfo {author} {\bibfnamefont {J.}~\bibnamefont
  {Liao}}, \bibinfo {author} {\bibfnamefont {D.}~\bibnamefont {Marfatia}},\
  and\ \bibinfo {author} {\bibfnamefont {K.}~\bibnamefont {Whisnant}},\
  }\bibfield  {title} {\bibinfo {title} {{Degeneracies in long-baseline
  neutrino experiments from nonstandard interactions}},\ }\href
  {https://doi.org/10.1103/PhysRevD.93.093016} {\bibfield  {journal} {\bibinfo
  {journal} {Phys. Rev. D}\ }\textbf {\bibinfo {volume} {93}},\ \bibinfo
  {pages} {093016} (\bibinfo {year} {2016})},\ \Eprint
  {https://arxiv.org/abs/1601.00927} {arXiv:1601.00927 [hep-ph]} \BibitemShut
  {NoStop}%
\bibitem [{\citenamefont {Coloma}\ and\ \citenamefont
  {Schwetz}(2016)}]{Coloma:2016gei}%
  \BibitemOpen
  \bibfield  {author} {\bibinfo {author} {\bibfnamefont {P.}~\bibnamefont
  {Coloma}}\ and\ \bibinfo {author} {\bibfnamefont {T.}~\bibnamefont
  {Schwetz}},\ }\bibfield  {title} {\bibinfo {title} {{Generalized mass
  ordering degeneracy in neutrino oscillation experiments}},\ }\href
  {https://doi.org/10.1103/PhysRevD.94.055005} {\bibfield  {journal} {\bibinfo
  {journal} {Phys. Rev. D}\ }\textbf {\bibinfo {volume} {94}},\ \bibinfo
  {pages} {055005} (\bibinfo {year} {2016})},\ \bibinfo {note} {[Erratum:
  Phys.Rev.D 95, 079903 (2017)]},\ \Eprint {https://arxiv.org/abs/1604.05772}
  {arXiv:1604.05772 [hep-ph]} \BibitemShut {NoStop}%
\bibitem [{\citenamefont {Chatterjee}\ \emph {et~al.}(2021)\citenamefont
  {Chatterjee}, \citenamefont {Dev},\ and\ \citenamefont
  {Machado}}]{Chatterjee:2021wac}%
  \BibitemOpen
  \bibfield  {author} {\bibinfo {author} {\bibfnamefont {S.~S.}\ \bibnamefont
  {Chatterjee}}, \bibinfo {author} {\bibfnamefont {P.~S.~B.}\ \bibnamefont
  {Dev}},\ and\ \bibinfo {author} {\bibfnamefont {P.~A.~N.}\ \bibnamefont
  {Machado}},\ }\bibfield  {title} {\bibinfo {title} {{Impact of improved
  energy resolution on DUNE sensitivity to neutrino non-standard
  interactions}},\ }\href {https://doi.org/10.1007/JHEP08(2021)163} {\bibfield
  {journal} {\bibinfo  {journal} {JHEP}\ }\textbf {\bibinfo {volume} {08}},\
  \bibinfo {pages} {163}},\ \Eprint {https://arxiv.org/abs/2106.04597}
  {arXiv:2106.04597 [hep-ph]} \BibitemShut {NoStop}%
\bibitem [{\citenamefont {Denton}\ \emph {et~al.}(2025)\citenamefont {Denton},
  \citenamefont {Gehrlein},\ and\ \citenamefont {Kong}}]{Denton:2025kvy}%
  \BibitemOpen
  \bibfield  {author} {\bibinfo {author} {\bibfnamefont {P.~B.}\ \bibnamefont
  {Denton}}, \bibinfo {author} {\bibfnamefont {J.}~\bibnamefont {Gehrlein}},\
  and\ \bibinfo {author} {\bibfnamefont {C.-F.}\ \bibnamefont {Kong}},\
  }\bibfield  {title} {\bibinfo {title} {{Testing new physics in oscillations
  at a neutrino factory}},\ }\href
  {https://doi.org/10.1016/j.nuclphysb.2025.117040} {\bibfield  {journal}
  {\bibinfo  {journal} {Nucl. Phys. B}\ }\textbf {\bibinfo {volume} {1018}},\
  \bibinfo {pages} {117040} (\bibinfo {year} {2025})},\ \Eprint
  {https://arxiv.org/abs/2502.14027} {arXiv:2502.14027 [hep-ph]} \BibitemShut
  {NoStop}%
\bibitem [{\citenamefont {Bakhti}\ and\ \citenamefont
  {Farzan}(2014)}]{Bakhti:2014pva}%
  \BibitemOpen
  \bibfield  {author} {\bibinfo {author} {\bibfnamefont {P.}~\bibnamefont
  {Bakhti}}\ and\ \bibinfo {author} {\bibfnamefont {Y.}~\bibnamefont
  {Farzan}},\ }\bibfield  {title} {\bibinfo {title} {{Shedding light on
  LMA-Dark solar neutrino solution by medium baseline reactor experiments: JUNO
  and RENO-50}},\ }\href {https://doi.org/10.1007/JHEP07(2014)064} {\bibfield
  {journal} {\bibinfo  {journal} {JHEP}\ }\textbf {\bibinfo {volume} {07}},\
  \bibinfo {pages} {064}},\ \Eprint {https://arxiv.org/abs/1403.0744}
  {arXiv:1403.0744 [hep-ph]} \BibitemShut {NoStop}%
\bibitem [{\citenamefont {Denton}\ \emph {et~al.}(2018)\citenamefont {Denton},
  \citenamefont {Farzan},\ and\ \citenamefont {Shoemaker}}]{Denton:2018xmq}%
  \BibitemOpen
  \bibfield  {author} {\bibinfo {author} {\bibfnamefont {P.~B.}\ \bibnamefont
  {Denton}}, \bibinfo {author} {\bibfnamefont {Y.}~\bibnamefont {Farzan}},\
  and\ \bibinfo {author} {\bibfnamefont {I.~M.}\ \bibnamefont {Shoemaker}},\
  }\bibfield  {title} {\bibinfo {title} {{Testing large non-standard neutrino
  interactions with arbitrary mediator mass after COHERENT data}},\ }\href
  {https://doi.org/10.1007/JHEP07(2018)037} {\bibfield  {journal} {\bibinfo
  {journal} {JHEP}\ }\textbf {\bibinfo {volume} {07}},\ \bibinfo {pages}
  {037}},\ \Eprint {https://arxiv.org/abs/1804.03660} {arXiv:1804.03660
  [hep-ph]} \BibitemShut {NoStop}%
\bibitem [{\citenamefont {Denton}\ and\ \citenamefont
  {Gehrlein}(2022)}]{Denton:2022nol}%
  \BibitemOpen
  \bibfield  {author} {\bibinfo {author} {\bibfnamefont {P.~B.}\ \bibnamefont
  {Denton}}\ and\ \bibinfo {author} {\bibfnamefont {J.}~\bibnamefont
  {Gehrlein}},\ }\bibfield  {title} {\bibinfo {title} {{New constraints on the
  dark side of non-standard interactions from reactor neutrino scattering
  data}},\ }\href {https://doi.org/10.1103/PhysRevD.106.015022} {\bibfield
  {journal} {\bibinfo  {journal} {Phys. Rev. D}\ }\textbf {\bibinfo {volume}
  {106}},\ \bibinfo {pages} {015022} (\bibinfo {year} {2022})},\ \Eprint
  {https://arxiv.org/abs/2204.09060} {arXiv:2204.09060 [hep-ph]} \BibitemShut
  {NoStop}%
\bibitem [{\citenamefont {Mitsuka}\ \emph {et~al.}(2011)\citenamefont {Mitsuka}
  \emph {et~al.}}]{Super-Kamiokande:2011dam}%
  \BibitemOpen
  \bibfield  {author} {\bibinfo {author} {\bibfnamefont {G.}~\bibnamefont
  {Mitsuka}} \emph {et~al.} (\bibinfo {collaboration} {Super-Kamiokande}),\
  }\bibfield  {title} {\bibinfo {title} {{Study of Non-Standard Neutrino
  Interactions with Atmospheric Neutrino Data in Super-Kamiokande I and II}},\
  }\href {https://doi.org/10.1103/PhysRevD.84.113008} {\bibfield  {journal}
  {\bibinfo  {journal} {Phys. Rev. D}\ }\textbf {\bibinfo {volume} {84}},\
  \bibinfo {pages} {113008} (\bibinfo {year} {2011})},\ \Eprint
  {https://arxiv.org/abs/1109.1889} {arXiv:1109.1889 [hep-ex]} \BibitemShut
  {NoStop}%
\bibitem [{\citenamefont {Salvado}\ \emph {et~al.}(2017)\citenamefont
  {Salvado}, \citenamefont {Mena}, \citenamefont {Palomares-Ruiz},\ and\
  \citenamefont {Rius}}]{Salvado:2016uqu}%
  \BibitemOpen
  \bibfield  {author} {\bibinfo {author} {\bibfnamefont {J.}~\bibnamefont
  {Salvado}}, \bibinfo {author} {\bibfnamefont {O.}~\bibnamefont {Mena}},
  \bibinfo {author} {\bibfnamefont {S.}~\bibnamefont {Palomares-Ruiz}},\ and\
  \bibinfo {author} {\bibfnamefont {N.}~\bibnamefont {Rius}},\ }\bibfield
  {title} {\bibinfo {title} {{Non-standard interactions with high-energy
  atmospheric neutrinos at IceCube}},\ }\href
  {https://doi.org/10.1007/JHEP01(2017)141} {\bibfield  {journal} {\bibinfo
  {journal} {JHEP}\ }\textbf {\bibinfo {volume} {01}},\ \bibinfo {pages}
  {141}},\ \Eprint {https://arxiv.org/abs/1609.03450} {arXiv:1609.03450
  [hep-ph]} \BibitemShut {NoStop}%
\bibitem [{\citenamefont {Abbasi}\ \emph {et~al.}(2021)\citenamefont {Abbasi}
  \emph {et~al.}}]{IceCubeCollaboration:2021euf}%
  \BibitemOpen
  \bibfield  {author} {\bibinfo {author} {\bibfnamefont {R.}~\bibnamefont
  {Abbasi}} \emph {et~al.} (\bibinfo {collaboration} {(IceCube Collaboration)*,
  IceCube}),\ }\bibfield  {title} {\bibinfo {title} {{All-flavor constraints on
  nonstandard neutrino interactions and generalized matter potential with three
  years of IceCube DeepCore data}},\ }\href
  {https://doi.org/10.1103/PhysRevD.104.072006} {\bibfield  {journal} {\bibinfo
   {journal} {Phys. Rev. D}\ }\textbf {\bibinfo {volume} {104}},\ \bibinfo
  {pages} {072006} (\bibinfo {year} {2021})},\ \Eprint
  {https://arxiv.org/abs/2106.07755} {arXiv:2106.07755 [hep-ex]} \BibitemShut
  {NoStop}%
\bibitem [{\citenamefont {Abbasi}\ \emph {et~al.}(2022)\citenamefont {Abbasi}
  \emph {et~al.}}]{IceCube:2022ubv}%
  \BibitemOpen
  \bibfield  {author} {\bibinfo {author} {\bibfnamefont {R.}~\bibnamefont
  {Abbasi}} \emph {et~al.} (\bibinfo {collaboration} {IceCube}),\ }\bibfield
  {title} {\bibinfo {title} {{Strong Constraints on Neutrino Nonstandard
  Interactions from TeV-Scale $\nu_u$ Disappearance at IceCube}},\ }\href
  {https://doi.org/10.1103/PhysRevLett.129.011804} {\bibfield  {journal}
  {\bibinfo  {journal} {Phys. Rev. Lett.}\ }\textbf {\bibinfo {volume} {129}},\
  \bibinfo {pages} {011804} (\bibinfo {year} {2022})},\ \Eprint
  {https://arxiv.org/abs/2201.03566} {arXiv:2201.03566 [hep-ex]} \BibitemShut
  {NoStop}%
\bibitem [{\citenamefont {Weatherly}\ \emph {et~al.}(2022)\citenamefont
  {Weatherly} \emph {et~al.}}]{Super-Kamiokande:2022lyl}%
  \BibitemOpen
  \bibfield  {author} {\bibinfo {author} {\bibfnamefont {P.}~\bibnamefont
  {Weatherly}} \emph {et~al.} (\bibinfo {collaboration} {Super-Kamiokande}),\
  }\bibfield  {title} {\bibinfo {title} {{Testing Non-Standard Interactions
  Between Solar Neutrinos and Quarks with Super-Kamiokande}},\ }\href@noop {}
  {\  (\bibinfo {year} {2022})},\ \Eprint {https://arxiv.org/abs/2203.11772}
  {arXiv:2203.11772 [hep-ex]} \BibitemShut {NoStop}%
\bibitem [{\citenamefont {Scholberg}(2006)}]{Scholberg:2005qs}%
  \BibitemOpen
  \bibfield  {author} {\bibinfo {author} {\bibfnamefont {K.}~\bibnamefont
  {Scholberg}},\ }\bibfield  {title} {\bibinfo {title} {{Prospects for
  measuring coherent neutrino-nucleus elastic scattering at a stopped-pion
  neutrino source}},\ }\href {https://doi.org/10.1103/PhysRevD.73.033005}
  {\bibfield  {journal} {\bibinfo  {journal} {Phys. Rev. D}\ }\textbf {\bibinfo
  {volume} {73}},\ \bibinfo {pages} {033005} (\bibinfo {year} {2006})},\
  \Eprint {https://arxiv.org/abs/hep-ex/0511042} {arXiv:hep-ex/0511042}
  \BibitemShut {NoStop}%
\bibitem [{\citenamefont {Shoemaker}(2017)}]{Shoemaker:2017lzs}%
  \BibitemOpen
  \bibfield  {author} {\bibinfo {author} {\bibfnamefont {I.~M.}\ \bibnamefont
  {Shoemaker}},\ }\bibfield  {title} {\bibinfo {title} {{COHERENT search
  strategy for beyond standard model neutrino interactions}},\ }\href
  {https://doi.org/10.1103/PhysRevD.95.115028} {\bibfield  {journal} {\bibinfo
  {journal} {Phys. Rev. D}\ }\textbf {\bibinfo {volume} {95}},\ \bibinfo
  {pages} {115028} (\bibinfo {year} {2017})},\ \Eprint
  {https://arxiv.org/abs/1703.05774} {arXiv:1703.05774 [hep-ph]} \BibitemShut
  {NoStop}%
\bibitem [{\citenamefont {Liao}\ and\ \citenamefont
  {Marfatia}(2017)}]{Liao:2017uzy}%
  \BibitemOpen
  \bibfield  {author} {\bibinfo {author} {\bibfnamefont {J.}~\bibnamefont
  {Liao}}\ and\ \bibinfo {author} {\bibfnamefont {D.}~\bibnamefont
  {Marfatia}},\ }\bibfield  {title} {\bibinfo {title} {{COHERENT constraints on
  nonstandard neutrino interactions}},\ }\href
  {https://doi.org/10.1016/j.physletb.2017.10.046} {\bibfield  {journal}
  {\bibinfo  {journal} {Phys. Lett. B}\ }\textbf {\bibinfo {volume} {775}},\
  \bibinfo {pages} {54} (\bibinfo {year} {2017})},\ \Eprint
  {https://arxiv.org/abs/1708.04255} {arXiv:1708.04255 [hep-ph]} \BibitemShut
  {NoStop}%
\bibitem [{\citenamefont {Papoulias}\ and\ \citenamefont
  {Kosmas}(2018)}]{Papoulias:2017qdn}%
  \BibitemOpen
  \bibfield  {author} {\bibinfo {author} {\bibfnamefont {D.~K.}\ \bibnamefont
  {Papoulias}}\ and\ \bibinfo {author} {\bibfnamefont {T.~S.}\ \bibnamefont
  {Kosmas}},\ }\bibfield  {title} {\bibinfo {title} {{COHERENT constraints to
  conventional and exotic neutrino physics}},\ }\href
  {https://doi.org/10.1103/PhysRevD.97.033003} {\bibfield  {journal} {\bibinfo
  {journal} {Phys. Rev. D}\ }\textbf {\bibinfo {volume} {97}},\ \bibinfo
  {pages} {033003} (\bibinfo {year} {2018})},\ \Eprint
  {https://arxiv.org/abs/1711.09773} {arXiv:1711.09773 [hep-ph]} \BibitemShut
  {NoStop}%
\bibitem [{\citenamefont {Farzan}\ \emph {et~al.}(2018)\citenamefont {Farzan},
  \citenamefont {Lindner}, \citenamefont {Rodejohann},\ and\ \citenamefont
  {Xu}}]{Farzan:2018gtr}%
  \BibitemOpen
  \bibfield  {author} {\bibinfo {author} {\bibfnamefont {Y.}~\bibnamefont
  {Farzan}}, \bibinfo {author} {\bibfnamefont {M.}~\bibnamefont {Lindner}},
  \bibinfo {author} {\bibfnamefont {W.}~\bibnamefont {Rodejohann}},\ and\
  \bibinfo {author} {\bibfnamefont {X.-J.}\ \bibnamefont {Xu}},\ }\bibfield
  {title} {\bibinfo {title} {{Probing neutrino coupling to a light scalar with
  coherent neutrino scattering}},\ }\href
  {https://doi.org/10.1007/JHEP05(2018)066} {\bibfield  {journal} {\bibinfo
  {journal} {JHEP}\ }\textbf {\bibinfo {volume} {05}},\ \bibinfo {pages}
  {066}},\ \Eprint {https://arxiv.org/abs/1802.05171} {arXiv:1802.05171
  [hep-ph]} \BibitemShut {NoStop}%
\bibitem [{\citenamefont {Abdullah}\ \emph {et~al.}(2018)\citenamefont
  {Abdullah}, \citenamefont {Dent}, \citenamefont {Dutta}, \citenamefont
  {Kane}, \citenamefont {Liao},\ and\ \citenamefont
  {Strigari}}]{Abdullah:2018ykz}%
  \BibitemOpen
  \bibfield  {author} {\bibinfo {author} {\bibfnamefont {M.}~\bibnamefont
  {Abdullah}}, \bibinfo {author} {\bibfnamefont {J.~B.}\ \bibnamefont {Dent}},
  \bibinfo {author} {\bibfnamefont {B.}~\bibnamefont {Dutta}}, \bibinfo
  {author} {\bibfnamefont {G.~L.}\ \bibnamefont {Kane}}, \bibinfo {author}
  {\bibfnamefont {S.}~\bibnamefont {Liao}},\ and\ \bibinfo {author}
  {\bibfnamefont {L.~E.}\ \bibnamefont {Strigari}},\ }\bibfield  {title}
  {\bibinfo {title} {{Coherent elastic neutrino nucleus scattering as a probe
  of a Z' through kinetic and mass mixing effects}},\ }\href
  {https://doi.org/10.1103/PhysRevD.98.015005} {\bibfield  {journal} {\bibinfo
  {journal} {Phys. Rev. D}\ }\textbf {\bibinfo {volume} {98}},\ \bibinfo
  {pages} {015005} (\bibinfo {year} {2018})},\ \Eprint
  {https://arxiv.org/abs/1803.01224} {arXiv:1803.01224 [hep-ph]} \BibitemShut
  {NoStop}%
\bibitem [{\citenamefont {Dutta}\ \emph {et~al.}(2019)\citenamefont {Dutta},
  \citenamefont {Liao}, \citenamefont {Sinha},\ and\ \citenamefont
  {Strigari}}]{Dutta:2019eml}%
  \BibitemOpen
  \bibfield  {author} {\bibinfo {author} {\bibfnamefont {B.}~\bibnamefont
  {Dutta}}, \bibinfo {author} {\bibfnamefont {S.}~\bibnamefont {Liao}},
  \bibinfo {author} {\bibfnamefont {S.}~\bibnamefont {Sinha}},\ and\ \bibinfo
  {author} {\bibfnamefont {L.~E.}\ \bibnamefont {Strigari}},\ }\bibfield
  {title} {\bibinfo {title} {{Searching for Beyond the Standard Model Physics
  with COHERENT Energy and Timing Data}},\ }\href
  {https://doi.org/10.1103/PhysRevLett.123.061801} {\bibfield  {journal}
  {\bibinfo  {journal} {Phys. Rev. Lett.}\ }\textbf {\bibinfo {volume} {123}},\
  \bibinfo {pages} {061801} (\bibinfo {year} {2019})},\ \Eprint
  {https://arxiv.org/abs/1903.10666} {arXiv:1903.10666 [hep-ph]} \BibitemShut
  {NoStop}%
\bibitem [{\citenamefont {Dorenbosch}\ \emph {et~al.}(1986)\citenamefont
  {Dorenbosch} \emph {et~al.}}]{CHARM:1986vuz}%
  \BibitemOpen
  \bibfield  {author} {\bibinfo {author} {\bibfnamefont {J.}~\bibnamefont
  {Dorenbosch}} \emph {et~al.} (\bibinfo {collaboration} {CHARM}),\ }\bibfield
  {title} {\bibinfo {title} {{Experimental Verification of the Universality of
  $\nu_e$ and $\nu_\mu$ Coupling to the Neutral Weak Current}},\ }\href
  {https://doi.org/10.1016/0370-2693(86)90315-1} {\bibfield  {journal}
  {\bibinfo  {journal} {Phys. Lett. B}\ }\textbf {\bibinfo {volume} {180}},\
  \bibinfo {pages} {303} (\bibinfo {year} {1986})}\BibitemShut {NoStop}%
\bibitem [{\citenamefont {Zeller}\ \emph {et~al.}(2002)\citenamefont {Zeller}
  \emph {et~al.}}]{NuTeV:2001whx}%
  \BibitemOpen
  \bibfield  {author} {\bibinfo {author} {\bibfnamefont {G.~P.}\ \bibnamefont
  {Zeller}} \emph {et~al.} (\bibinfo {collaboration} {NuTeV}),\ }\bibfield
  {title} {\bibinfo {title} {{A Precise Determination of Electroweak Parameters
  in Neutrino Nucleon Scattering}},\ }\href
  {https://doi.org/10.1103/PhysRevLett.88.091802} {\bibfield  {journal}
  {\bibinfo  {journal} {Phys. Rev. Lett.}\ }\textbf {\bibinfo {volume} {88}},\
  \bibinfo {pages} {091802} (\bibinfo {year} {2002})},\ \bibinfo {note}
  {[Erratum: Phys.Rev.Lett. 90, 239902 (2003)]},\ \Eprint
  {https://arxiv.org/abs/hep-ex/0110059} {arXiv:hep-ex/0110059} \BibitemShut
  {NoStop}%
\bibitem [{\citenamefont {Kamada}\ and\ \citenamefont
  {Yu}(2015)}]{Kamada:2015era}%
  \BibitemOpen
  \bibfield  {author} {\bibinfo {author} {\bibfnamefont {A.}~\bibnamefont
  {Kamada}}\ and\ \bibinfo {author} {\bibfnamefont {H.-B.}\ \bibnamefont
  {Yu}},\ }\bibfield  {title} {\bibinfo {title} {{Coherent Propagation of PeV
  Neutrinos and the Dip in the Neutrino Spectrum at IceCube}},\ }\href
  {https://doi.org/10.1103/PhysRevD.92.113004} {\bibfield  {journal} {\bibinfo
  {journal} {Phys. Rev. D}\ }\textbf {\bibinfo {volume} {92}},\ \bibinfo
  {pages} {113004} (\bibinfo {year} {2015})},\ \Eprint
  {https://arxiv.org/abs/1504.00711} {arXiv:1504.00711 [hep-ph]} \BibitemShut
  {NoStop}%
\bibitem [{\citenamefont {Huang}\ \emph {et~al.}(2018)\citenamefont {Huang},
  \citenamefont {Ohlsson},\ and\ \citenamefont {Zhou}}]{Huang:2017egl}%
  \BibitemOpen
  \bibfield  {author} {\bibinfo {author} {\bibfnamefont {G.-y.}\ \bibnamefont
  {Huang}}, \bibinfo {author} {\bibfnamefont {T.}~\bibnamefont {Ohlsson}},\
  and\ \bibinfo {author} {\bibfnamefont {S.}~\bibnamefont {Zhou}},\ }\bibfield
  {title} {\bibinfo {title} {{Observational Constraints on Secret Neutrino
  Interactions from Big Bang Nucleosynthesis}},\ }\href
  {https://doi.org/10.1103/PhysRevD.97.075009} {\bibfield  {journal} {\bibinfo
  {journal} {Phys. Rev. D}\ }\textbf {\bibinfo {volume} {97}},\ \bibinfo
  {pages} {075009} (\bibinfo {year} {2018})},\ \Eprint
  {https://arxiv.org/abs/1712.04792} {arXiv:1712.04792 [hep-ph]} \BibitemShut
  {NoStop}%
\bibitem [{\citenamefont {Freedman}(1974)}]{Freedman:1973yd}%
  \BibitemOpen
  \bibfield  {author} {\bibinfo {author} {\bibfnamefont {D.~Z.}\ \bibnamefont
  {Freedman}},\ }\bibfield  {title} {\bibinfo {title} {{Coherent Neutrino
  Nucleus Scattering as a Probe of the Weak Neutral Current}},\ }\href
  {https://doi.org/10.1103/PhysRevD.9.1389} {\bibfield  {journal} {\bibinfo
  {journal} {Phys. Rev. D}\ }\textbf {\bibinfo {volume} {9}},\ \bibinfo {pages}
  {1389} (\bibinfo {year} {1974})}\BibitemShut {NoStop}%
\bibitem [{\citenamefont {Akimov}\ \emph {et~al.}(2017)\citenamefont {Akimov}
  \emph {et~al.}}]{COHERENT:2017ipa}%
  \BibitemOpen
  \bibfield  {author} {\bibinfo {author} {\bibfnamefont {D.}~\bibnamefont
  {Akimov}} \emph {et~al.} (\bibinfo {collaboration} {COHERENT}),\ }\bibfield
  {title} {\bibinfo {title} {{Observation of Coherent Elastic Neutrino-Nucleus
  Scattering}},\ }\href {https://doi.org/10.1126/science.aao0990} {\bibfield
  {journal} {\bibinfo  {journal} {Science}\ }\textbf {\bibinfo {volume}
  {357}},\ \bibinfo {pages} {1123} (\bibinfo {year} {2017})},\ \Eprint
  {https://arxiv.org/abs/1708.01294} {arXiv:1708.01294 [nucl-ex]} \BibitemShut
  {NoStop}%
\bibitem [{\citenamefont {Monroe}\ and\ \citenamefont
  {Fisher}(2007)}]{Monroe:2007xp}%
  \BibitemOpen
  \bibfield  {author} {\bibinfo {author} {\bibfnamefont {J.}~\bibnamefont
  {Monroe}}\ and\ \bibinfo {author} {\bibfnamefont {P.}~\bibnamefont
  {Fisher}},\ }\bibfield  {title} {\bibinfo {title} {{Neutrino Backgrounds to
  Dark Matter Searches}},\ }\href {https://doi.org/10.1103/PhysRevD.76.033007}
  {\bibfield  {journal} {\bibinfo  {journal} {Phys. Rev. D}\ }\textbf {\bibinfo
  {volume} {76}},\ \bibinfo {pages} {033007} (\bibinfo {year} {2007})},\
  \Eprint {https://arxiv.org/abs/0706.3019} {arXiv:0706.3019 [astro-ph]}
  \BibitemShut {NoStop}%
\bibitem [{\citenamefont {Vergados}\ and\ \citenamefont
  {Ejiri}(2008)}]{Vergados:2008jp}%
  \BibitemOpen
  \bibfield  {author} {\bibinfo {author} {\bibfnamefont {J.~D.}\ \bibnamefont
  {Vergados}}\ and\ \bibinfo {author} {\bibfnamefont {H.}~\bibnamefont
  {Ejiri}},\ }\bibfield  {title} {\bibinfo {title} {{Can Solar Neutrinos be a
  Serious Background in Direct Dark Matter Searches?}},\ }\href
  {https://doi.org/10.1016/j.nuclphysb.2008.06.004} {\bibfield  {journal}
  {\bibinfo  {journal} {Nucl. Phys. B}\ }\textbf {\bibinfo {volume} {804}},\
  \bibinfo {pages} {144} (\bibinfo {year} {2008})},\ \Eprint
  {https://arxiv.org/abs/0805.2583} {arXiv:0805.2583 [hep-ph]} \BibitemShut
  {NoStop}%
\bibitem [{\citenamefont {Strigari}(2009)}]{Strigari:2009bq}%
  \BibitemOpen
  \bibfield  {author} {\bibinfo {author} {\bibfnamefont {L.~E.}\ \bibnamefont
  {Strigari}},\ }\bibfield  {title} {\bibinfo {title} {{Neutrino Coherent
  Scattering Rates at Direct Dark Matter Detectors}},\ }\href
  {https://doi.org/10.1088/1367-2630/11/10/105011} {\bibfield  {journal}
  {\bibinfo  {journal} {New J. Phys.}\ }\textbf {\bibinfo {volume} {11}},\
  \bibinfo {pages} {105011} (\bibinfo {year} {2009})},\ \Eprint
  {https://arxiv.org/abs/0903.3630} {arXiv:0903.3630 [astro-ph.CO]}
  \BibitemShut {NoStop}%
\bibitem [{\citenamefont {Gutlein}\ \emph {et~al.}(2010)\citenamefont {Gutlein}
  \emph {et~al.}}]{Gutlein:2010tq}%
  \BibitemOpen
  \bibfield  {author} {\bibinfo {author} {\bibfnamefont {A.}~\bibnamefont
  {Gutlein}} \emph {et~al.},\ }\bibfield  {title} {\bibinfo {title} {{Solar and
  atmospheric neutrinos: Background sources for the direct dark matter
  search}},\ }\href {https://doi.org/10.1016/j.astropartphys.2010.06.002}
  {\bibfield  {journal} {\bibinfo  {journal} {Astropart. Phys.}\ }\textbf
  {\bibinfo {volume} {34}},\ \bibinfo {pages} {90} (\bibinfo {year} {2010})},\
  \Eprint {https://arxiv.org/abs/1003.5530} {arXiv:1003.5530 [hep-ph]}
  \BibitemShut {NoStop}%
\bibitem [{\citenamefont {O'Hare}(2021)}]{OHare:2021utq}%
  \BibitemOpen
  \bibfield  {author} {\bibinfo {author} {\bibfnamefont {C.~A.~J.}\
  \bibnamefont {O'Hare}},\ }\bibfield  {title} {\bibinfo {title} {{New
  Definition of the Neutrino Floor for Direct Dark Matter Searches}},\ }\href
  {https://doi.org/10.1103/PhysRevLett.127.251802} {\bibfield  {journal}
  {\bibinfo  {journal} {Phys. Rev. Lett.}\ }\textbf {\bibinfo {volume} {127}},\
  \bibinfo {pages} {251802} (\bibinfo {year} {2021})},\ \Eprint
  {https://arxiv.org/abs/2109.03116} {arXiv:2109.03116 [hep-ph]} \BibitemShut
  {NoStop}%
\bibitem [{\citenamefont {Carew}\ \emph {et~al.}(2024)\citenamefont {Carew},
  \citenamefont {Caddell}, \citenamefont {Maity},\ and\ \citenamefont
  {O'Hare}}]{Carew:2023qrj}%
  \BibitemOpen
  \bibfield  {author} {\bibinfo {author} {\bibfnamefont {B.}~\bibnamefont
  {Carew}}, \bibinfo {author} {\bibfnamefont {A.~R.}\ \bibnamefont {Caddell}},
  \bibinfo {author} {\bibfnamefont {T.~N.}\ \bibnamefont {Maity}},\ and\
  \bibinfo {author} {\bibfnamefont {C.~A.~J.}\ \bibnamefont {O'Hare}},\
  }\bibfield  {title} {\bibinfo {title} {{Neutrino fog for dark matter-electron
  scattering experiments}},\ }\href
  {https://doi.org/10.1103/PhysRevD.109.083016} {\bibfield  {journal} {\bibinfo
   {journal} {Phys. Rev. D}\ }\textbf {\bibinfo {volume} {109}},\ \bibinfo
  {pages} {083016} (\bibinfo {year} {2024})},\ \Eprint
  {https://arxiv.org/abs/2312.04303} {arXiv:2312.04303 [hep-ph]} \BibitemShut
  {NoStop}%
\bibitem [{\citenamefont {Harnik}\ \emph {et~al.}(2012)\citenamefont {Harnik},
  \citenamefont {Kopp},\ and\ \citenamefont {Machado}}]{Harnik:2012ni}%
  \BibitemOpen
  \bibfield  {author} {\bibinfo {author} {\bibfnamefont {R.}~\bibnamefont
  {Harnik}}, \bibinfo {author} {\bibfnamefont {J.}~\bibnamefont {Kopp}},\ and\
  \bibinfo {author} {\bibfnamefont {P.~A.~N.}\ \bibnamefont {Machado}},\
  }\bibfield  {title} {\bibinfo {title} {{Exploring nu Signals in Dark Matter
  Detectors}},\ }\href {https://doi.org/10.1088/1475-7516/2012/07/026}
  {\bibfield  {journal} {\bibinfo  {journal} {JCAP}\ }\textbf {\bibinfo
  {volume} {07}},\ \bibinfo {pages} {026}},\ \Eprint
  {https://arxiv.org/abs/1202.6073} {arXiv:1202.6073 [hep-ph]} \BibitemShut
  {NoStop}%
\bibitem [{\citenamefont {Baudis}\ \emph {et~al.}(2014)\citenamefont {Baudis},
  \citenamefont {Ferella}, \citenamefont {Kish}, \citenamefont {Manalaysay},
  \citenamefont {Marrodan~Undagoitia},\ and\ \citenamefont
  {Schumann}}]{Baudis:2013qla}%
  \BibitemOpen
  \bibfield  {author} {\bibinfo {author} {\bibfnamefont {L.}~\bibnamefont
  {Baudis}}, \bibinfo {author} {\bibfnamefont {A.}~\bibnamefont {Ferella}},
  \bibinfo {author} {\bibfnamefont {A.}~\bibnamefont {Kish}}, \bibinfo {author}
  {\bibfnamefont {A.}~\bibnamefont {Manalaysay}}, \bibinfo {author}
  {\bibfnamefont {T.}~\bibnamefont {Marrodan~Undagoitia}},\ and\ \bibinfo
  {author} {\bibfnamefont {M.}~\bibnamefont {Schumann}},\ }\bibfield  {title}
  {\bibinfo {title} {{Neutrino physics with multi-ton scale liquid xenon
  detectors}},\ }\href {https://doi.org/10.1088/1475-7516/2014/01/044}
  {\bibfield  {journal} {\bibinfo  {journal} {JCAP}\ }\textbf {\bibinfo
  {volume} {01}},\ \bibinfo {pages} {044}},\ \Eprint
  {https://arxiv.org/abs/1309.7024} {arXiv:1309.7024 [physics.ins-det]}
  \BibitemShut {NoStop}%
\bibitem [{\citenamefont {Billard}\ \emph {et~al.}(2015)\citenamefont
  {Billard}, \citenamefont {Strigari},\ and\ \citenamefont
  {Figueroa-Feliciano}}]{Billard:2014yka}%
  \BibitemOpen
  \bibfield  {author} {\bibinfo {author} {\bibfnamefont {J.}~\bibnamefont
  {Billard}}, \bibinfo {author} {\bibfnamefont {L.~E.}\ \bibnamefont
  {Strigari}},\ and\ \bibinfo {author} {\bibfnamefont {E.}~\bibnamefont
  {Figueroa-Feliciano}},\ }\bibfield  {title} {\bibinfo {title} {{Solar
  neutrino physics with low-threshold dark matter detectors}},\ }\href
  {https://doi.org/10.1103/PhysRevD.91.095023} {\bibfield  {journal} {\bibinfo
  {journal} {Phys. Rev. D}\ }\textbf {\bibinfo {volume} {91}},\ \bibinfo
  {pages} {095023} (\bibinfo {year} {2015})},\ \Eprint
  {https://arxiv.org/abs/1409.0050} {arXiv:1409.0050 [astro-ph.CO]}
  \BibitemShut {NoStop}%
\bibitem [{\citenamefont {Cerde{\~n}o}\ \emph {et~al.}(2016)\citenamefont
  {Cerde{\~n}o}, \citenamefont {Fairbairn}, \citenamefont {Jubb}, \citenamefont
  {Machado}, \citenamefont {Vincent},\ and\ \citenamefont
  {B{\oe}hm}}]{Cerdeno:2016sfi}%
  \BibitemOpen
  \bibfield  {author} {\bibinfo {author} {\bibfnamefont {D.~G.}\ \bibnamefont
  {Cerde{\~n}o}}, \bibinfo {author} {\bibfnamefont {M.}~\bibnamefont
  {Fairbairn}}, \bibinfo {author} {\bibfnamefont {T.}~\bibnamefont {Jubb}},
  \bibinfo {author} {\bibfnamefont {P.~A.~N.}\ \bibnamefont {Machado}},
  \bibinfo {author} {\bibfnamefont {A.~C.}\ \bibnamefont {Vincent}},\ and\
  \bibinfo {author} {\bibfnamefont {C.}~\bibnamefont {B{\oe}hm}},\ }\bibfield
  {title} {\bibinfo {title} {{Physics from solar neutrinos in dark matter
  direct detection experiments}},\ }\href
  {https://doi.org/10.1007/JHEP09(2016)048} {\bibfield  {journal} {\bibinfo
  {journal} {JHEP}\ }\textbf {\bibinfo {volume} {05}},\ \bibinfo {pages}
  {118}},\ \bibinfo {note} {[Erratum: JHEP 09, 048 (2016)]},\ \Eprint
  {https://arxiv.org/abs/1604.01025} {arXiv:1604.01025 [hep-ph]} \BibitemShut
  {NoStop}%
\bibitem [{\citenamefont {Bertuzzo}\ \emph {et~al.}(2017)\citenamefont
  {Bertuzzo}, \citenamefont {Deppisch}, \citenamefont {Kulkarni}, \citenamefont
  {Perez~Gonzalez},\ and\ \citenamefont
  {Zukanovich~Funchal}}]{Bertuzzo:2017tuf}%
  \BibitemOpen
  \bibfield  {author} {\bibinfo {author} {\bibfnamefont {E.}~\bibnamefont
  {Bertuzzo}}, \bibinfo {author} {\bibfnamefont {F.~F.}\ \bibnamefont
  {Deppisch}}, \bibinfo {author} {\bibfnamefont {S.}~\bibnamefont {Kulkarni}},
  \bibinfo {author} {\bibfnamefont {Y.~F.}\ \bibnamefont {Perez~Gonzalez}},\
  and\ \bibinfo {author} {\bibfnamefont {R.}~\bibnamefont
  {Zukanovich~Funchal}},\ }\bibfield  {title} {\bibinfo {title} {{Dark Matter
  and Exotic Neutrino Interactions in Direct Detection Searches}},\ }\href
  {https://doi.org/10.1007/JHEP04(2017)073} {\bibfield  {journal} {\bibinfo
  {journal} {JHEP}\ }\textbf {\bibinfo {volume} {04}},\ \bibinfo {pages}
  {073}},\ \Eprint {https://arxiv.org/abs/1701.07443} {arXiv:1701.07443
  [hep-ph]} \BibitemShut {NoStop}%
\bibitem [{\citenamefont {Cui}\ \emph {et~al.}(2018)\citenamefont {Cui},
  \citenamefont {Pospelov},\ and\ \citenamefont {Pradler}}]{Cui:2017ytb}%
  \BibitemOpen
  \bibfield  {author} {\bibinfo {author} {\bibfnamefont {Y.}~\bibnamefont
  {Cui}}, \bibinfo {author} {\bibfnamefont {M.}~\bibnamefont {Pospelov}},\ and\
  \bibinfo {author} {\bibfnamefont {J.}~\bibnamefont {Pradler}},\ }\bibfield
  {title} {\bibinfo {title} {{Signatures of Dark Radiation in Neutrino and Dark
  Matter Detectors}},\ }\href {https://doi.org/10.1103/PhysRevD.97.103004}
  {\bibfield  {journal} {\bibinfo  {journal} {Phys. Rev. D}\ }\textbf {\bibinfo
  {volume} {97}},\ \bibinfo {pages} {103004} (\bibinfo {year} {2018})},\
  \Eprint {https://arxiv.org/abs/1711.04531} {arXiv:1711.04531 [hep-ph]}
  \BibitemShut {NoStop}%
\bibitem [{\citenamefont {Dutta}\ \emph {et~al.}(2017)\citenamefont {Dutta},
  \citenamefont {Liao}, \citenamefont {Strigari},\ and\ \citenamefont
  {Walker}}]{Dutta:2017nht}%
  \BibitemOpen
  \bibfield  {author} {\bibinfo {author} {\bibfnamefont {B.}~\bibnamefont
  {Dutta}}, \bibinfo {author} {\bibfnamefont {S.}~\bibnamefont {Liao}},
  \bibinfo {author} {\bibfnamefont {L.~E.}\ \bibnamefont {Strigari}},\ and\
  \bibinfo {author} {\bibfnamefont {J.~W.}\ \bibnamefont {Walker}},\ }\bibfield
   {title} {\bibinfo {title} {{Non-standard interactions of solar neutrinos in
  dark matter experiments}},\ }\href
  {https://doi.org/10.1016/j.physletb.2017.08.031} {\bibfield  {journal}
  {\bibinfo  {journal} {Phys. Lett. B}\ }\textbf {\bibinfo {volume} {773}},\
  \bibinfo {pages} {242} (\bibinfo {year} {2017})},\ \Eprint
  {https://arxiv.org/abs/1705.00661} {arXiv:1705.00661 [hep-ph]} \BibitemShut
  {NoStop}%
\bibitem [{\citenamefont {Newstead}\ \emph {et~al.}(2019)\citenamefont
  {Newstead}, \citenamefont {Strigari},\ and\ \citenamefont
  {Lang}}]{Newstead:2018muu}%
  \BibitemOpen
  \bibfield  {author} {\bibinfo {author} {\bibfnamefont {J.~L.}\ \bibnamefont
  {Newstead}}, \bibinfo {author} {\bibfnamefont {L.~E.}\ \bibnamefont
  {Strigari}},\ and\ \bibinfo {author} {\bibfnamefont {R.~F.}\ \bibnamefont
  {Lang}},\ }\bibfield  {title} {\bibinfo {title} {{Detecting CNO solar
  neutrinos in next-generation xenon dark matter experiments}},\ }\href
  {https://doi.org/10.1103/PhysRevD.99.043006} {\bibfield  {journal} {\bibinfo
  {journal} {Phys. Rev. D}\ }\textbf {\bibinfo {volume} {99}},\ \bibinfo
  {pages} {043006} (\bibinfo {year} {2019})},\ \Eprint
  {https://arxiv.org/abs/1807.07169} {arXiv:1807.07169 [astro-ph.SR]}
  \BibitemShut {NoStop}%
\bibitem [{\citenamefont {Essig}\ \emph {et~al.}(2018)\citenamefont {Essig},
  \citenamefont {Sholapurkar},\ and\ \citenamefont {Yu}}]{Essig:2018tss}%
  \BibitemOpen
  \bibfield  {author} {\bibinfo {author} {\bibfnamefont {R.}~\bibnamefont
  {Essig}}, \bibinfo {author} {\bibfnamefont {M.}~\bibnamefont {Sholapurkar}},\
  and\ \bibinfo {author} {\bibfnamefont {T.-T.}\ \bibnamefont {Yu}},\
  }\bibfield  {title} {\bibinfo {title} {{Solar Neutrinos as a Signal and
  Background in Direct-Detection Experiments Searching for Sub-GeV Dark Matter
  With Electron Recoils}},\ }\href {https://doi.org/10.1103/PhysRevD.97.095029}
  {\bibfield  {journal} {\bibinfo  {journal} {Phys. Rev. D}\ }\textbf {\bibinfo
  {volume} {97}},\ \bibinfo {pages} {095029} (\bibinfo {year} {2018})},\
  \Eprint {https://arxiv.org/abs/1801.10159} {arXiv:1801.10159 [hep-ph]}
  \BibitemShut {NoStop}%
\bibitem [{\citenamefont {Aristizabal~Sierra}\ \emph
  {et~al.}(2019)\citenamefont {Aristizabal~Sierra}, \citenamefont {Dutta},
  \citenamefont {Liao},\ and\ \citenamefont
  {Strigari}}]{AristizabalSierra:2019ykk}%
  \BibitemOpen
  \bibfield  {author} {\bibinfo {author} {\bibfnamefont {D.}~\bibnamefont
  {Aristizabal~Sierra}}, \bibinfo {author} {\bibfnamefont {B.}~\bibnamefont
  {Dutta}}, \bibinfo {author} {\bibfnamefont {S.}~\bibnamefont {Liao}},\ and\
  \bibinfo {author} {\bibfnamefont {L.~E.}\ \bibnamefont {Strigari}},\
  }\bibfield  {title} {\bibinfo {title} {{Coherent elastic neutrino-nucleus
  scattering in multi-ton scale dark matter experiments: Classification of
  vector and scalar interactions new physics signals}},\ }\href
  {https://doi.org/10.1007/JHEP12(2019)124} {\bibfield  {journal} {\bibinfo
  {journal} {JHEP}\ }\textbf {\bibinfo {volume} {12}},\ \bibinfo {pages}
  {124}},\ \Eprint {https://arxiv.org/abs/1910.12437} {arXiv:1910.12437
  [hep-ph]} \BibitemShut {NoStop}%
\bibitem [{\citenamefont {Dutta}\ and\ \citenamefont
  {Strigari}(2019)}]{Dutta:2019oaj}%
  \BibitemOpen
  \bibfield  {author} {\bibinfo {author} {\bibfnamefont {B.}~\bibnamefont
  {Dutta}}\ and\ \bibinfo {author} {\bibfnamefont {L.~E.}\ \bibnamefont
  {Strigari}},\ }\bibfield  {title} {\bibinfo {title} {{Neutrino physics with
  dark matter detectors}},\ }\href
  {https://doi.org/10.1146/annurev-nucl-101918-023450} {\bibfield  {journal}
  {\bibinfo  {journal} {Ann. Rev. Nucl. Part. Sci.}\ }\textbf {\bibinfo
  {volume} {69}},\ \bibinfo {pages} {137} (\bibinfo {year} {2019})},\ \Eprint
  {https://arxiv.org/abs/1901.08876} {arXiv:1901.08876 [hep-ph]} \BibitemShut
  {NoStop}%
\bibitem [{\citenamefont {Amaral}\ \emph {et~al.}(2020)\citenamefont {Amaral},
  \citenamefont {Cerde{\~n}o}, \citenamefont {Foldenauer},\ and\ \citenamefont
  {Reid}}]{Amaral:2020tga}%
  \BibitemOpen
  \bibfield  {author} {\bibinfo {author} {\bibfnamefont {D.~W. P.~d.}\
  \bibnamefont {Amaral}}, \bibinfo {author} {\bibfnamefont {D.~G.}\
  \bibnamefont {Cerde{\~n}o}}, \bibinfo {author} {\bibfnamefont
  {P.}~\bibnamefont {Foldenauer}},\ and\ \bibinfo {author} {\bibfnamefont
  {E.}~\bibnamefont {Reid}},\ }\bibfield  {title} {\bibinfo {title} {{Solar
  neutrino probes of the muon anomalous magnetic moment in the gauged $
  \mathrm{U}{(1)}_{L_{\mu }-{L}_{\tau }} $}},\ }\href
  {https://doi.org/10.1007/JHEP12(2020)155} {\bibfield  {journal} {\bibinfo
  {journal} {JHEP}\ }\textbf {\bibinfo {volume} {12}},\ \bibinfo {pages}
  {155}},\ \Eprint {https://arxiv.org/abs/2006.11225} {arXiv:2006.11225
  [hep-ph]} \BibitemShut {NoStop}%
\bibitem [{\citenamefont {Abdullah}\ \emph {et~al.}(2020)\citenamefont
  {Abdullah}, \citenamefont {Aristizabal~Sierra}, \citenamefont {Dutta},\ and\
  \citenamefont {Strigari}}]{Abdullah:2020iiv}%
  \BibitemOpen
  \bibfield  {author} {\bibinfo {author} {\bibfnamefont {M.}~\bibnamefont
  {Abdullah}}, \bibinfo {author} {\bibfnamefont {D.}~\bibnamefont
  {Aristizabal~Sierra}}, \bibinfo {author} {\bibfnamefont {B.}~\bibnamefont
  {Dutta}},\ and\ \bibinfo {author} {\bibfnamefont {L.~E.}\ \bibnamefont
  {Strigari}},\ }\bibfield  {title} {\bibinfo {title} {{Coherent Elastic
  Neutrino-Nucleus Scattering with directional detectors}},\ }\href
  {https://doi.org/10.1103/PhysRevD.102.015009} {\bibfield  {journal} {\bibinfo
   {journal} {Phys. Rev. D}\ }\textbf {\bibinfo {volume} {102}},\ \bibinfo
  {pages} {015009} (\bibinfo {year} {2020})},\ \Eprint
  {https://arxiv.org/abs/2003.11510} {arXiv:2003.11510 [hep-ph]} \BibitemShut
  {NoStop}%
\bibitem [{\citenamefont {Amaral}\ \emph {et~al.}(2021)\citenamefont {Amaral},
  \citenamefont {Cerde{\~n}o}, \citenamefont {Cheek},\ and\ \citenamefont
  {Foldenauer}}]{Amaral:2021rzw}%
  \BibitemOpen
  \bibfield  {author} {\bibinfo {author} {\bibfnamefont {D.~W.~P.}\
  \bibnamefont {Amaral}}, \bibinfo {author} {\bibfnamefont {D.~G.}\
  \bibnamefont {Cerde{\~n}o}}, \bibinfo {author} {\bibfnamefont
  {A.}~\bibnamefont {Cheek}},\ and\ \bibinfo {author} {\bibfnamefont
  {P.}~\bibnamefont {Foldenauer}},\ }\bibfield  {title} {\bibinfo {title}
  {{Confirming $U(1)_{L_\mu -L_{\tau }}$ as a solution for $(g-2)_\mu $ with
  neutrinos}},\ }\href {https://doi.org/10.1140/epjc/s10052-021-09670-z}
  {\bibfield  {journal} {\bibinfo  {journal} {Eur. Phys. J. C}\ }\textbf
  {\bibinfo {volume} {81}},\ \bibinfo {pages} {861} (\bibinfo {year} {2021})},\
  \Eprint {https://arxiv.org/abs/2104.03297} {arXiv:2104.03297 [hep-ph]}
  \BibitemShut {NoStop}%
\bibitem [{\citenamefont {de~Gouv{\^e}a}\ \emph
  {et~al.}(2022{\natexlab{b}})\citenamefont {de~Gouv{\^e}a}, \citenamefont
  {McGinness}, \citenamefont {Martinez-Soler},\ and\ \citenamefont
  {Perez-Gonzalez}}]{deGouvea:2021ymm}%
  \BibitemOpen
  \bibfield  {author} {\bibinfo {author} {\bibfnamefont {A.}~\bibnamefont
  {de~Gouv{\^e}a}}, \bibinfo {author} {\bibfnamefont {E.}~\bibnamefont
  {McGinness}}, \bibinfo {author} {\bibfnamefont {I.}~\bibnamefont
  {Martinez-Soler}},\ and\ \bibinfo {author} {\bibfnamefont {Y.~F.}\
  \bibnamefont {Perez-Gonzalez}},\ }\bibfield  {title} {\bibinfo {title} {{pp
  solar neutrinos at DARWIN}},\ }\href
  {https://doi.org/10.1103/PhysRevD.106.096017} {\bibfield  {journal} {\bibinfo
   {journal} {Phys. Rev. D}\ }\textbf {\bibinfo {volume} {106}},\ \bibinfo
  {pages} {096017} (\bibinfo {year} {2022}{\natexlab{b}})},\ \Eprint
  {https://arxiv.org/abs/2111.02421} {arXiv:2111.02421 [hep-ph]} \BibitemShut
  {NoStop}%
\bibitem [{\citenamefont {Akerib}\ \emph {et~al.}(2020)\citenamefont {Akerib}
  \emph {et~al.}}]{LZ:2019sgr}%
  \BibitemOpen
  \bibfield  {author} {\bibinfo {author} {\bibfnamefont {D.~S.}\ \bibnamefont
  {Akerib}} \emph {et~al.} (\bibinfo {collaboration} {LZ}),\ }\bibfield
  {title} {\bibinfo {title} {{The LUX-ZEPLIN (LZ) Experiment}},\ }\href
  {https://doi.org/10.1016/j.nima.2019.163047} {\bibfield  {journal} {\bibinfo
  {journal} {Nucl. Instrum. Meth. A}\ }\textbf {\bibinfo {volume} {953}},\
  \bibinfo {pages} {163047} (\bibinfo {year} {2020})},\ \Eprint
  {https://arxiv.org/abs/1910.09124} {arXiv:1910.09124 [physics.ins-det]}
  \BibitemShut {NoStop}%
\bibitem [{\citenamefont {Aprile}\ \emph {et~al.}(2024)\citenamefont {Aprile}
  \emph {et~al.}}]{XENON:2024wpa}%
  \BibitemOpen
  \bibfield  {author} {\bibinfo {author} {\bibfnamefont {E.}~\bibnamefont
  {Aprile}} \emph {et~al.} (\bibinfo {collaboration} {XENON}),\ }\bibfield
  {title} {\bibinfo {title} {{The XENONnT dark matter experiment}},\ }\href
  {https://doi.org/10.1140/epjc/s10052-024-12982-5} {\bibfield  {journal}
  {\bibinfo  {journal} {Eur. Phys. J. C}\ }\textbf {\bibinfo {volume} {84}},\
  \bibinfo {pages} {784} (\bibinfo {year} {2024})},\ \Eprint
  {https://arxiv.org/abs/2402.10446} {arXiv:2402.10446 [physics.ins-det]}
  \BibitemShut {NoStop}%
\bibitem [{\citenamefont {Cao}\ \emph {et~al.}(2014)\citenamefont {Cao} \emph
  {et~al.}}]{PandaX:2014mem}%
  \BibitemOpen
  \bibfield  {author} {\bibinfo {author} {\bibfnamefont {X.}~\bibnamefont
  {Cao}} \emph {et~al.} (\bibinfo {collaboration} {PandaX}),\ }\bibfield
  {title} {\bibinfo {title} {{PandaX: A Liquid Xenon Dark Matter Experiment at
  CJPL}},\ }\href {https://doi.org/10.1007/s11433-014-5521-2} {\bibfield
  {journal} {\bibinfo  {journal} {Sci. China Phys. Mech. Astron.}\ }\textbf
  {\bibinfo {volume} {57}},\ \bibinfo {pages} {1476} (\bibinfo {year}
  {2014})},\ \Eprint {https://arxiv.org/abs/1405.2882} {arXiv:1405.2882
  [physics.ins-det]} \BibitemShut {NoStop}%
\bibitem [{\citenamefont {Akerib}\ \emph
  {et~al.}(2025{\natexlab{a}})\citenamefont {Akerib} \emph
  {et~al.}}]{LZ:2025igz}%
  \BibitemOpen
  \bibfield  {author} {\bibinfo {author} {\bibfnamefont {D.~S.}\ \bibnamefont
  {Akerib}} \emph {et~al.} (\bibinfo {collaboration} {LZ}),\ }\bibfield
  {title} {\bibinfo {title} {{Searches for Light Dark Matter and Evidence of
  Coherent Elastic Neutrino-Nucleus Scattering of Solar Neutrinos with the
  LUX-ZEPLIN (LZ) Experiment}},\ }\href@noop {} {\  (\bibinfo {year}
  {2025}{\natexlab{a}})},\ \Eprint {https://arxiv.org/abs/2512.08065}
  {arXiv:2512.08065 [hep-ex]} \BibitemShut {NoStop}%
\bibitem [{\citenamefont {Aprile}\ \emph
  {et~al.}(2026{\natexlab{a}})\citenamefont {Aprile} \emph
  {et~al.}}]{XENON:2026ydt}%
  \BibitemOpen
  \bibfield  {author} {\bibinfo {author} {\bibfnamefont {E.}~\bibnamefont
  {Aprile}} \emph {et~al.} (\bibinfo {collaboration} {XENON}),\ }\bibfield
  {title} {\bibinfo {title} {{Probing the Solar $^8$B Neutrino Fog with
  XENONnT}},\ }\href@noop {} {\  (\bibinfo {year} {2026}{\natexlab{a}})},\
  \Eprint {https://arxiv.org/abs/2604.06002} {arXiv:2604.06002 [hep-ex]}
  \BibitemShut {NoStop}%
\bibitem [{\citenamefont {Bo}\ \emph {et~al.}(2024)\citenamefont {Bo} \emph
  {et~al.}}]{PandaX:2024muv}%
  \BibitemOpen
  \bibfield  {author} {\bibinfo {author} {\bibfnamefont {Z.}~\bibnamefont {Bo}}
  \emph {et~al.} (\bibinfo {collaboration} {PandaX}),\ }\bibfield  {title}
  {\bibinfo {title} {{First Indication of Solar B8 Neutrinos through Coherent
  Elastic Neutrino-Nucleus Scattering in PandaX-4T}},\ }\href
  {https://doi.org/10.1103/PhysRevLett.133.191001} {\bibfield  {journal}
  {\bibinfo  {journal} {Phys. Rev. Lett.}\ }\textbf {\bibinfo {volume} {133}},\
  \bibinfo {pages} {191001} (\bibinfo {year} {2024})},\ \Eprint
  {https://arxiv.org/abs/2407.10892} {arXiv:2407.10892 [hep-ex]} \BibitemShut
  {NoStop}%
\bibitem [{\citenamefont {Chen}\ \emph {et~al.}(2026)\citenamefont {Chen} \emph
  {et~al.}}]{PandaX:2026qdp}%
  \BibitemOpen
  \bibfield  {author} {\bibinfo {author} {\bibfnamefont {P.}~\bibnamefont
  {Chen}} \emph {et~al.} (\bibinfo {collaboration} {PandaX}),\ }\bibfield
  {title} {\bibinfo {title} {{Measurement of solar $pp$ neutrino flux with the
  new PandaX-4T data}},\ }\href@noop {} {\  (\bibinfo {year} {2026})},\ \Eprint
  {https://arxiv.org/abs/2607.02405} {arXiv:2607.02405 [hep-ex]} \BibitemShut
  {NoStop}%
\bibitem [{\citenamefont {Aprile}\ \emph
  {et~al.}(2026{\natexlab{b}})\citenamefont {Aprile} \emph
  {et~al.}}]{XENON:2026rbz}%
  \BibitemOpen
  \bibfield  {author} {\bibinfo {author} {\bibfnamefont {E.}~\bibnamefont
  {Aprile}} \emph {et~al.} (\bibinfo {collaboration} {XENON}),\ }\bibfield
  {title} {\bibinfo {title} {{First Measurement of Solar Neutrinos through
  Elastic Neutrino-Electron Scattering at the keV Scale}},\ }\href@noop {} {\
  (\bibinfo {year} {2026}{\natexlab{b}})},\ \Eprint
  {https://arxiv.org/abs/2608.29450} {arXiv:2608.29450 [hep-ex]} \BibitemShut
  {NoStop}%
\bibitem [{\citenamefont {Aalbers}\ \emph
  {et~al.}(2025{\natexlab{a}})\citenamefont {Aalbers} \emph
  {et~al.}}]{XLZD:2024nsu}%
  \BibitemOpen
  \bibfield  {author} {\bibinfo {author} {\bibfnamefont {J.}~\bibnamefont
  {Aalbers}} \emph {et~al.} (\bibinfo {collaboration} {XLZD}),\ }\bibfield
  {title} {\bibinfo {title} {{The XLZD Design Book: towards the next-generation
  liquid xenon observatory for dark matter and neutrino physics}},\ }\href
  {https://doi.org/10.1140/epjc/s10052-025-14810-w} {\bibfield  {journal}
  {\bibinfo  {journal} {Eur. Phys. J. C}\ }\textbf {\bibinfo {volume} {85}},\
  \bibinfo {pages} {1192} (\bibinfo {year} {2025}{\natexlab{a}})},\ \Eprint
  {https://arxiv.org/abs/2410.17137} {arXiv:2410.17137 [hep-ex]} \BibitemShut
  {NoStop}%
\bibitem [{\citenamefont {Abdukerim}\ \emph {et~al.}(2025)\citenamefont
  {Abdukerim} \emph {et~al.}}]{PANDA-X:2024dlo}%
  \BibitemOpen
  \bibfield  {author} {\bibinfo {author} {\bibfnamefont {A.}~\bibnamefont
  {Abdukerim}} \emph {et~al.} (\bibinfo {collaboration} {PANDA-X, PandaX}),\
  }\bibfield  {title} {\bibinfo {title} {{PandaX-xT{\textemdash}A deep
  underground multi-ten-tonne liquid xenon observatory}},\ }\href
  {https://doi.org/10.1007/s11433-024-2539-y} {\bibfield  {journal} {\bibinfo
  {journal} {Sci. China Phys. Mech. Astron.}\ }\textbf {\bibinfo {volume}
  {68}},\ \bibinfo {pages} {221011} (\bibinfo {year} {2025})},\ \Eprint
  {https://arxiv.org/abs/2402.03596} {arXiv:2402.03596 [hep-ex]} \BibitemShut
  {NoStop}%
\bibitem [{\citenamefont {Amaral}\ \emph {et~al.}(2023)\citenamefont {Amaral},
  \citenamefont {Cerde{\~n}o}, \citenamefont {Cheek},\ and\ \citenamefont
  {Foldenauer}}]{Amaral:2023tbs}%
  \BibitemOpen
  \bibfield  {author} {\bibinfo {author} {\bibfnamefont {D.~W.~P.}\
  \bibnamefont {Amaral}}, \bibinfo {author} {\bibfnamefont {D.}~\bibnamefont
  {Cerde{\~n}o}}, \bibinfo {author} {\bibfnamefont {A.}~\bibnamefont {Cheek}},\
  and\ \bibinfo {author} {\bibfnamefont {P.}~\bibnamefont {Foldenauer}},\
  }\bibfield  {title} {\bibinfo {title} {{A direct detection view of the
  neutrino NSI landscape}},\ }\href {https://doi.org/10.1007/JHEP07(2023)071}
  {\bibfield  {journal} {\bibinfo  {journal} {JHEP}\ }\textbf {\bibinfo
  {volume} {07}},\ \bibinfo {pages} {071}},\ \Eprint
  {https://arxiv.org/abs/2302.12846} {arXiv:2302.12846 [hep-ph]} \BibitemShut
  {NoStop}%
\bibitem [{\citenamefont {De~Romeri}\ \emph {et~al.}(2025)\citenamefont
  {De~Romeri}, \citenamefont {Papoulias},\ and\ \citenamefont
  {Ternes}}]{DeRomeri:2024iaw}%
  \BibitemOpen
  \bibfield  {author} {\bibinfo {author} {\bibfnamefont {V.}~\bibnamefont
  {De~Romeri}}, \bibinfo {author} {\bibfnamefont {D.~K.}\ \bibnamefont
  {Papoulias}},\ and\ \bibinfo {author} {\bibfnamefont {C.~A.}\ \bibnamefont
  {Ternes}},\ }\bibfield  {title} {\bibinfo {title} {{Bounds on new neutrino
  interactions from the first CE{\ensuremath{\nu}}NS data at direct detection
  experiments}},\ }\href {https://doi.org/10.1088/1475-7516/2025/05/012}
  {\bibfield  {journal} {\bibinfo  {journal} {JCAP}\ }\textbf {\bibinfo
  {volume} {05}},\ \bibinfo {pages} {012}},\ \Eprint
  {https://arxiv.org/abs/2411.11749} {arXiv:2411.11749 [hep-ph]} \BibitemShut
  {NoStop}%
\bibitem [{\citenamefont {Atzori~Corona}\ \emph {et~al.}(2026)\citenamefont
  {Atzori~Corona}, \citenamefont {Cadeddu}, \citenamefont {Cargioli},
  \citenamefont {Dordei},\ and\ \citenamefont {Sestu}}]{AtzoriCorona:2025gyz}%
  \BibitemOpen
  \bibfield  {author} {\bibinfo {author} {\bibfnamefont {M.}~\bibnamefont
  {Atzori~Corona}}, \bibinfo {author} {\bibfnamefont {M.}~\bibnamefont
  {Cadeddu}}, \bibinfo {author} {\bibfnamefont {N.}~\bibnamefont {Cargioli}},
  \bibinfo {author} {\bibfnamefont {F.}~\bibnamefont {Dordei}},\ and\ \bibinfo
  {author} {\bibfnamefont {M.}~\bibnamefont {Sestu}},\ }\bibfield  {title}
  {\bibinfo {title} {{When backgrounds become signals: neutrino interactions in
  xenon-based dark matter detectors}},\ }\href
  {https://doi.org/10.1088/1475-7516/2026/05/065} {\bibfield  {journal}
  {\bibinfo  {journal} {JCAP}\ }\textbf {\bibinfo {volume} {05}},\ \bibinfo
  {pages} {065}},\ \Eprint {https://arxiv.org/abs/2509.22178} {arXiv:2509.22178
  [hep-ph]} \BibitemShut {NoStop}%
\bibitem [{\citenamefont {De~Romeri}\ \emph
  {et~al.}(2026{\natexlab{a}})\citenamefont {De~Romeri}, \citenamefont
  {Papoulias}, \citenamefont {Pompa}, \citenamefont {Sanchez~Garcia},\ and\
  \citenamefont {Ternes}}]{DeRomeri:2026prc}%
  \BibitemOpen
  \bibfield  {author} {\bibinfo {author} {\bibfnamefont {V.}~\bibnamefont
  {De~Romeri}}, \bibinfo {author} {\bibfnamefont {D.~K.}\ \bibnamefont
  {Papoulias}}, \bibinfo {author} {\bibfnamefont {F.}~\bibnamefont {Pompa}},
  \bibinfo {author} {\bibfnamefont {G.}~\bibnamefont {Sanchez~Garcia}},\ and\
  \bibinfo {author} {\bibfnamefont {C.~A.}\ \bibnamefont {Ternes}},\ }\bibfield
   {title} {\bibinfo {title} {{Testing light and heavy vector mediators with
  solar CE$\nu$NS measurements}},\ }\href@noop {} {\  (\bibinfo {year}
  {2026}{\natexlab{a}})},\ \Eprint {https://arxiv.org/abs/2603.00554}
  {arXiv:2603.00554 [hep-ph]} \BibitemShut {NoStop}%
\bibitem [{\citenamefont {Amaral}\ \emph {et~al.}(2026)\citenamefont {Amaral},
  \citenamefont {Cerde{\~n}o}, \citenamefont {Cheek}, \citenamefont {Costa},\
  and\ \citenamefont {Foldenauer}}]{Amaral:2026gml}%
  \BibitemOpen
  \bibfield  {author} {\bibinfo {author} {\bibfnamefont {D.~W.~P.}\
  \bibnamefont {Amaral}}, \bibinfo {author} {\bibfnamefont {D.}~\bibnamefont
  {Cerde{\~n}o}}, \bibinfo {author} {\bibfnamefont {A.}~\bibnamefont {Cheek}},
  \bibinfo {author} {\bibfnamefont {V.}~\bibnamefont {Costa}},\ and\ \bibinfo
  {author} {\bibfnamefont {P.}~\bibnamefont {Foldenauer}},\ }\bibfield  {title}
  {\bibinfo {title} {{$\texttt{SNuDD}$: Solar Neutrinos for Direct
  Detection}},\ }\href@noop {} {\  (\bibinfo {year} {2026})},\ \Eprint
  {https://arxiv.org/abs/2607.22817} {arXiv:2607.22817 [hep-ph]} \BibitemShut
  {NoStop}%
\bibitem [{\citenamefont {Leone}(2026)}]{Leone:2026knr}%
  \BibitemOpen
  \bibfield  {author} {\bibinfo {author} {\bibfnamefont {G.}~\bibnamefont
  {Leone}},\ }\bibfield  {title} {\bibinfo {title} {{XENONnT $pp$-Dominated
  Solar Neutrinos Constrain the $\nu_1$ Pseudo-Dirac Splitting}},\ }\href@noop
  {} {\  (\bibinfo {year} {2026})},\ \Eprint {https://arxiv.org/abs/2609.38233}
  {arXiv:2609.38233 [hep-ph]} \BibitemShut {NoStop}%
\bibitem [{\citenamefont {Davidson}\ \emph {et~al.}(2003)\citenamefont
  {Davidson}, \citenamefont {Pena-Garay}, \citenamefont {Rius},\ and\
  \citenamefont {Santamaria}}]{Davidson:2003ha}%
  \BibitemOpen
  \bibfield  {author} {\bibinfo {author} {\bibfnamefont {S.}~\bibnamefont
  {Davidson}}, \bibinfo {author} {\bibfnamefont {C.}~\bibnamefont
  {Pena-Garay}}, \bibinfo {author} {\bibfnamefont {N.}~\bibnamefont {Rius}},\
  and\ \bibinfo {author} {\bibfnamefont {A.}~\bibnamefont {Santamaria}},\
  }\bibfield  {title} {\bibinfo {title} {{Present and future bounds on
  nonstandard neutrino interactions}},\ }\href
  {https://doi.org/10.1088/1126-6708/2003/03/011} {\bibfield  {journal}
  {\bibinfo  {journal} {JHEP}\ }\textbf {\bibinfo {volume} {03}},\ \bibinfo
  {pages} {011}},\ \Eprint {https://arxiv.org/abs/hep-ph/0302093}
  {arXiv:hep-ph/0302093} \BibitemShut {NoStop}%
\bibitem [{\citenamefont {Miranda}\ and\ \citenamefont
  {Nunokawa}(2015)}]{Miranda:2015dra}%
  \BibitemOpen
  \bibfield  {author} {\bibinfo {author} {\bibfnamefont {O.~G.}\ \bibnamefont
  {Miranda}}\ and\ \bibinfo {author} {\bibfnamefont {H.}~\bibnamefont
  {Nunokawa}},\ }\bibfield  {title} {\bibinfo {title} {{Non standard neutrino
  interactions: current status and future prospects}},\ }\href
  {https://doi.org/10.1088/1367-2630/17/9/095002} {\bibfield  {journal}
  {\bibinfo  {journal} {New J. Phys.}\ }\textbf {\bibinfo {volume} {17}},\
  \bibinfo {pages} {095002} (\bibinfo {year} {2015})},\ \Eprint
  {https://arxiv.org/abs/1505.06254} {arXiv:1505.06254 [hep-ph]} \BibitemShut
  {NoStop}%
\bibitem [{\citenamefont {Dziewonski}\ and\ \citenamefont
  {Anderson}(1981)}]{DZIEWONSKI1981297}%
  \BibitemOpen
  \bibfield  {author} {\bibinfo {author} {\bibfnamefont {A.~M.}\ \bibnamefont
  {Dziewonski}}\ and\ \bibinfo {author} {\bibfnamefont {D.~L.}\ \bibnamefont
  {Anderson}},\ }\bibfield  {title} {\bibinfo {title} {Preliminary reference
  earth model},\ }\href
  {https://doi.org/https://doi.org/10.1016/0031-9201(81)90046-7} {\bibfield
  {journal} {\bibinfo  {journal} {Physics of the Earth and Planetary
  Interiors}\ }\textbf {\bibinfo {volume} {25}},\ \bibinfo {pages} {297}
  (\bibinfo {year} {1981})}\BibitemShut {NoStop}%
\bibitem [{\citenamefont {Coloma}\ \emph {et~al.}(2022)\citenamefont {Coloma},
  \citenamefont {Gonzalez-Garcia}, \citenamefont {Maltoni}, \citenamefont
  {Pinheiro},\ and\ \citenamefont {Urrea}}]{Coloma:2022umy}%
  \BibitemOpen
  \bibfield  {author} {\bibinfo {author} {\bibfnamefont {P.}~\bibnamefont
  {Coloma}}, \bibinfo {author} {\bibfnamefont {M.~C.}\ \bibnamefont
  {Gonzalez-Garcia}}, \bibinfo {author} {\bibfnamefont {M.}~\bibnamefont
  {Maltoni}}, \bibinfo {author} {\bibfnamefont {J.~P.}\ \bibnamefont
  {Pinheiro}},\ and\ \bibinfo {author} {\bibfnamefont {S.}~\bibnamefont
  {Urrea}},\ }\bibfield  {title} {\bibinfo {title} {{Constraining new physics
  with Borexino Phase-II spectral data}},\ }\href
  {https://doi.org/10.1007/JHEP07(2022)138} {\bibfield  {journal} {\bibinfo
  {journal} {JHEP}\ }\textbf {\bibinfo {volume} {07}},\ \bibinfo {pages}
  {138}},\ \bibinfo {note} {[Erratum: JHEP 11, 138 (2022)]},\ \Eprint
  {https://arxiv.org/abs/2204.03011} {arXiv:2204.03011 [hep-ph]} \BibitemShut
  {NoStop}%
\bibitem [{\citenamefont {Helm}(1956)}]{Helm:1956zz}%
  \BibitemOpen
  \bibfield  {author} {\bibinfo {author} {\bibfnamefont {R.~H.}\ \bibnamefont
  {Helm}},\ }\bibfield  {title} {\bibinfo {title} {{Inelastic and Elastic
  Scattering of 187-Mev Electrons from Selected Even-Even Nuclei}},\ }\href
  {https://doi.org/10.1103/PhysRev.104.1466} {\bibfield  {journal} {\bibinfo
  {journal} {Phys. Rev.}\ }\textbf {\bibinfo {volume} {104}},\ \bibinfo {pages}
  {1466} (\bibinfo {year} {1956})}\BibitemShut {NoStop}%
\bibitem [{\citenamefont {Bahcall}\ \emph {et~al.}(2006)\citenamefont
  {Bahcall}, \citenamefont {Serenelli},\ and\ \citenamefont
  {Basu}}]{Bahcall:2005va}%
  \BibitemOpen
  \bibfield  {author} {\bibinfo {author} {\bibfnamefont {J.~N.}\ \bibnamefont
  {Bahcall}}, \bibinfo {author} {\bibfnamefont {A.~M.}\ \bibnamefont
  {Serenelli}},\ and\ \bibinfo {author} {\bibfnamefont {S.}~\bibnamefont
  {Basu}},\ }\bibfield  {title} {\bibinfo {title} {{10,000 standard solar
  models: a Monte Carlo simulation}},\ }\href {https://doi.org/10.1086/504043}
  {\bibfield  {journal} {\bibinfo  {journal} {Astrophys. J. Suppl.}\ }\textbf
  {\bibinfo {volume} {165}},\ \bibinfo {pages} {400} (\bibinfo {year}
  {2006})},\ \Eprint {https://arxiv.org/abs/astro-ph/0511337}
  {arXiv:astro-ph/0511337} \BibitemShut {NoStop}%
\bibitem [{\citenamefont {Akerib}\ \emph {et~al.}(2017)\citenamefont {Akerib}
  \emph {et~al.}}]{LUX:2016rfb}%
  \BibitemOpen
  \bibfield  {author} {\bibinfo {author} {\bibfnamefont {D.~S.}\ \bibnamefont
  {Akerib}} \emph {et~al.} (\bibinfo {collaboration} {LUX}),\ }\bibfield
  {title} {\bibinfo {title} {{Signal yields, energy resolution, and
  recombination fluctuations in liquid xenon}},\ }\href
  {https://doi.org/10.1103/PhysRevD.95.012008} {\bibfield  {journal} {\bibinfo
  {journal} {Phys. Rev. D}\ }\textbf {\bibinfo {volume} {95}},\ \bibinfo
  {pages} {012008} (\bibinfo {year} {2017})},\ \Eprint
  {https://arxiv.org/abs/1610.02076} {arXiv:1610.02076 [physics.ins-det]}
  \BibitemShut {NoStop}%
\bibitem [{\citenamefont {Aalbers}\ \emph {et~al.}(2023)\citenamefont {Aalbers}
  \emph {et~al.}}]{LZ:2022lsv}%
  \BibitemOpen
  \bibfield  {author} {\bibinfo {author} {\bibfnamefont {J.}~\bibnamefont
  {Aalbers}} \emph {et~al.} (\bibinfo {collaboration} {LZ}),\ }\bibfield
  {title} {\bibinfo {title} {{First Dark Matter Search Results from the
  LUX-ZEPLIN (LZ) Experiment}},\ }\href
  {https://doi.org/10.1103/PhysRevLett.131.041002} {\bibfield  {journal}
  {\bibinfo  {journal} {Phys. Rev. Lett.}\ }\textbf {\bibinfo {volume} {131}},\
  \bibinfo {pages} {041002} (\bibinfo {year} {2023})},\ \Eprint
  {https://arxiv.org/abs/2207.03764} {arXiv:2207.03764 [hep-ex]} \BibitemShut
  {NoStop}%
\bibitem [{\citenamefont {Aalbers}\ \emph
  {et~al.}(2025{\natexlab{b}})\citenamefont {Aalbers} \emph
  {et~al.}}]{LZ:2024zvo}%
  \BibitemOpen
  \bibfield  {author} {\bibinfo {author} {\bibfnamefont {J.}~\bibnamefont
  {Aalbers}} \emph {et~al.} (\bibinfo {collaboration} {LZ}),\ }\bibfield
  {title} {\bibinfo {title} {{Dark Matter Search Results from
  4.2{\,}{\,}Tonne-Years of Exposure of the LUX-ZEPLIN (LZ) Experiment}},\
  }\href {https://doi.org/10.1103/4dyc-z8zf} {\bibfield  {journal} {\bibinfo
  {journal} {Phys. Rev. Lett.}\ }\textbf {\bibinfo {volume} {135}},\ \bibinfo
  {pages} {011802} (\bibinfo {year} {2025}{\natexlab{b}})},\ \Eprint
  {https://arxiv.org/abs/2410.17036} {arXiv:2410.17036 [hep-ex]} \BibitemShut
  {NoStop}%
\bibitem [{\citenamefont {Akerib}\ \emph
  {et~al.}(2025{\natexlab{b}})\citenamefont {Akerib} \emph
  {et~al.}}]{LZ:2025zpw}%
  \BibitemOpen
  \bibfield  {author} {\bibinfo {author} {\bibfnamefont {D.~S.}\ \bibnamefont
  {Akerib}} \emph {et~al.} (\bibinfo {collaboration} {LZ}),\ }\bibfield
  {title} {\bibinfo {title} {{Search for New Physics via Low-Energy Electron
  Recoils with a 4.2 Tonne{\texttimes} Year Exposure from the LZ Experiment}},\
  }\href@noop {} {\  (\bibinfo {year} {2025}{\natexlab{b}})},\ \Eprint
  {https://arxiv.org/abs/2511.17350} {arXiv:2511.17350 [hep-ex]} \BibitemShut
  {NoStop}%
\bibitem [{\citenamefont {XU}(2025)}]{xu_2025_17660135}%
  \BibitemOpen
  \bibfield  {author} {\bibinfo {author} {\bibfnamefont {Y.}~\bibnamefont
  {XU}},\ }\href {https://doi.org/10.5281/zenodo.17660135} {\bibinfo {title}
  {Lz ws2022+ws2024 low energy electronic recoil analysis}} (\bibinfo {year}
  {2025})\BibitemShut {NoStop}%
\bibitem [{\citenamefont {Xu}(2026)}]{dacheng_xu_2026_20576156}%
  \BibitemOpen
  \bibfield  {author} {\bibinfo {author} {\bibfnamefont {D.}~\bibnamefont
  {Xu}},\ }\href {https://doi.org/10.5281/zenodo.20576156} {\bibinfo {title}
  {Xenonnt/cevns\_data\_release: Sr0+sr1+sr2}} (\bibinfo {year}
  {2026})\BibitemShut {NoStop}%
\bibitem [{\citenamefont {Aprile}\ \emph {et~al.}(2020)\citenamefont {Aprile}
  \emph {et~al.}}]{XENON:2020rca}%
  \BibitemOpen
  \bibfield  {author} {\bibinfo {author} {\bibfnamefont {E.}~\bibnamefont
  {Aprile}} \emph {et~al.} (\bibinfo {collaboration} {XENON}),\ }\bibfield
  {title} {\bibinfo {title} {{Excess electronic recoil events in XENON1T}},\
  }\href {https://doi.org/10.1103/PhysRevD.102.072004} {\bibfield  {journal}
  {\bibinfo  {journal} {Phys. Rev. D}\ }\textbf {\bibinfo {volume} {102}},\
  \bibinfo {pages} {072004} (\bibinfo {year} {2020})},\ \Eprint
  {https://arxiv.org/abs/2006.09721} {arXiv:2006.09721 [hep-ex]} \BibitemShut
  {NoStop}%
\bibitem [{\citenamefont {Zhang}\ \emph {et~al.}(2022)\citenamefont {Zhang}
  \emph {et~al.}}]{PandaX:2022ood}%
  \BibitemOpen
  \bibfield  {author} {\bibinfo {author} {\bibfnamefont {D.}~\bibnamefont
  {Zhang}} \emph {et~al.} (\bibinfo {collaboration} {PandaX}),\ }\bibfield
  {title} {\bibinfo {title} {{Search for Light Fermionic Dark Matter Absorption
  on Electrons in PandaX-4T}},\ }\href
  {https://doi.org/10.1103/PhysRevLett.129.161804} {\bibfield  {journal}
  {\bibinfo  {journal} {Phys. Rev. Lett.}\ }\textbf {\bibinfo {volume} {129}},\
  \bibinfo {pages} {161804} (\bibinfo {year} {2022})},\ \Eprint
  {https://arxiv.org/abs/2206.02339} {arXiv:2206.02339 [hep-ex]} \BibitemShut
  {NoStop}%
\bibitem [{\citenamefont {Aalbers}\ \emph {et~al.}(2020)\citenamefont {Aalbers}
  \emph {et~al.}}]{DARWIN:2020bnc}%
  \BibitemOpen
  \bibfield  {author} {\bibinfo {author} {\bibfnamefont {J.}~\bibnamefont
  {Aalbers}} \emph {et~al.} (\bibinfo {collaboration} {DARWIN}),\ }\bibfield
  {title} {\bibinfo {title} {{Solar neutrino detection sensitivity in DARWIN
  via electron scattering}},\ }\href
  {https://doi.org/10.1140/epjc/s10052-020-08602-7} {\bibfield  {journal}
  {\bibinfo  {journal} {Eur. Phys. J. C}\ }\textbf {\bibinfo {volume} {80}},\
  \bibinfo {pages} {1133} (\bibinfo {year} {2020})},\ \Eprint
  {https://arxiv.org/abs/2006.03114} {arXiv:2006.03114 [physics.ins-det]}
  \BibitemShut {NoStop}%
\bibitem [{\citenamefont {Cowan}\ \emph {et~al.}(2011)\citenamefont {Cowan},
  \citenamefont {Cranmer}, \citenamefont {Gross},\ and\ \citenamefont
  {Vitells}}]{Cowan:2010js}%
  \BibitemOpen
  \bibfield  {author} {\bibinfo {author} {\bibfnamefont {G.}~\bibnamefont
  {Cowan}}, \bibinfo {author} {\bibfnamefont {K.}~\bibnamefont {Cranmer}},
  \bibinfo {author} {\bibfnamefont {E.}~\bibnamefont {Gross}},\ and\ \bibinfo
  {author} {\bibfnamefont {O.}~\bibnamefont {Vitells}},\ }\bibfield  {title}
  {\bibinfo {title} {{Asymptotic formulae for likelihood-based tests of new
  physics}},\ }\href {https://doi.org/10.1140/epjc/s10052-011-1554-0}
  {\bibfield  {journal} {\bibinfo  {journal} {Eur. Phys. J. C}\ }\textbf
  {\bibinfo {volume} {71}},\ \bibinfo {pages} {1554} (\bibinfo {year}
  {2011})},\ \bibinfo {note} {[Erratum: Eur.Phys.J.C 73, 2501 (2013)]},\
  \Eprint {https://arxiv.org/abs/1007.1727} {arXiv:1007.1727 [physics.data-an]}
  \BibitemShut {NoStop}%
\bibitem [{\citenamefont {Vinyoles}\ \emph {et~al.}(2017)\citenamefont
  {Vinyoles}, \citenamefont {Serenelli}, \citenamefont {Villante},
  \citenamefont {Basu}, \citenamefont {Bergstr{\"o}m}, \citenamefont
  {Gonzalez-Garcia}, \citenamefont {Maltoni}, \citenamefont {Pe{\~n}a-Garay},\
  and\ \citenamefont {Song}}]{Vinyoles:2016djt}%
  \BibitemOpen
  \bibfield  {author} {\bibinfo {author} {\bibfnamefont {N.}~\bibnamefont
  {Vinyoles}}, \bibinfo {author} {\bibfnamefont {A.~M.}\ \bibnamefont
  {Serenelli}}, \bibinfo {author} {\bibfnamefont {F.~L.}\ \bibnamefont
  {Villante}}, \bibinfo {author} {\bibfnamefont {S.}~\bibnamefont {Basu}},
  \bibinfo {author} {\bibfnamefont {J.}~\bibnamefont {Bergstr{\"o}m}}, \bibinfo
  {author} {\bibfnamefont {M.~C.}\ \bibnamefont {Gonzalez-Garcia}}, \bibinfo
  {author} {\bibfnamefont {M.}~\bibnamefont {Maltoni}}, \bibinfo {author}
  {\bibfnamefont {C.}~\bibnamefont {Pe{\~n}a-Garay}},\ and\ \bibinfo {author}
  {\bibfnamefont {N.}~\bibnamefont {Song}},\ }\bibfield  {title} {\bibinfo
  {title} {{A new Generation of Standard Solar Models}},\ }\href
  {https://doi.org/10.3847/1538-4357/835/2/202} {\bibfield  {journal} {\bibinfo
   {journal} {Astrophys. J.}\ }\textbf {\bibinfo {volume} {835}},\ \bibinfo
  {pages} {202} (\bibinfo {year} {2017})},\ \Eprint
  {https://arxiv.org/abs/1611.09867} {arXiv:1611.09867 [astro-ph.SR]}
  \BibitemShut {NoStop}%
\bibitem [{\citenamefont {Aharmim}\ \emph {et~al.}(2013)\citenamefont {Aharmim}
  \emph {et~al.}}]{SNO:2011hxd}%
  \BibitemOpen
  \bibfield  {author} {\bibinfo {author} {\bibfnamefont {B.}~\bibnamefont
  {Aharmim}} \emph {et~al.} (\bibinfo {collaboration} {SNO}),\ }\bibfield
  {title} {\bibinfo {title} {{Combined Analysis of all Three Phases of Solar
  Neutrino Data from the Sudbury Neutrino Observatory}},\ }\href
  {https://doi.org/10.1103/PhysRevC.88.025501} {\bibfield  {journal} {\bibinfo
  {journal} {Phys. Rev. C}\ }\textbf {\bibinfo {volume} {88}},\ \bibinfo
  {pages} {025501} (\bibinfo {year} {2013})},\ \Eprint
  {https://arxiv.org/abs/1109.0763} {arXiv:1109.0763 [nucl-ex]} \BibitemShut
  {NoStop}%
\bibitem [{\citenamefont {B{\oe}hm}\ \emph {et~al.}(2019)\citenamefont
  {B{\oe}hm}, \citenamefont {Cerde{\~n}o}, \citenamefont {Machado},
  \citenamefont {Olivares-Del~Campo}, \citenamefont {Perdomo},\ and\
  \citenamefont {Reid}}]{Boehm:2018sux}%
  \BibitemOpen
  \bibfield  {author} {\bibinfo {author} {\bibfnamefont {C.}~\bibnamefont
  {B{\oe}hm}}, \bibinfo {author} {\bibfnamefont {D.~G.}\ \bibnamefont
  {Cerde{\~n}o}}, \bibinfo {author} {\bibfnamefont {P.~A.~N.}\ \bibnamefont
  {Machado}}, \bibinfo {author} {\bibfnamefont {A.}~\bibnamefont
  {Olivares-Del~Campo}}, \bibinfo {author} {\bibfnamefont {E.}~\bibnamefont
  {Perdomo}},\ and\ \bibinfo {author} {\bibfnamefont {E.}~\bibnamefont
  {Reid}},\ }\bibfield  {title} {\bibinfo {title} {{How high is the neutrino
  floor?}},\ }\href {https://doi.org/10.1088/1475-7516/2019/01/043} {\bibfield
  {journal} {\bibinfo  {journal} {JCAP}\ }\textbf {\bibinfo {volume} {01}},\
  \bibinfo {pages} {043}},\ \Eprint {https://arxiv.org/abs/1809.06385}
  {arXiv:1809.06385 [hep-ph]} \BibitemShut {NoStop}%
\bibitem [{\citenamefont {De~Romeri}\ \emph
  {et~al.}(2026{\natexlab{b}})\citenamefont {De~Romeri}, \citenamefont
  {Majumdar}, \citenamefont {Papoulias},\ and\ \citenamefont
  {Srivastava}}]{DeRomeri:2025nkx}%
  \BibitemOpen
  \bibfield  {author} {\bibinfo {author} {\bibfnamefont {V.}~\bibnamefont
  {De~Romeri}}, \bibinfo {author} {\bibfnamefont {A.}~\bibnamefont {Majumdar}},
  \bibinfo {author} {\bibfnamefont {D.~K.}\ \bibnamefont {Papoulias}},\ and\
  \bibinfo {author} {\bibfnamefont {R.}~\bibnamefont {Srivastava}},\ }\bibfield
   {title} {\bibinfo {title} {{New light mediators and the neutrino fog:
  Implications from XENONnT nuclear recoil data}},\ }\href
  {https://doi.org/10.1088/1475-7516/2026/05/093} {\bibfield  {journal}
  {\bibinfo  {journal} {JCAP}\ }\textbf {\bibinfo {volume} {05}},\ \bibinfo
  {pages} {093}},\ \Eprint {https://arxiv.org/abs/2512.08853} {arXiv:2512.08853
  [hep-ph]} \BibitemShut {NoStop}%
\bibitem [{\citenamefont {Abdurashitov}\ \emph {et~al.}(2009)\citenamefont
  {Abdurashitov} \emph {et~al.}}]{SAGE:2009eeu}%
  \BibitemOpen
  \bibfield  {author} {\bibinfo {author} {\bibfnamefont {J.~N.}\ \bibnamefont
  {Abdurashitov}} \emph {et~al.} (\bibinfo {collaboration} {SAGE}),\ }\bibfield
   {title} {\bibinfo {title} {{Measurement of the solar neutrino capture rate
  with gallium metal. III: Results for the 2002--2007 data-taking period}},\
  }\href {https://doi.org/10.1103/PhysRevC.80.015807} {\bibfield  {journal}
  {\bibinfo  {journal} {Phys. Rev. C}\ }\textbf {\bibinfo {volume} {80}},\
  \bibinfo {pages} {015807} (\bibinfo {year} {2009})},\ \Eprint
  {https://arxiv.org/abs/0901.2200} {arXiv:0901.2200 [nucl-ex]} \BibitemShut
  {NoStop}%
\bibitem [{\citenamefont {Agostini}\ \emph {et~al.}(2018)\citenamefont
  {Agostini} \emph {et~al.}}]{BOREXINO:2018ohr}%
  \BibitemOpen
  \bibfield  {author} {\bibinfo {author} {\bibfnamefont {M.}~\bibnamefont
  {Agostini}} \emph {et~al.} (\bibinfo {collaboration} {BOREXINO}),\ }\bibfield
   {title} {\bibinfo {title} {{Comprehensive measurement of $pp$-chain solar
  neutrinos}},\ }\href {https://doi.org/10.1038/s41586-018-0624-y} {\bibfield
  {journal} {\bibinfo  {journal} {Nature}\ }\textbf {\bibinfo {volume} {562}},\
  \bibinfo {pages} {505} (\bibinfo {year} {2018})}\BibitemShut {NoStop}%
\bibitem [{\citenamefont {Grevesse}\ and\ \citenamefont
  {Sauval}(1998)}]{Grevesse:1998bj}%
  \BibitemOpen
  \bibfield  {author} {\bibinfo {author} {\bibfnamefont {N.}~\bibnamefont
  {Grevesse}}\ and\ \bibinfo {author} {\bibfnamefont {A.~J.}\ \bibnamefont
  {Sauval}},\ }\bibfield  {title} {\bibinfo {title} {{Standard Solar
  Composition}},\ }\href {https://doi.org/10.1023/A:1005161325181} {\bibfield
  {journal} {\bibinfo  {journal} {Space Sci. Rev.}\ }\textbf {\bibinfo {volume}
  {85}},\ \bibinfo {pages} {161} (\bibinfo {year} {1998})}\BibitemShut
  {NoStop}%
\bibitem [{\citenamefont {Asplund}\ \emph {et~al.}(2009)\citenamefont
  {Asplund}, \citenamefont {Grevesse}, \citenamefont {Sauval},\ and\
  \citenamefont {Scott}}]{Asplund:2009fu}%
  \BibitemOpen
  \bibfield  {author} {\bibinfo {author} {\bibfnamefont {M.}~\bibnamefont
  {Asplund}}, \bibinfo {author} {\bibfnamefont {N.}~\bibnamefont {Grevesse}},
  \bibinfo {author} {\bibfnamefont {A.~J.}\ \bibnamefont {Sauval}},\ and\
  \bibinfo {author} {\bibfnamefont {P.}~\bibnamefont {Scott}},\ }\bibfield
  {title} {\bibinfo {title} {{The chemical composition of the Sun}},\ }\href
  {https://doi.org/10.1146/annurev.astro.46.060407.145222} {\bibfield
  {journal} {\bibinfo  {journal} {Ann. Rev. Astron. Astrophys.}\ }\textbf
  {\bibinfo {volume} {47}},\ \bibinfo {pages} {481} (\bibinfo {year} {2009})},\
  \Eprint {https://arxiv.org/abs/0909.0948} {arXiv:0909.0948 [astro-ph.SR]}
  \BibitemShut {NoStop}%
\bibitem [{\citenamefont {Haxton}\ \emph {et~al.}(2013)\citenamefont {Haxton},
  \citenamefont {Hamish~Robertson},\ and\ \citenamefont
  {Serenelli}}]{Haxton:2012wfz}%
  \BibitemOpen
  \bibfield  {author} {\bibinfo {author} {\bibfnamefont {W.~C.}\ \bibnamefont
  {Haxton}}, \bibinfo {author} {\bibfnamefont {R.~G.}\ \bibnamefont
  {Hamish~Robertson}},\ and\ \bibinfo {author} {\bibfnamefont {A.~M.}\
  \bibnamefont {Serenelli}},\ }\bibfield  {title} {\bibinfo {title} {{Solar
  Neutrinos: Status and Prospects}},\ }\href
  {https://doi.org/10.1146/annurev-astro-081811-125539} {\bibfield  {journal}
  {\bibinfo  {journal} {Ann. Rev. Astron. Astrophys.}\ }\textbf {\bibinfo
  {volume} {51}},\ \bibinfo {pages} {21} (\bibinfo {year} {2013})},\ \Eprint
  {https://arxiv.org/abs/1208.5723} {arXiv:1208.5723 [astro-ph.SR]}
  \BibitemShut {NoStop}%
\bibitem [{\citenamefont {Foot}(1991)}]{Foot:1990mn}%
  \BibitemOpen
  \bibfield  {author} {\bibinfo {author} {\bibfnamefont {R.}~\bibnamefont
  {Foot}},\ }\bibfield  {title} {\bibinfo {title} {{New Physics From Electric
  Charge Quantization?}},\ }\href {https://doi.org/10.1142/S0217732391000543}
  {\bibfield  {journal} {\bibinfo  {journal} {Mod. Phys. Lett. A}\ }\textbf
  {\bibinfo {volume} {6}},\ \bibinfo {pages} {527} (\bibinfo {year}
  {1991})}\BibitemShut {NoStop}%
\bibitem [{\citenamefont {He}\ \emph {et~al.}(1991{\natexlab{a}})\citenamefont
  {He}, \citenamefont {Joshi}, \citenamefont {Lew},\ and\ \citenamefont
  {Volkas}}]{He:1990pn}%
  \BibitemOpen
  \bibfield  {author} {\bibinfo {author} {\bibfnamefont {X.~G.}\ \bibnamefont
  {He}}, \bibinfo {author} {\bibfnamefont {G.~C.}\ \bibnamefont {Joshi}},
  \bibinfo {author} {\bibfnamefont {H.}~\bibnamefont {Lew}},\ and\ \bibinfo
  {author} {\bibfnamefont {R.~R.}\ \bibnamefont {Volkas}},\ }\bibfield  {title}
  {\bibinfo {title} {{NEW Z-prime PHENOMENOLOGY}},\ }\href
  {https://doi.org/10.1103/PhysRevD.43.R22} {\bibfield  {journal} {\bibinfo
  {journal} {Phys. Rev. D}\ }\textbf {\bibinfo {volume} {43}},\ \bibinfo
  {pages} {22} (\bibinfo {year} {1991}{\natexlab{a}})}\BibitemShut {NoStop}%
\bibitem [{\citenamefont {He}\ \emph {et~al.}(1991{\natexlab{b}})\citenamefont
  {He}, \citenamefont {Joshi}, \citenamefont {Lew},\ and\ \citenamefont
  {Volkas}}]{He:1991qd}%
  \BibitemOpen
  \bibfield  {author} {\bibinfo {author} {\bibfnamefont {X.-G.}\ \bibnamefont
  {He}}, \bibinfo {author} {\bibfnamefont {G.~C.}\ \bibnamefont {Joshi}},
  \bibinfo {author} {\bibfnamefont {H.}~\bibnamefont {Lew}},\ and\ \bibinfo
  {author} {\bibfnamefont {R.~R.}\ \bibnamefont {Volkas}},\ }\bibfield  {title}
  {\bibinfo {title} {{Simplest Z-prime model}},\ }\href
  {https://doi.org/10.1103/PhysRevD.44.2118} {\bibfield  {journal} {\bibinfo
  {journal} {Phys. Rev. D}\ }\textbf {\bibinfo {volume} {44}},\ \bibinfo
  {pages} {2118} (\bibinfo {year} {1991}{\natexlab{b}})}\BibitemShut {NoStop}%
\bibitem [{\citenamefont {Foot}\ \emph {et~al.}(1994)\citenamefont {Foot},
  \citenamefont {He}, \citenamefont {Lew},\ and\ \citenamefont
  {Volkas}}]{Foot:1994vd}%
  \BibitemOpen
  \bibfield  {author} {\bibinfo {author} {\bibfnamefont {R.}~\bibnamefont
  {Foot}}, \bibinfo {author} {\bibfnamefont {X.~G.}\ \bibnamefont {He}},
  \bibinfo {author} {\bibfnamefont {H.}~\bibnamefont {Lew}},\ and\ \bibinfo
  {author} {\bibfnamefont {R.~R.}\ \bibnamefont {Volkas}},\ }\bibfield  {title}
  {\bibinfo {title} {{Model for a light Z-prime boson}},\ }\href
  {https://doi.org/10.1103/PhysRevD.50.4571} {\bibfield  {journal} {\bibinfo
  {journal} {Phys. Rev. D}\ }\textbf {\bibinfo {volume} {50}},\ \bibinfo
  {pages} {4571} (\bibinfo {year} {1994})},\ \Eprint
  {https://arxiv.org/abs/hep-ph/9401250} {arXiv:hep-ph/9401250} \BibitemShut
  {NoStop}%
\bibitem [{\citenamefont {Bauer}\ and\ \citenamefont
  {Foldenauer}(2022)}]{Bauer:2022nwt}%
  \BibitemOpen
  \bibfield  {author} {\bibinfo {author} {\bibfnamefont {M.}~\bibnamefont
  {Bauer}}\ and\ \bibinfo {author} {\bibfnamefont {P.}~\bibnamefont
  {Foldenauer}},\ }\bibfield  {title} {\bibinfo {title} {{Consistent Theory of
  Kinetic Mixing and the Higgs Low-Energy Theorem}},\ }\href
  {https://doi.org/10.1103/PhysRevLett.129.171801} {\bibfield  {journal}
  {\bibinfo  {journal} {Phys. Rev. Lett.}\ }\textbf {\bibinfo {volume} {129}},\
  \bibinfo {pages} {171801} (\bibinfo {year} {2022})},\ \Eprint
  {https://arxiv.org/abs/2207.00023} {arXiv:2207.00023 [hep-ph]} \BibitemShut
  {NoStop}%
\bibitem [{\citenamefont {Andreev}\ \emph {et~al.}(2024)\citenamefont {Andreev}
  \emph {et~al.}}]{NA64:2024klw}%
  \BibitemOpen
  \bibfield  {author} {\bibinfo {author} {\bibfnamefont {Y.~M.}\ \bibnamefont
  {Andreev}} \emph {et~al.} (\bibinfo {collaboration} {NA64}),\ }\bibfield
  {title} {\bibinfo {title} {{First Results in the Search for Dark Sectors at
  NA64 with the CERN SPS High Energy Muon Beam}},\ }\href
  {https://doi.org/10.1103/PhysRevLett.132.211803} {\bibfield  {journal}
  {\bibinfo  {journal} {Phys. Rev. Lett.}\ }\textbf {\bibinfo {volume} {132}},\
  \bibinfo {pages} {211803} (\bibinfo {year} {2024})},\ \Eprint
  {https://arxiv.org/abs/2401.01708} {arXiv:2401.01708 [hep-ex]} \BibitemShut
  {NoStop}%
\bibitem [{\citenamefont {Escudero}\ \emph {et~al.}(2019)\citenamefont
  {Escudero}, \citenamefont {Hooper}, \citenamefont {Krnjaic},\ and\
  \citenamefont {Pierre}}]{Escudero:2019gzq}%
  \BibitemOpen
  \bibfield  {author} {\bibinfo {author} {\bibfnamefont {M.}~\bibnamefont
  {Escudero}}, \bibinfo {author} {\bibfnamefont {D.}~\bibnamefont {Hooper}},
  \bibinfo {author} {\bibfnamefont {G.}~\bibnamefont {Krnjaic}},\ and\ \bibinfo
  {author} {\bibfnamefont {M.}~\bibnamefont {Pierre}},\ }\bibfield  {title}
  {\bibinfo {title} {{Cosmology with A Very Light L$_{\mu}$ {\ensuremath{-}}
  L$_{\tau}$ Gauge Boson}},\ }\href {https://doi.org/10.1007/JHEP03(2019)071}
  {\bibfield  {journal} {\bibinfo  {journal} {JHEP}\ }\textbf {\bibinfo
  {volume} {03}},\ \bibinfo {pages} {071}},\ \Eprint
  {https://arxiv.org/abs/1901.02010} {arXiv:1901.02010 [hep-ph]} \BibitemShut
  {NoStop}%
\bibitem [{\citenamefont {Foldenauer}\ and\ \citenamefont
  {Hoefken~Zink}(2024)}]{Foldenauer:2024cdp}%
  \BibitemOpen
  \bibfield  {author} {\bibinfo {author} {\bibfnamefont {P.}~\bibnamefont
  {Foldenauer}}\ and\ \bibinfo {author} {\bibfnamefont {J.}~\bibnamefont
  {Hoefken~Zink}},\ }\bibfield  {title} {\bibinfo {title} {{How to rule out (g
  {\ensuremath{-}} 2)$_{\mu}$ in $ \textrm{U}{(1)}_{L_{\mu }-{L}_{\tau }} $
  with white dwarf cooling}},\ }\href {https://doi.org/10.1007/JHEP07(2024)096}
  {\bibfield  {journal} {\bibinfo  {journal} {JHEP}\ }\textbf {\bibinfo
  {volume} {07}},\ \bibinfo {pages} {096}},\ \Eprint
  {https://arxiv.org/abs/2405.00094} {arXiv:2405.00094 [hep-ph]} \BibitemShut
  {NoStop}%
\bibitem [{\citenamefont {Gninenko}\ and\ \citenamefont
  {Gorbunov}(2021)}]{Gninenko:2020xys}%
  \BibitemOpen
  \bibfield  {author} {\bibinfo {author} {\bibfnamefont {S.}~\bibnamefont
  {Gninenko}}\ and\ \bibinfo {author} {\bibfnamefont {D.}~\bibnamefont
  {Gorbunov}},\ }\bibfield  {title} {\bibinfo {title} {{Refining constraints
  from Borexino measurements on a light Z'-boson coupled to
  L{\ensuremath{\mu}}-L{\ensuremath{\tau}} current}},\ }\href
  {https://doi.org/10.1016/j.physletb.2021.136739} {\bibfield  {journal}
  {\bibinfo  {journal} {Phys. Lett. B}\ }\textbf {\bibinfo {volume} {823}},\
  \bibinfo {pages} {136739} (\bibinfo {year} {2021})},\ \Eprint
  {https://arxiv.org/abs/2007.16098} {arXiv:2007.16098 [hep-ph]} \BibitemShut
  {NoStop}%
\end{thebibliography}%

\end{document}